\documentclass[twocolumn,english,aps,pra,10pt,superscriptaddress,floatfix]{revtex4-2}
\usepackage{times}
\usepackage{xcolor}
\usepackage{amsmath}
\usepackage{amsfonts}
\usepackage{stmaryrd}
\usepackage{amssymb}
\usepackage{graphicx}
\usepackage{dcolumn}
\usepackage{multirow}
\usepackage{bm}

\usepackage{chemformula}
\usepackage{braket}
\usepackage[english]{babel}
\usepackage{url}
\usepackage[mode=text]{siunitx}
\DeclareSIUnit \dbc {dBc}
\usepackage{soul}

\DeclareSIUnit \dbm {dBm}
\DeclareSIQualifier\peak{p}
\usepackage[colorlinks=true,citecolor=blue,linkcolor=magenta]{hyperref}
\usepackage[capitalise]{cleveref} 
\usepackage{chemformula,xspace} 
\newcommand{\Fref}[1]{Figure~\ref{#1}}

\newcommand{\LN}[0]{\ch{LiNbO3}\xspace} 
\newcommand{\LT}[0]{\ch{LiTaO3}\xspace}

\begin{document}





\title{An integrated photonic millimeter-wave receiver with sub-ambient noise temperature}

\author{Junyin Zhang}
\thanks{These authors contributed equally.}
\affiliation{Institute of Physics, Swiss Federal Institute of Technology Lausanne (EPFL), CH-1015 Lausanne, Switzerland}
\affiliation{Institute of Electrical and Micro Engineering, EPFL, CH-1015 Lausanne, Switzerland}

\author{Shuhang Zheng}
\thanks{These authors contributed equally.}
\affiliation{Institute of Physics, Swiss Federal Institute of Technology Lausanne (EPFL), CH-1015 Lausanne, Switzerland}
\affiliation{Institute of Electrical and Micro Engineering, EPFL, CH-1015 Lausanne, Switzerland}

\author{Jiachen Cai}
\thanks{These authors contributed equally.}
\affiliation{Institute of Physics, Swiss Federal Institute of Technology Lausanne (EPFL), CH-1015 Lausanne, Switzerland}
\affiliation{Institute of Electrical and Micro Engineering, EPFL, CH-1015 Lausanne, Switzerland}
\affiliation{Shanghai Institute of Microsystem and Information Technology, Chinese Academy of Sciences, Shanghai, China}

\author{Connor Denney}
\affiliation{Department of Electrical Engineering, Colorado School of Mines, Golden, Colorado 80401, United States}

\author{Zihan Li}
\affiliation{Institute of Physics, Swiss Federal Institute of Technology Lausanne (EPFL), CH-1015 Lausanne, Switzerland}
\affiliation{Institute of Electrical and Micro Engineering, EPFL, CH-1015 Lausanne, Switzerland}

\author{Yichi Zhang}
\affiliation{Institute of Physics, Swiss Federal Institute of Technology Lausanne (EPFL), CH-1015 Lausanne, Switzerland}
\affiliation{Institute of Electrical and Micro Engineering, EPFL, CH-1015 Lausanne, Switzerland}

\author{Changhao Han}
\affiliation{Institute of Physics, Swiss Federal Institute of Technology Lausanne (EPFL), CH-1015 Lausanne, Switzerland}
\affiliation{Institute of Electrical and Micro Engineering, EPFL, CH-1015 Lausanne, Switzerland}

\author{Xin Ou}
\affiliation{Shanghai Institute of Microsystem and Information Technology, Chinese Academy of Sciences, Shanghai, China}

\author{Gabriel Santamaria-Botello}
\email{gabriel.santamariabotello@mines.edu}
\affiliation{Department of Electrical Engineering, Colorado School of Mines, Golden, Colorado 80401, United States}

\author{Tobias J. Kippenberg}
\email{tobias.kippenberg@epfl.ch}
\affiliation{Institute of Physics, Swiss Federal Institute of Technology Lausanne (EPFL), CH-1015 Lausanne, Switzerland}
\affiliation{Institute of Electrical and Micro Engineering, EPFL, CH-1015 Lausanne, Switzerland}

\maketitle

\textbf{\boldmath
Growing demands for timing precision in positioning systems, and resolution in atmospheric sensing and automotive radar, have extended receiver frontends into the millimetre-wave and THz regimes. At these frequencies, the noise performance of field-effect transistors used in low-noise amplifiers degrades rapidly due to parasitic effects, limited carrier mobility, hot electrons, and shot noise. Parametric transducers that couple electromagnetic signals to optical fields can offer quantum-limited sensitivity at room temperature. Electro-optic materials can enable receivers that transduce radiofrequency signals into optical phase shifts. Yet to date, room-temperature receivers using electro-optical detection have not attained a noise figure comparable to the electronic amplifiers. Here we demonstrate a room-temperature cavity electro-optic-based millimetre-wave receiver on a lithium tantalate photonic integrated circuit, achieving an on-chip photon-number transduction efficiency of 10\%. The equivalent input-referred receiver-added noise temperatures reach $110~\mathrm{K}$ on-chip and $202^{+35}_{-33}~\mathrm{K}$ for the full system at 60.5~GHz, with noise figure matching state-of-the-art V-band electronic low-noise amplifiers. We resolve the thermal noise in cavity electro-optic transducers, showing that the system is limited by thermal photon occupation ($\sim$100) in the millimetre-wave cavity and optical vacuum noise, with no measurable excess noise. Our work demonstrates that integrated photonics can reach state-of-the-art electronic receiver noise performance while offering high-radiofrequency-power survivability and compatibility with remote optical readout. This establishes cavity electro-optics as a low-noise, electromagnetic interference-resilient, chip-scale receiver frontend for millimetre-wave applications, enabling scalable analogue processing in the optical domain.
}

\noindent Detectors of radio frequency signals, i.e. RF receivers, are ubiquitous in science and technology, including radar, Earth observation, radio astronomy, navigation, and deep-space sensing
.  Driven by the need for higher resolution and timing precision, the frequency bands have continuously been increasing, and receiver frontends extended from the microwave to the millimeter and even sub-millimeter wave bands\cite{dang2020should, DeLima2021,rappaport2019wireless,ghelfiFullyPhotonicsbasedCoherent2014}.
At the heart of these microwave systems lies a low noise amplifier (LNA), a critical component that ensures weak signals can be reliably detected, decoded, and processed. The LNA fundamentally sets the sensitivity and dynamic range of the entire system; signals degraded or lost at this stage cannot, in general, be recovered by any level of subsequent processing.

\begin{figure*}[htbp]
	\centering
	\includegraphics[width = 1.0\textwidth]{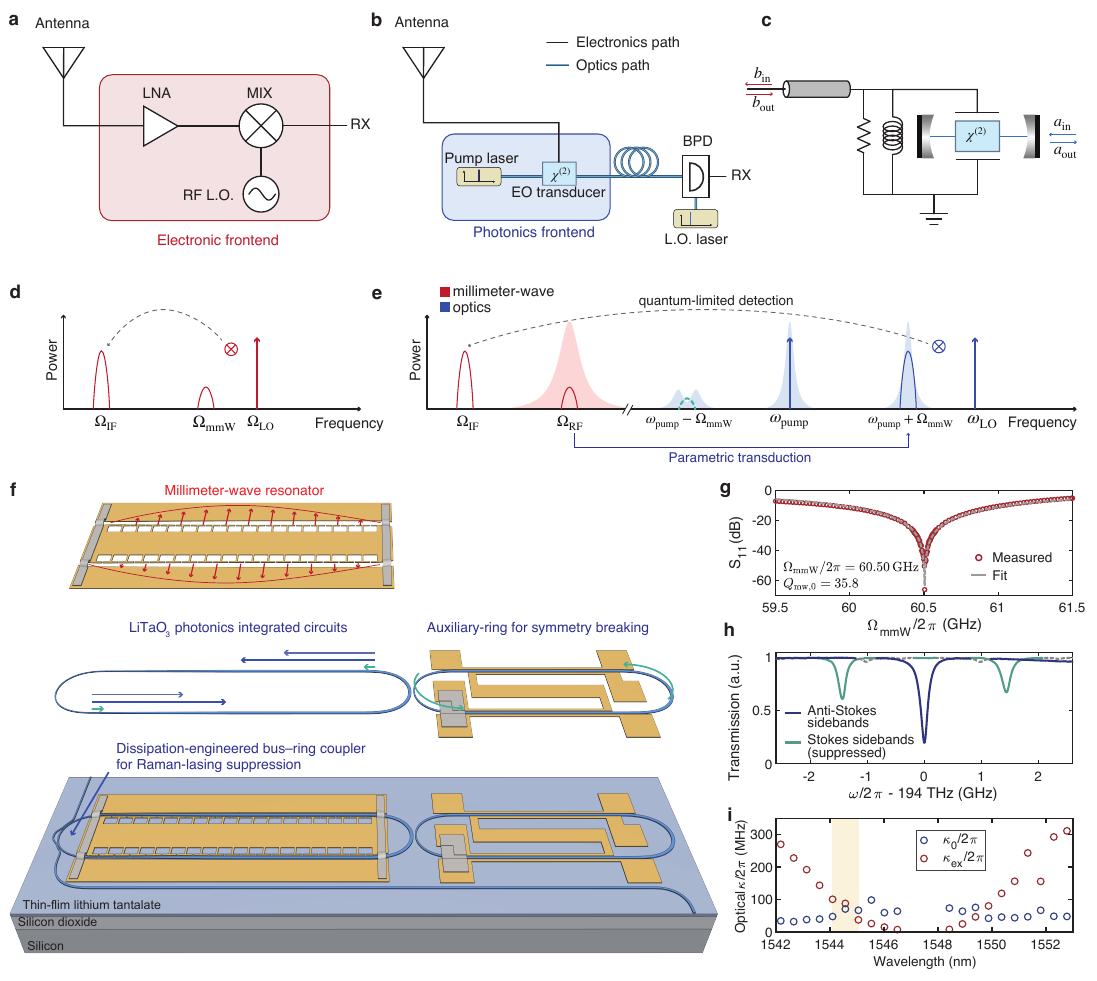}
	\caption{\textbf{Cavity electro-optics based mmWave-photonics interface.} 
		(a) Typical schematic of an electronic receiver. LNA: Low-noise amplifier; MIX: mixer; L.O.: local oscillator; RX: receive.
		(b) Operating principle of a photonic receiver. EO: electro-optics; BPD: Balanced photodiode.
		(c) Electro-optics transducer by room temperature cavity electro-optics. The \(\chi^{(2)}\) interaction enables bidirectional transduction between optical and mmWave modes, resonantly enhanced by cavity coupling. 
		(d,e) Spectrum diagram for the electronics and photonics receivers.
		(f)  Schematic of the hybrid integrated cavity electro-optic (EO) transducer, comprising a monolithically integrated $\lambda/2$ mmW microwave resonator operating in the fundamental mode, and a thin-film \LT photonic racetrack microresonator. An auxiliary ring is employed to suppress the Stokes sideband in the transduction process.
		(g) Measured reflection coefficient \(S_{11}\) of the mmWave resonator.
		(h) Measured optical transmission with and without sideband suppression; the calibration sidebands are shown in gray.
		(i) Fitted wavelength dependence of the external coupling rate \(\kappa_\mathrm{ex}\) and intrinsic loss rate \(\kappa_0\). The Mach--Zehnder-interferometer-type bus--ring coupler produces a strongly wavelength-dependent \(\kappa_\mathrm{ex}\), strongly overcoupling the cavity near the lithium-tantalate Raman-gain region around \(1600\,\mathrm{nm}\) and thereby suppressing Raman lasing.}
	\label{fig1}
\end{figure*}

\noindent \Fref{fig1}.(a) shows a typical receiver front end, where weak signals collected by the antenna are first amplified and then downconverted to an intermediate frequency. 
As the very first component in the receiver chain, the LNA sets the noise figure and thus the minimum detectable signal level for the entire system.
Unlike other microwave elements that can be shielded or isolated, the LNA connects directly to the antenna and cannot be effectively protected against strong electromagnetic inputs. 
Methods like shunt diode protection degrade high-frequency performance due to parasitic capacitance\cite{gong2002study, rudolph2007analysis}, leaving LNAs at microwave and millimeter-wave frequencies generally unprotected. 
Moreover, pushing LNAs to higher frequencies requires reducing the transistor gate length to maintain the necessary cut-off frequency, constrained by the transit time of charge carriers. 
This scaling increases shot and hot electron noise effects\cite{Esho2022, Gabritchidze2024}, 
degrading the noise figure and making devices more susceptible to electrical breakdown. These effects impose fundamental and practical limits on extending traditional receivers to higher frequencies\cite{pospieszalskiLimitsNoisePerformance2017,Pospieszalski2010}.

\noindent Addressing these limitations requires fundamentally rethinking the receiver front end. One approach is to use optical detection. It has long been recognized that parametric transducers that couple electromagnetic (or other signals, such as displacements) \cite{bagci2014optical,van2025optical} to optical fields can offer extraordinarily sensitive detection, that is quantum limited at room temperature \cite{bagci2014optical}. This scheme is in fact the basis of gravitational wave antenna (such as LIGO), and is the physical basis of quantum optomechanics\cite{aspelmeyerCavityOptomechanics2014a}. Similarly, receivers can be realized by transducing radio-frequency signals via electro-optical materials into optical phase shifts (\Fref{fig1}.(b)).
Pioneering works\cite{Ilchenko2003, hsuAlldielectricPhotonicassistedRadio2007,cohen2001microphotonic,hossein200614} have demonstrated such receivers using resonant electro-optical devices, where the optical pump, optical sidebands, and the microwave modes are resonantly enhanced (\Fref{fig1}.(c)). More recently such devices have received significant attention for cryogenic microwave to optical conversion\cite{warnerCoherentControlSuperconducting2025, arnoldAllopticalSuperconductingQubit2025,Xu2021,Holzgrafe20}. 
Traditional RF receivers down-convert the input millimeter-wave signal at frequency \( \Omega_{\mathrm{mmW}} \) by mixing it with a local oscillator at \( \Omega_{\mathrm{LO}} \), producing an intermediate frequency \( \Omega_{\mathrm{IF}} \) (see \Fref{fig1}(d)). In contrast, optical receivers based on a triply resonant electro-optic architecture\cite{tsangCavityQuantumElectrooptics2010b,javerzac-galyOnchipMicrowavetoopticalQuantum2016c} directly transduce the microwave signal into optical sidebands through an electro-optic interaction. A microwave carrier at frequency \( \Omega_{\mathrm{mmW}} \) modulates the optical pump at frequency \( \omega_{\mathrm{p}} \), generating optical sidebands at \( \omega_{\mathrm{p}} \pm \Omega_{\mathrm{mmW}} \). These sidebands are then detected using optical heterodyne detection, which recovers the signal at an intermediate frequency with quantum-limited added noise at room temperature, owing to the higher energy of optical photons relative to microwave photons.
Such optical `dielectric' receivers have unique advantages, including resilience to strong bursts of electromagnetic radiation (due e.g., solar bursts, lightnings, or electromagnetic warfare scenarios \cite{hsuAlldielectricPhotonicassistedRadio2007}), mitigation of RF saturation or signal distortion effects, large dynamic range, and chief among all, can theoretically operate at the quantum limit. 
Yet, despite these attractive properties, to date, a room temperature receiver using electro-optical detection has not attained a noise figure comparable to state of the art electronic amplifiers. The approach has been compounded by the insufficient coupling strength between the optical and RF fields\cite{Ilchenko2003,hsuAlldielectricPhotonicassistedRadio2007,cohen2001microphotonic,hossein200614}.
Here we overcome these challenges and demonstrate a photonics-based millimeter-wave receiver that employs a lithium tantalate (\LT) integrated photonics resonator, co-integrated with a millimeter wave cavity (\Fref{fig1}.(f)).
The cavity electro-optics system can be described by the interaction Hamiltonian \cite{aspelmeyerCavityOptomechanics2014a}:
\begin{equation*}
	\hat{H} = \hbar\omega_\mathrm{p} \hat{a}_\mathrm{p}^\dagger \hat{a}_\mathrm{p} + \hbar\omega_\mathrm{as} \hat{a}_\mathrm{as}^\dagger \hat{a}_\mathrm{as} + \hbar\Omega \hat{b}^\dagger \hat{b} + \hbar g_0 (\hat{a}_\mathrm{as}^\dagger \hat{a}_\mathrm{p} \hat{b} + \hat{a}_\mathrm{p}^\dagger \hat{a}_\mathrm{as} \hat{b}^\dagger),
\end{equation*}
where a strong optical pump field at frequency \(\omega_\mathrm{p}\) (mode \(\hat{a}_\mathrm{p}\)) mediates a parametric interaction between the anti-Stokes optical mode \(\hat{a}_\mathrm{as}\) and the microwave mode \(\hat{b}\), where  $g_0$ is the vacuum EO coupling rate, quantifying the frequency shift induced by a single microwave photon, and given by:
\begin{equation*}
	\begin{aligned}
		g_0 &= \frac{\varepsilon_0}{4} \sqrt{ \frac{\hbar  \omega_\mathrm{p}^2 \Omega}{W_{\mathrm{m}} W_{\mathrm{p}} W_{\mathrm{s}}} } \int_{\mathrm{LiTaO_3}} \chi^{(2)}_{\mathrm{ijk}} E_{\mathrm{p,i}} E^{*}_{\mathrm{s,j}} E_{\mathrm{m,k}} \, \mathrm{d}V\\
		&\propto \Omega,
	\end{aligned}\label{eq:g0}
\end{equation*}
where $\varepsilon_0$ is the vacuum permittivity and $\hbar$ is the reduced Planck's constant. The three-wave mixing process here involves a microwave field $\mathbf{E}_{\mathrm{m}}$ of frequency 
$\Omega$, mediating interaction between two optical modes ($\mathbf{E}_{\text{p,s}}$) such that $\omega_{\text{s}} = \omega_{\text{p}} \pm \Omega$. The mode fields are normalized to their respective energies using $W_{\text{m,p,s}} = \frac{\varepsilon_0}{2} \int \varepsilon_r(\mathbf{r},\omega_{\text{m,p,s}}) |\mathbf{E}_{\text{m,p,s}}(\mathbf{r})|^2 \, \mathrm{d}V$, where $\varepsilon_r(\mathbf{r},\omega_{\text{m,p,s}})$ is the space-dependent relative permittivity of the medium at the corresponding mode frequency, and $\chi^{(2)}_{ijk}$ refers to the Pockels susceptibility tensor.
Importantly, unlike electronic transistors, which degrade at higher frequencies, the cavity EO interaction exhibits an advantageous scaling ($g_0\propto \Omega$) when moving to millimeter-wave bands—benefiting from the increased zero point fluctuations of the electrical field from both reduced modal volumes and increased photon energy\cite{multani2026integrated,gaier2025wirelessmillimeterwaveelectroopticsfilm}. As a result, while electronic receivers become noisier with frequency, cavity EO transducers improve.
These opposing trends are expected to cross, opening the door for integrated photonics receivers to outperform electronics in the millimeter-wave regime.
Interestingly, owing to the enhanced vacuum EO coupling rate at millimeter-wave frequencies and the increased pump photon occupation allowed at room temperature, our device exceeds current state-of-the-art integrated cryogenic transducers in terms of photon conversion efficiency\cite{warnerCoherentControlSuperconducting2025,holzgrafeCavityElectroopticsThinfilm2020d,fanSuperconductingCavityElectrooptics2018b}, despite not using superconducting microwave resonators.

\subsection{Integrated room-temperature cavity electro-optic device and bidirectional transduction}
\begin{figure*}[htbp]
	\centering
	\includegraphics[width = 0.90\textwidth]{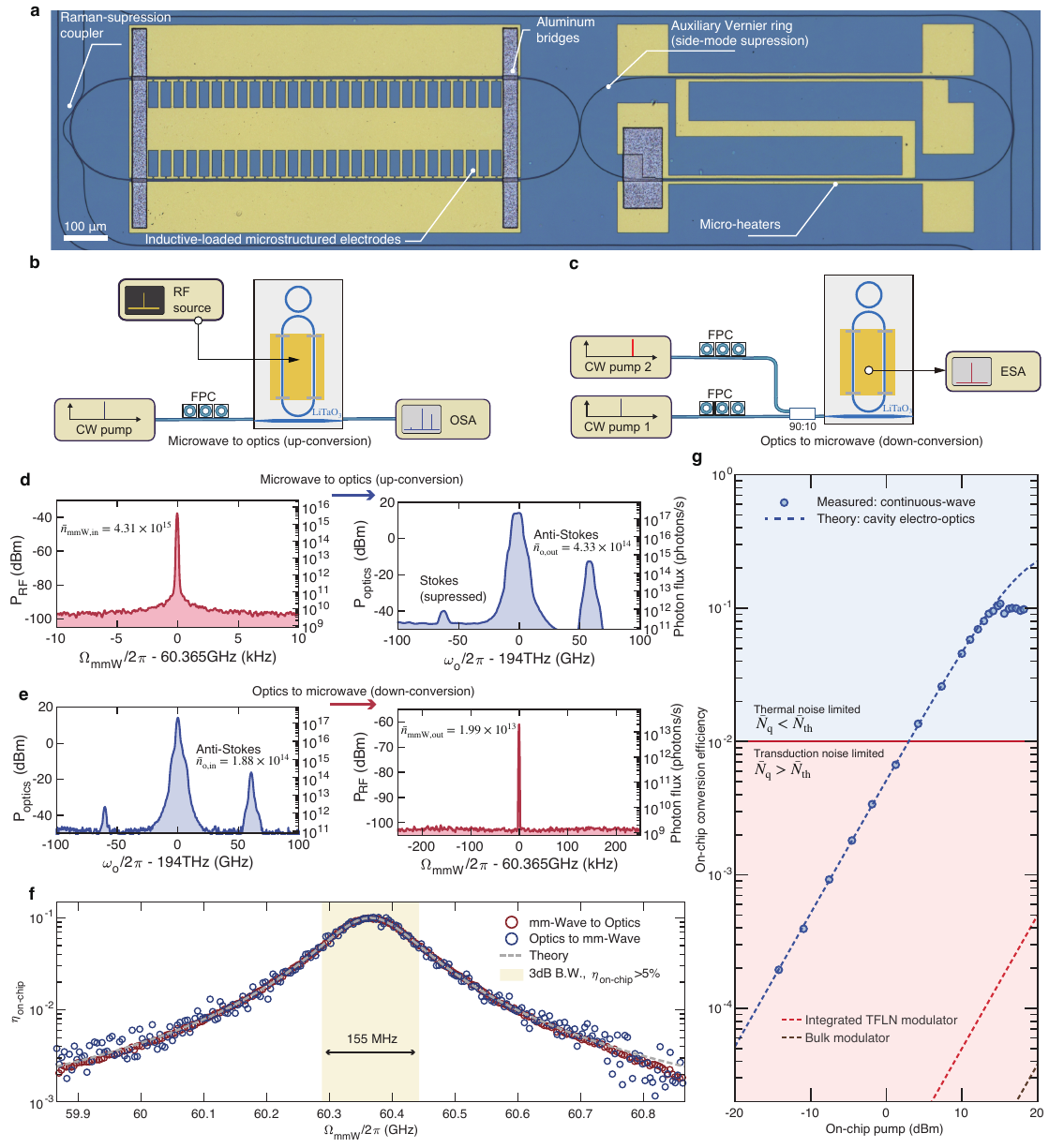}
	\caption{\textbf{Characterization of millimeter-wave-to-optical bidirectional transduction.} 
		(a) Microscope image of the fabricated cavity electro-optic device.
		(b) Microwave-to-optics (up-conversion) setup: An RF signal and continuous wave optical pump are coupled into the device; the optical output is measured with an optical spectrum analyzer (OSA).
		(c) Optics-to-microwave (down-conversion) setup. The generated mmW spectrum is measured by an electronic spectrum analyzer (ESA).
		(d) Measured up-conversion spectra of 60.365 GHz mmWave input and optical output.
		(e) Measured down-conversion spectra of optical input and mmWave output.
		(f) The measured conversion efficiency spectra for both down-conversion (optics to mmWave) and up-conversion (mmWave to optics) processes with an on-chip optical pump power of 26 mW (below the optical parametric oscillation threshold). The shaded region marks the 155 MHz 3-dB conversion bandwidth.
		(g) On-chip conversion efficiency for mmWave to optics in the continuous-pump regime using the Raman-suppressed device. At higher pump powers, the response is limited by Kerr-mediated optical parametric oscillation.
	}
	\label{fig2}
\end{figure*}

Our device integrates a monolithically fabricated microwave resonator---based on an inductively loaded coplanar waveguide (CPW), designed to match optical and microwave mode sizes---with a lithium tantalate (\LT) photonic racetrack resonator, as illustrated in Fig.~\ref{fig1}.(f). 
The resulting transduction efficiency corresponds to an effective half-wave voltage of $V_{\pi} = 100\,\mathrm{mV}$, which is orders of magnitude lower than commercial bulk lithium niobate modulators and thin-film lithium niobate modulators.
The microwave cavity is formed by short-circuited gold electrodes operating in the fundamental half-wave mode, with periodic loading of the coplanar waveguide to reduce microwave losses and achieve phase matching with the optical mode (\Fref{fig1}.(f)) \cite{zhangUltrabroadbandIntegratedElectrooptic2025a}.
The photonic layer incorporates a coupled-ring structure: the primary racetrack resonator facilitates the electro-optic transduction, while an auxiliary ring suppresses unwanted Stokes mode (\Fref{fig1}.(h)).
We incorporate a Mach--Zehnder-interferometer-type bus--ring coupling section to achieve a strongly wavelength-dependent external coupling rate \(\kappa_\mathrm{ex}\). \(\kappa_\mathrm{ex}/2\pi\) is approximately \(70\,\mathrm{MHz}\) for the pump and anti-Stokes modes (measured) and \(3.9\,\mathrm{GHz}\) at \(1605\,\mathrm{nm}\) (simulated), an approximately 50-fold increase that strongly overcouples the resonator in the lithium-tantalate Raman-gain band to suppress Raman lasing and support higher intracavity optical pump powers~\cite{song2024octave} (\Fref{fig1}.(i)).
Under strong pumping, the interaction is linearized to:
$\hat{H}_{\mathrm{eff}} = \hbar g_0 \langle \hat{a}_\mathrm{p} \rangle \hat{a}_\mathrm{as}^\dagger \hat{b} + \mathrm{h.c.},$
enabling coherent frequency conversion via a three-wave mixing process.
%
This coplanar cavity electro-optic platform ensures strong spatial mode overlap, yielding an enhanced vacuum coupling rate of \( g_0/2\pi = 6.48\pm0.04\,\mathrm{kHz} \) at 60.5~GHz (Supplementary Fig. S39).
For critically coupled optical and microwave modes, the on-chip photon-number conversion efficiency, including cavity electro-optic backaction, is \(\eta=\mathcal{C}/(1+\mathcal{C})^2\). Here, the cooperativity is defined as \(\mathcal{C}=4g_0^2n_p/(\kappa\Gamma)\).
Here, \(n_p = |\langle a_\mathrm{p} \rangle|^2\) is the intra-cavity pump photon number, and \(\kappa\), \(\Gamma\) are the total loss rates of the optical and microwave modes, respectively. Our device exhibits loaded microwave and optical linewidths on the order of $\Gamma/2\pi\approx 3.4\,\mathrm{GHz}$ (\Fref{fig1}(g)) and $\kappa/2\pi\approx 155\,\mathrm{MHz}$ (\Fref{fig1}(h)), respectively. Unlike cryogenic cavity EO schemes\cite{warnerCoherentControlSuperconducting2025,holzgrafeCavityElectroopticsThinfilm2020d,fanSuperconductingCavityElectrooptics2018b}, Ohmic losses in the microwave resonator dominate at room temperature ($\Gamma \gg \kappa$) placing the system in the reversed dissipation regime (RDR)\cite{nunnenkampQuantumLimitedAmplificationParametric2014, tothDissipativeQuantumReservoir2017}.
	\noindent \Fref{fig2}.(a) shows the fabricated device. 
	We first characterize the transduction efficiency and bandwidth by bidirectional transduction experiments (setup shown in \Fref{fig2}.(b,c)).
	We feed the CPW resonator with an external calibrated mmW source, and observe the generation of optical sidebands via EO modulation (\Fref{fig2}.(d)). 
	The power ratio between the optical anti-Stokes sideband and the mmW tone defines the photon number conversion efficiency $\eta$ (with normalization to the frequency difference $\omega/\Omega$). 
	Vice-versa, as a symmetric process, when applying two laser tones simultaneously inside the optical resonator, the microwave tone generated from the down-conversion process can be observed (\Fref{fig2}.(e)).
	Physically, down-conversion is a consequence of optical rectification of the beat note produced by the two optical modes, whose envelope induces a polarization vector oscillating at mmW frequencies that generates a detectable signal. 
	We performed coherent response measurement to determine the transduction bandwidth (Supplementary Figs. S47--S48).
	In both cases, the measured 155~MHz conversion bandwidth is determined by the optical linewidth as it corresponds to the lowest dissipation rate of this cavity electro-optics system in room temperature.
	As up- and down-conversion are symmetric, the conversion efficiency spectra overlap with each other within measurement error, as shown in \Fref{fig2}(f).
	The cooperativity can be enhanced by the photon occupation of the pump $n_\mathrm{p}$, and thus photon conversion efficiency scales proportionally with the pump power. 
	A calibrated efficiency measurement as a function of pump power shown in \Fref{fig2}.(g) agrees well with the theoretical prediction and reaches $\eta=10\%$ at 26.1~mW of on-chip pump power, compared with $\eta=2.5\%$ for the original device, whose maximum pump power was limited by Raman lasing (Supplementary Fig. S43).
	The revised device suppresses Raman lasing through wavelength-selective dissipation engineering; at higher pump powers, Kerr-mediated optical parametric oscillation rather than Raman lasing limits the pump photon occupation number and induces pump depletion once the oscillation threshold is exceeded (Supplementary Figs. S44--S46).
	The choice of using \LT over \LN is also motivated by its relatively higher Raman threshold\cite{wangLithiumTantalatePhotonic2024b}.
	The cavity electro-optic scheme achieves significantly higher photon-number conversion efficiency than conventional modulators used in radio-over-fiber and microwave photonics, with an exceptionally low equivalent half-wave voltage of $V_{\pi} = 100\,\mathrm{mV}$, compared to the Volt-level $V_{\pi}$ typical of integrated thin-film and bulk modulators.
	For photonic-enabled receiver schemes, the input-referred added noise contains thermal and quantum contributions. At room temperature, the microwave cavity has an equilibrium thermal occupation \(\bar n_{\mathrm{th}}=k_BT/(\hbar\Omega)\approx100\). For quantum-limited phase-preserving detection schemes such as heterodyne, the noise floor is one photon including the vacuum contribution, leading to an input-referred quantum noise of \( N_q^{\mathrm{in}} = {1}/{\eta} \), inversely proportional to the transduction efficiency $\eta$.
	As shown in \Fref{fig2}(g), the room-temperature cavity electro-optic system surpasses the efficiency threshold where the added quantum noise falls below the thermal noise level:
	\begin{equation*}
		N_q^{\mathrm{in}} < \bar n_{\mathrm{th}} .
	\end{equation*}
	This enables sub-ambient noise temperature detection through further radiative cooling of the microwave modes 
	\cite{Matsko2008, SantamaraBotello2018, Xu2020} (Supplementary Section 3.5 and Supplementary Figs. S16 and S18).
	
	\subsection{Direct observation of Johnson noise in mmWave cavities}
\begin{figure*}[htbp]
	\centering
	\includegraphics[width = 1.0\textwidth]{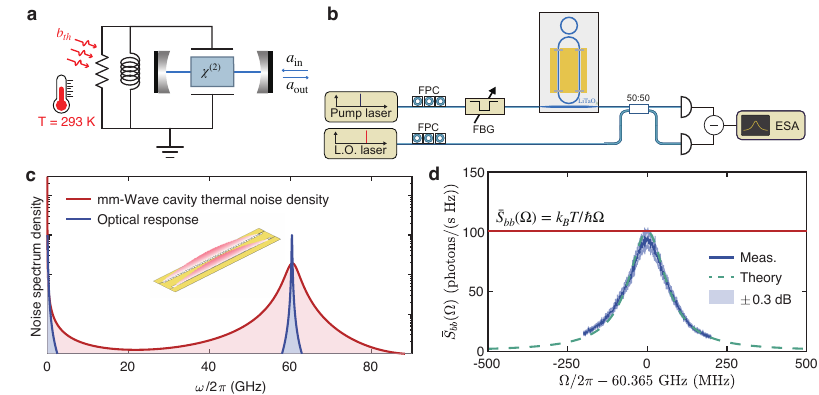}
	\caption{\textbf{{Probing millimeter-wave Johnson noise with cavity electro-optic transducer.}} 
		(a) The thermally induced motion of electrons in the microwave cavity can be transduced into the optical domain by the electro-optic effect in the cavity EO system.
		(b) Experiment setup for weak optical noise characterization: quantum-limited heterodyne measurement. FBG: fiber-Bragg-grating; FPC: fiber-polarization controller; ESA: electronic spectrum analyzer.
        (c) Illustration of the millimeter-wave cavity thermal-noise spectrum and the optical response near the 60 GHz mode. In the reversed dissipation regime ($\Gamma >\kappa$), the measured noise spectrum is shaped by the optical response. The inset shows the numerically simulated voltage distribution along the coplanar waveguide resonator.
        (d) Input-referred noise for the 60.365 GHz mode matches the predicted thermal occupation \(k_B T / \hbar \Omega\) at room temperature. The shaded band indicates a \(\pm0.3\,\mathrm{dB}\) measurement uncertainty.
	}
	\label{fig3}
\end{figure*}

	To further demonstrate the utility of efficient transduction and directly analyze the noise processes involved, we performed a calibrated transduction noise measurement. 
	As shown in \Fref{fig3}(a), at thermal equilibrium, 
	each mmW mode exhibits a photon occupation \( n_{\mathrm{cav}} = k_B T / \hbar \Omega \), where \( T \) is the physical temperature. 
	\Fref{fig3}(b) shows a simplified optical heterodyne setup for characterizing transduction noise at quantum-limited sensitivity.
	In the reversed dissipation regime, this thermal noise is transduced with efficiency $\eta$ and spectrally shaped by the response of the optical cavity (\Fref{fig3}.(c)). 
	Thermal noise has previously been obscured in cryogenic implementations due to the limited transduction efficiency and parasitic noise sources such as optically induced microwave noise and superconducting instabilities
	\cite{warnerCoherentControlSuperconducting2025, xuLightinducedMicrowaveNoise2024}. 
	In contrast, our system achieves a transduction efficiency allowing the transduced thermal noise to exceed the optical vacuum fluctuation noise floor of heterodyne detection, thereby enabling its effective direct resolution. 
	The calibrated input-referred noise spectrum at \(60.365\,\mathrm{GHz}\) is shown in Fig.~\ref{fig3}(d), along with the theoretical prediction for thermal noise in the microwave cavity at room temperature, analogous to thermal noise in optomechanical systems\cite{saulsonThermalNoiseMechanical1990a, aspelmeyerCavityOptomechanics2014a}.
	These measurements directly resolve thermal noise in the millimetre-wave cavity through electro-optic transduction.
	The results confirm that the transduction process is fundamentally limited by microwave thermal fluctuations and optical vacuum noise, with no measurable excess noise from parasitic channels within the \(\pm0.3\,\mathrm{dB}\) measurement uncertainty.

	\subsection{Demonstration of ultra-sensitive millimeter-wave signal detection}
\begin{figure*}[htbp]
	\centering
	\includegraphics[width = 0.97\textwidth]{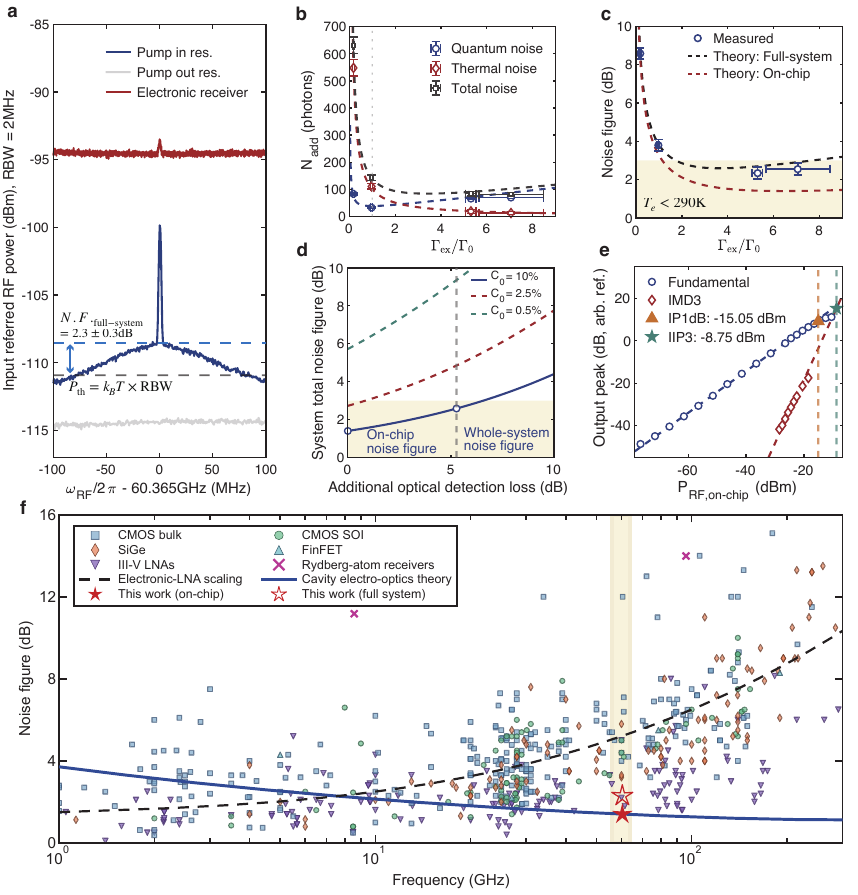}
	\caption{\textbf{Ultra-sensitive millimeter-wave signal detection.} 
		(a) Detection of calibrated \(-100\,\mathrm{dBm}\) input tones at \(60.365\,\mathrm{GHz}\) with the photonic receiver and at \(60.0\,\mathrm{GHz}\) with a commercial electronic receiver (R\&S FSW67 with FSW-B24 LNA option), using a \(2\,\mathrm{MHz}\) resolution bandwidth (RBW). For visual comparison, the electronic trace is shifted horizontally to align its carrier with the photonic trace; the measured powers are unchanged. The peak-bin power includes both the coherent signal and the in-band receiver noise.
		(b) Quantum, thermal, and total added noise as a function of the normalized microwave external coupling \(\Gamma_\mathrm{ex}/\Gamma_0\). Markers show the experimental values and dashed curves the theoretical predictions.
		(c) Measured full-system noise figure and theoretical full-system and on-chip noise figures as functions of the normalized microwave external coupling \(\Gamma_\mathrm{ex}/\Gamma_0\). Error bars represent fitting and calibration uncertainties (Supplementary Sections 3.3 and 3.5).
		(d) Calculated full-system noise figure versus additional optical detection loss for representative critical-coupling cooperativities \(C_0\). The vertical dashed line indicates the measured optical detection loss. The shaded regions in (c,d) correspond to an equivalent input-referred added-noise temperature \(T_e<290\,\mathrm{K}\).
		(e) Linearity and input-referred 1\,dB compression point (IP1dB) and third-order intercept point (IIP3) for the photonic receiver scheme.
		(f) Theoretical on-chip noise figure of the cavity electro-optic front-end as a function of frequency based on the measured device parameters, together with the inferred on-chip result and the directly measured full-system result at \(60.365\,\mathrm{GHz}\), compared with state-of-the-art on-chip electronic low-noise amplifiers and Rydberg-atom receivers reported in the literature.}
\label{fig4}
\end{figure*}

	The absence of excess noise processes, combined with efficient transduction, enables sensitive detection of weak mmW signals. 
	At room temperature, the total noise is dominated by two contributions: the up-converted thermal noise introduced by the microwave cavity due to its physical temperature and the nearly quantum-limited optical shot noise. 
	At high-offset frequencies ($60.5\pm0.3\,\mathrm{GHz}$ from the optical pump in the present measurement), the intrinsic refractive noise from the ferroelectric cavity can be omitted \cite{zhangFundamentalChargeNoise2025a}.
	Figure~\ref{fig4}(a) compares the photonic detection of a calibrated $-100\,\mathrm{dBm}$ mmWave tone at $60.365\,\mathrm{GHz}$ with an electronic-receiver measurement of the same calibrated input power at $60.0\,\mathrm{GHz}$, both using a $2\ \mathrm{MHz}$ resolution bandwidth. For visual comparison, the electronic trace is shifted horizontally to align its carrier with the photonic trace; the measured powers are unchanged. At the carrier frequency, each trace reports the total input-referred power within the $2\,\mathrm{MHz}$ resolution bandwidth, including both the coherent tone and the in-band receiver noise. Consequently, the higher noise floor of the electronic receiver produces a larger apparent peak, even though both measurements use the same calibrated input power.
	Our system achieves a $14\ \mathrm{dB}$ lower input-referred noise floor than the electronic receiver and near thermal-noise-limited sensitivity.
	Figure~\ref{fig4}(b) shows the quantum contribution and the thermal contribution from intrinsic microwave dissipation to the input-referred added noise, $N_{\mathrm{add}}^{\mathrm{in}} = N_q^{\mathrm{in}} + N_{\mathrm{th}}^{\mathrm{in}}$, and their sum as functions of the normalized microwave external coupling $\Gamma_{\mathrm{ex}}/\Gamma_0$. Increasing the external coupling suppresses the thermal contribution, while reducing the conversion efficiency and increasing the input-referred heterodyne quantum noise.
	For a standard thermal input at $T_0=290\,\mathrm{K}$ with occupation $\bar{n}_{\mathrm{th}} = k_B T_0/(\hbar\Omega)$, the receiver noise figure is defined by
	\begin{equation*}
		\mathrm{NF}_{\mathrm{sys}}^{(\mathrm{phot})}
		= 10\log_{10}\!\left(1+\frac{N_{\mathrm{add}}^{\mathrm{in}}}{\bar{n}_{\mathrm{th}}}\right).
	\end{equation*}
	Here, $N_q^{\mathrm{in}} = 1/\eta_{\mathrm{sys}}$ and $N_{\mathrm{th}}^{\mathrm{in}} = (\Gamma_0/\Gamma_{\mathrm{ex}})\bar{n}_{\mathrm{th}}$ are the input-referred quantum and thermal noise contributions, respectively. The overall system efficiency $\eta_{\mathrm{sys}}$ accounts for both the on-chip transduction efficiency and the optical detection-chain loss, and is given by $\eta_{\mathrm{sys}} = \eta_{\mathrm{on\text{-}chip}} 10^{-L_{\mathrm{sys}}^{(\mathrm{opt})}/10}$, where $\eta_{\mathrm{on\text{-}chip}}$ is the coupling-dependent on-chip photon-number conversion efficiency and $L_{\mathrm{sys}}^{(\mathrm{opt})}$ is the equivalent optical detection-chain penalty in dB. The rates $\Gamma_{\mathrm{ex}}$ and $\Gamma_0$ denote the external and intrinsic microwave loss rates, respectively. The resulting noise figure is shown in Fig.~\ref{fig4}(c).
	At the optimal microwave coupling, the inferred on-chip noise figure is $1.4\,\mathrm{dB}$ at $60.5\,\mathrm{GHz}$, corresponding to an equivalent input noise temperature of $110\,\mathrm{K}$. Including the measured $5.27\,\mathrm{dB}$ optical detection-chain penalty increases the calculated full-system noise figure to $2.58\,\mathrm{dB}$ (\Fref{fig4}.(d)), consistent with the directly measured value of $2.3\pm0.3\,\mathrm{dB}$ (\Fref{fig4}.(c)). Such external detection losses can be reduced by employing co-integrated photodiodes.
	The transduction from the input mmWave signal to the optical sideband remains highly linear up to an input power of \(-15.1\,\mathrm{dBm}\) (\Fref{fig4}.(e)), corresponding to the 1\,dB input compression point (IP1dB). A calibrated two-tone measurement gives an on-chip input third-order intercept point (IIP3) of \(-8.8\,\mathrm{dBm}\) (\Fref{fig4}.(e)). Beyond this point, optical pump depletion and electro-optic comb formation introduce nonlinearity and saturation\cite{zhangUltrabroadbandIntegratedElectrooptic2025a}. This corresponds to an input-referred dynamic range of 156.7\,dB, normalized to a 1\,Hz bandwidth.
	\Fref{fig4}.(f) compares the theoretically predicted noise figure of the photonic receiver, based on experimental parameters, with the inferred on-chip data point and state-of-the-art electronic LNAs reported in the literature\cite{belostotskiLNA,wang2025lna} and also Rydberg-atom based receivers \cite{legaie2024millimeter}.
	Owing to the favorable frequency scaling of electro-optic transduction, 
	the proposed photonic receiver is expected to exhibit an increasingly pronounced noise-figure advantage toward higher-frequency mmWave and THz bands; already at 60.5~GHz, its 1.4~dB on-chip noise figure is, to the best of our knowledge, below those of the best reported room-temperature electronic LNAs.
	The room-temperature cavity electro-optic platform demonstrated here functions not only as a coherent transducer but also as a fully-photonic low-noise front-end that could replace 
	electronic pre-selectors, low-noise amplifiers and subsequent mixer chains in narrowband applications. 

\begin{figure*}[htbp]
	\centering
	\includegraphics[width = 0.8\textwidth]{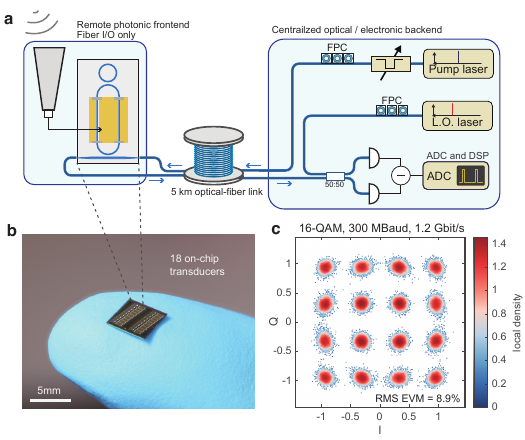}
	\caption{\textbf{Demonstration of wireless 16-state quadrature amplitude modulation (16-QAM) signal reception with the cavity electro-optic receiver.}
		(a) Experimental receiver configuration for a 60\,GHz wireless link. The remote antenna-side photonic frontend contains the cavity electro-optic transducer and fiber I/O only, while the pump and local-oscillator lasers, coherent detector, analog-to-digital converter (ADC), and digital signal processor (DSP) are located in a centralized optical/electronic backend connected through a 5\,km optical-fiber link. The received mmWave signal is transduced into the optical domain at the frontend and coherently detected at the backend.
		(b) Photograph of the photonic receiver chip on a gloved fingertip. The chip integrates 18 individual transducers. Scale bar, 5\,mm.
		(c) Recovered 16-QAM constellation at 300\,MBaud (1.2\,Gbit\,s$^{-1}$), with a measured root-mean-square error vector magnitude (RMS EVM) of 8.9\%, below the equivalent hard-decision forward-error-correction (HD-FEC) threshold.}
	\label{fig5}
\end{figure*}

	This is demonstrated in the proof-of-concept 60~GHz wireless photonic receiver experiment shown in \Fref{fig5}(a).
	A unique advantage of the photonic receiver scheme is that the antenna-side frontend, containing only the cavity electro-optic transducer and fiber I/O, is separated from the centralized optical/electronic backend.
	The transducer chip itself also has a small footprint and integrates 18 individual transducers (\Fref{fig5}(b)).
	In this proof-of-concept demonstration, the receiver recovers a 300~MBaud (1.2~Gbit/s) 16-state quadrature amplitude modulation (16-QAM) signal with a root-mean-square error vector magnitude (RMS EVM) of 8.9\%, below the hard-decision forward error correction (HD-FEC) threshold (\Fref{fig5}(c)).

	\subsection{Conclusion and outlook}
	Our work demonstrates that  \LT electro-optical photonic integrated circuits co-integrated with millimeter wave cavities can operate as millimeter wave low-noise frontends, offering high dynamic range, an EMI-resilient architecture, and, above all, a noise figure that matches state-of-the-art room-temperature electronic low-noise amplifiers.
	Improvements in transduction efficiency toward unity cooperativity could be achieved through further suppression of residual optical nonlinearities, including Kerr-mediated optical parametric oscillation, enhancement of the microwave resonator quality factor, and operation at higher carrier frequencies.
	While cavity EO systems inherently trade instantaneous bandwidth for enhanced efficiency and low-noise performance, the present device has a measured 155 MHz 3-dB transduction bandwidth.
	By combining multiple resonators through wavelength-division multiplexing, the architecture could process broader bandwidths in parallel (e.g., 7 and 13 channels for 1 and 2~GHz of contiguous coverage, respectively) while retaining the efficiency of each resonant channel.
	
	\noindent Moreover, electro-optic transduction introduces a fundamentally different detection paradigm, distinct from conventional schemes based on photon absorption. Its parametric nature enables rethinking core limits in electromagnetic sensing—for example, it could be configured as an electrically-small field sensor not subject to the
	Chu-Wheeler-Harrington bound for small antennas\cite{chuPhysicalLimitationsOmniDirectional1948,harringtonEffectAntennaSize1960,PhysRevLett.121.110502}. The favorable frequency scaling of cavity electro-optic systems also offers a promising pathway for interconnecting high-frequency/ mmWave qubits \cite{anferov2025millimeter}, paving the way for heterogeneous quantum systems.
	With further increase in cooperativity, cavity electro-optic systems are expected to access regimes of stronger dynamical back-action and parametric instability\cite{tsangCavityQuantumElectrooptics2010b}, enabling parametric amplification/squeezing\cite{safavi-naeiniSqueezedLightSilicon2013}, tunable slow-light effects\cite{weisOptomechanicallyInducedTransparency2010, safavi-naeiniElectromagneticallyInducedTransparency2011}, direct optical control of microwave signals\cite{qiu2023coherent}, and access to a millimeter-wave maser\cite{tsangCavityQuantumElectrooptics2010b}. These capabilities may also enable the exploration of new physical phenomena in the reversed dissipation regime at cooperativity greater than unity, where a quantum-limited optical mode governs the dynamics of a thermally populated microwave field, even under room-temperature conditions.
	Viewed more broadly, these results point to a crossover where photonic technologies could reach higher frequencies with lower noise than conventional electronics.

\section*{Acknowledgments:}
We thank Wil Kao and Terence Blésin for their valuable discussions, and Xinru Ji for assistance during the fabrication process.
The samples were fabricated in the EPFL Center of MicroNanoTechnology (CMi) and the Institute of Physics (IPHYS) cleanroom.
The LTOI wafers were fabricated in Shanghai Novel Si Integration Technology (NSIT) and the SIMIT-CAS.

\section*{Funding Statement}
This work has received funding from Swiss National Science Foundation grant no. 216493 (HEROIC) and by the EU Horizon Europe research and innovation programme under grant No. 101187515 (ELLIPTIC), with further support from the Swiss State Secretariat for Education, Research and Innovation (SERI).

\section*{Author contributions:}
J.Z., G.S.B. and T.J.K. conceived the concept and experiments.
J.Z. designed the photonics structures, C.D. and G.S.B. designed the microwave structures.
J.C. and X.O. prepared the LTOI substrates.
J.C. and J.Z. fabricated the devices with help from S.Z. and Y.Z.
Z.L. developed the etching hard-mask deposition recipe.
J.Z., S.Z., J.C. and C.H. performed the measurements.
J.Z., S.Z., J.C., C.H. and G.S.B. analyzed the data.
J.Z., C.D., S.Z., J.C., G.S.B. and T.J.K. wrote the manuscript with contributions from all authors.
G.S.B. and T.J.K. supervised the project.

\section*{Competing interests:}
 The authors declare no competing financial interests.

\section*{Methods}

\subsection{Device fabrication:} 
Devices were fabricated on x-cut thin-film lithium tantalate-on-insulator (LTOI) wafers supplied by SIMIT, with a nominal lithium tantalate device-layer thickness of 600~nm.
The \LT waveguides are patterned using deep-UV steppers and etched with an ion-beam etcher. Diamond-like carbon serves as the etching hardmask. The total etching depth is 400~nm, leaving a 200~nm-thick slab. After the ion-beam etching, redeposition is removed by chemical cleaning with a mixture of 40\% \ch{KOH} and 30\% H\ch{H2O2} (volume ratio 3:1) at 80$^\circ$C for 10~minutes.
The electrode layer comprises 800~nm Au with a 20~nm Al adhesion layer. Metal bridges are fabricated using 800~nm thick angle-deposited Al in an electron-beam evaporator with a 20~nm Ti adhesion layer. The device is oxide-cladded. Additional fabrication details are provided in Supplementary Section 6.

\subsection{The millimeter-wave resonator and the tunable external coupling:}
We employed a coplanar waveguide resonator in ground-signal-ground configuration as our millimeter-wave resonator, following the approach in Ref.~\cite{zhangUltrabroadbandIntegratedElectrooptic2025a}. Air bridges provide short-circuit termination at both ends of the resonator. The resonator is designed to support a fundamental half-wave mode near 60.5~GHz, which is used throughout this work.
Since voltage reaches a minimum at both termination ends while current density achieves a maximum, the effective input impedance of the microwave cavity can be tuned by adjusting the probe landing position, thereby controlling the external coupling rate. We monitor the microwave external coupling rate through $S_{11}$ measurements. For bidirectional conversion measurements, we tune the external coupling rate to achieve critical coupling. Overcoupling is obtained by further adjusting probe positions for noise figure characterization and radiative cooling measurements (Supplementary Sections 5.1 and 3.5, and Supplementary Figs. S17--S18).

\subsection{Bidirectional transduction efficiency measurement:}
We characterized the photon-number conversion efficiency by both the up-conversion process and the down-conversion process. In the up-conversion measurements, we applied a calibrated microwave tone to the chip and measured the output optical spectrum using the optical spectrum analyzer (YOKOGAWA AQ6370D). The microwave tone is generated by a vector-network analyzer (VNA) working in the continuous-wave mode (Rohde \& Schwarz ZNA67). To obtain the absolute microwave power on-chip, we calibrated the microwave power by an RF power-meter (Rohde \& Schwarz NRP67TN) for power above -30dBm, to obtain power below -30dBm,  we use fixed attenuators. The RF cable losses, RF probe losses, and the attenuation values are measured by the same vector-network analyzer. In order to have on-chip referred optical power, we measured a fiber-to-fiber transmission of 3.7dB, reproducible in multiple chips, and as the input and output facets are symmetrical, we assume a 1.85dB loss per facet. In the down-conversion measurements, two separate lasers (Toptica CTL) are tuned into resonances (pump mode and the anti-Stokes sidebands), and the generated microwave is measured by an electronic spectrum analyzer (Rohde \& Schwarz FSW67) with a low noise amplifier option (FSW-B24). For the transduction spectra measurements, the signal detection is also done with a vector network analyzer. In the up-conversion spectra measurement, a fast-photodiode is used to transduce the optical sidebands to an electronic signal and as input for the  VNA for $S_{\mathrm{OE}}$ measurement. In the down-conversion spectrum measurement, a commercial electro-optics phase modulator is directly driven by the VNA output port to generate phase-locked optical sidebands, and the down-converted microwave signal is directly detected by the VNA to measure $S_{\mathrm{EO}}$. In the down-conversion measurement, an erbium-doped fiber amplifier (EDFA) is used to boost the sideband power generated by the phase modulator. The detailed schematics for the transduction measurements are shown in Supplementary Figs. S47--S48.

\subsection{Noise measurement and small signal detection:}
We use optical heterodyne to measure the noises in the transduction process and also for the demonstration of small -100dBm mmWave signal detection. 
We use a 600MHz bandwidth InGaAs balanced photodiode to do the detection (Wieserlabs, WL-BPD600MA). We use a total local oscillator laser power of 18mW incident on the two balanced-photodiode ports to ensure the system is shot-noise limited. A fiber Bragg grating (FBG) is placed before the chip to suppress input amplified spontaneous emission (ASE), while the near-critically coupled pump resonance provides on-chip pump rejection, eliminating the need for a post-chip FBG. The measured residual-pump and ASE penalty is included in the reported full-system noise figure. To calibrate the response of the measurement chain, we applied a calibrated microwave signal to the chip as a calibration tone. The microwave signal is generated by a vector-network analyzer (VNA) working in the continuous-wave mode (Rohde \& Schwarz ZNA67). As the noise floor of the VNA output is above room temperature, we use a $\sim$70dB attenuator chain to thermalize the microwave tone to room temperature.  The attenuator loss, cable losses, and the probe losses are calibrated by the same VNA. The detailed heterodyne setup and the calibration procedure are shown in Supplementary Figs. S49--S50.

\section*{Data Availability Statement} The simulation files and original data used to produce the plots in this work will be available at Zenodo upon publication of the preprint (\href{https://doi.org/10.5281/zenodo.16748426}{https://doi.org/10.5281/zenodo.16748426}).

\section*{Code Availability Statement} The processing codes  used to produce the plots in this work will be available at Zenodo upon publication of the preprint (\href{https://doi.org/10.5281/zenodo.16748426}{https://doi.org/10.5281/zenodo.16748426}).

\section*{Materials \& Correspondence:}
Correspondence and requests for materials should be addressed to G.S.B., or T.J.K.

\section*{Supplementary materials}
Materials and Methods\\
Supplementary theory and data\\
Figs. S1 to S52\\
Tables I to VII\\
References \textit{(1-59)}


\bibliography{refs_no_url}
\bibliographystyle{apsrev4-2}

\end{document}


\renewcommand{\thefigure}{S\arabic{figure}}
\renewcommand{\thetable}{\Roman{table}}

\begin{center}
  \vspace*{0.15in}
  {\LARGE Supplementary information:
  \textnormal{\textit{An integrated photonic millimeter-wave receiver with sub-ambient noise temperature}}\par}

  \vspace{1.3em}
  {\normalsize
  Junyin Zhang\textsuperscript{1,2,*},
  Shuhang Zheng\textsuperscript{1,2,*},
  Jiachen Cai\textsuperscript{1,2,3,*},
  Connor Denney\textsuperscript{4},\\[0.25em]
  Zihan Li\textsuperscript{1,2},
  Yichi Zhang\textsuperscript{1,2},
  Changhao Han\textsuperscript{1,2},
  Xin Ou\textsuperscript{3},\\[0.25em]
  Gabriel Santamaria-Botello\textsuperscript{4,\ensuremath{\dagger}},
  Tobias J. Kippenberg\textsuperscript{1,2,\ensuremath{\ddagger}}\par}

  \vspace{1.1em}
  \begin{minipage}{0.88\textwidth}
    \footnotesize
    \setlength{\parindent}{0pt}
    \setlength{\parskip}{0.25em}
    \hangindent=1.2em\hangafter=1\textsuperscript{1}Institute of Physics, Swiss Federal Institute of Technology Lausanne (EPFL), CH-1015 Lausanne, Switzerland\par
    \hangindent=1.2em\hangafter=1\textsuperscript{2}Institute of Electrical and Micro Engineering, EPFL, CH-1015 Lausanne, Switzerland\par
    \hangindent=1.2em\hangafter=1\textsuperscript{3}Shanghai Institute of Microsystem and Information Technology, Chinese Academy of Sciences, Shanghai, China\par
    \hangindent=1.2em\hangafter=1\textsuperscript{4}Department of Electrical Engineering, Colorado School of Mines, Golden, Colorado 80401, United States\par
  \end{minipage}

  \vspace{0.8em}
  {\scriptsize
  \textsuperscript{*}These authors contributed equally.\par
  \textsuperscript{\ensuremath{\dagger}}\href{mailto:gabriel.santamariabotello@mines.edu}{gabriel.santamariabotello@mines.edu}
  \quad
  \textsuperscript{\ensuremath{\ddagger}}\href{mailto:tobias.kippenberg@epfl.ch}{tobias.kippenberg@epfl.ch}\par}
\end{center}

\vspace{1.4em}
\compacttableofcontents
\newpage

\section{Comparison with the state of the art}

In this section, we present an overview of the conversion efficiencies and bandwidths of cavity electro-optic transducers, together with a comparison of noise figures for microwave photonics systems, electronic low-noise amplifiers, and Rydberg atom receivers, to the best of our knowledge.
\subsection{Cavity electro-optic transduction efficiency and bandwidth}

\begin{figure}[htbp]
	\centering
	\includegraphics[width=0.9\linewidth]{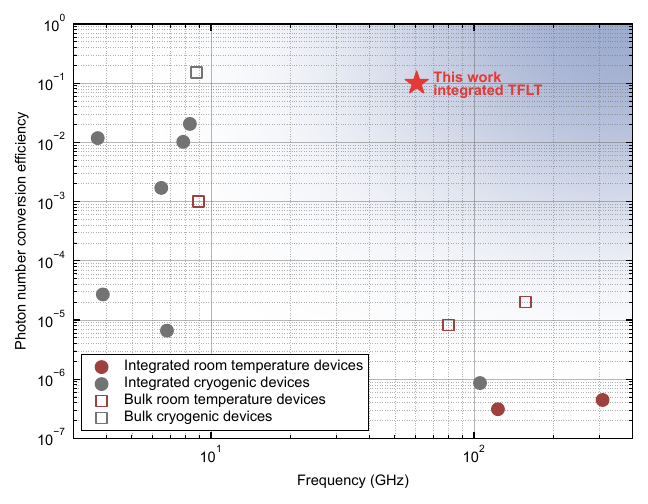}
	\caption{Photon number conversion efficiency of cavity electro-optic transducers. Grey markers indicate cryogenic devices using superconducting microwave resonators~\cite{fanSuperconductingCavityElectrooptics2018b,Holzgrafe20,McKenna_20,Xu2021,PhysRevApplied.18.064045,warnerCoherentControlSuperconducting2025,multani2026integrated}. Red markers represent room-temperature devices~\cite{gaier2025wirelessmillimeterwaveelectroopticsfilm,Rueda2016,Matsko2008,santamaria2018sensitivity,Suresh2025}. Filled circles denote integrated devices, while unfilled squares correspond to bulk devices based on whispering-gallery mode (WGM) resonators~\cite{sahu2022quantum,Rueda2016,strekalov2009efficient,santamaria2018sensitivity,Suresh2025}.}
	\label{fig:SI_comparison}
\end{figure}

\Fref{fig:SI_comparison} compares reported photon-number conversion efficiencies. Table~\ref{tab:SI_quality_factor_comparison} reports the corresponding microwave and optical quality factors, conversion efficiencies, and bandwidths for the representative devices. Quality factors are specified as intrinsic or loaded and are included only when reported or calculated directly from a reported cavity linewidth.

\begin{figure}[htbp]
  \centering
  \includegraphics[width=0.75\linewidth]{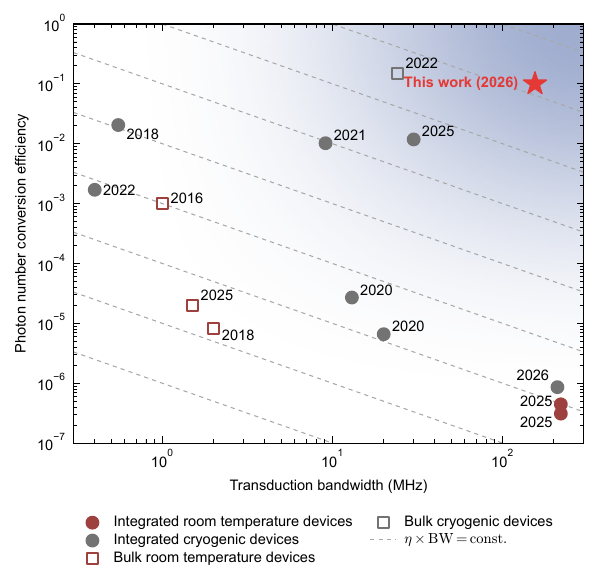}
  \caption{\textbf{Bandwidth--efficiency comparison of cavity electro-optic transducers.} Reported photon-number conversion efficiency versus transduction bandwidth for representative integrated and bulk devices operated at room and cryogenic temperatures. Filled and open symbols distinguish integrated and bulk implementations, while red and gray distinguish room-temperature and cryogenic operation. Dashed lines indicate constant efficiency--bandwidth product $\eta\times\mathrm{BW}$.}
  \label{fig:SI_bandwidth_efficiency}
\end{figure}

\begin{table}[htbp]
\centering
\scriptsize
\setlength{\tabcolsep}{2pt}
\renewcommand{\arraystretch}{1.15}
\setlength{\extrarowheight}{1.5pt}
\begin{tabular}{|>{\raggedright\arraybackslash}p{0.29\linewidth}|>{\centering\arraybackslash}p{0.07\linewidth}|>{\centering\arraybackslash}p{0.145\linewidth}|>{\centering\arraybackslash}p{0.165\linewidth}|>{\centering\arraybackslash}p{0.12\linewidth}|>{\centering\arraybackslash}p{0.08\linewidth}|}
\hline
\textbf{Reference} & \textbf{$f_{\mathrm{mw}}$} & \textbf{$Q_{\mathrm{mw}}$} & \textbf{$Q_{\mathrm{opt}}$} & \textbf{$\eta$} & \textbf{BW} \\
 & (GHz) & & & & (MHz) \\
\hline
Strekalov et al., \textit{Opt. Lett.} \textbf{34}, 713--715 (2009) & 101.12 & N.A. & Loaded $\sim1.0\times10^{7}$ & $3.7\times10^{-9}$ & 20\textsuperscript{b} \\
\hline
Rueda et al., \textit{Optica} \textbf{3}, 597--604 (2016) & 8.94 & Loaded $2.3\times10^{2}$; Intrinsic $2.8\times10^{2}$ & Loaded $1.4\times10^{8}$; Intrinsic $2.8\times10^{8}$ & $1.0\times10^{-3}$ & 1.0 \\
\hline
Fan et al., \textit{Sci. Adv.} \textbf{4}, eaar4994 (2018) & 8.31 & Loaded $1.5\times10^{4}$ & Loaded $4.0\times10^{5}$ & $2.05\times10^{-2}$ & 0.55 \\
\hline
Santamar\'{\i}a Botello et al., \textit{Optica} \textbf{5}, 1210--1219 (2018) & 80 & Intrinsic $4.0\times10^{2}$ & Loaded $7.9\times10^{7}$; Intrinsic $1.6\times10^{8}$ & $8.2\times10^{-6}$ & 2.0 \\
\hline
Holzgrafe et al., \textit{Optica} \textbf{7}, 1714--1720 (2020) & 3.88 & Loaded $3.9\times10^{2}$; Intrinsic $>1.0\times10^{3}$ & Intrinsic $1.4\times10^{6}$ & $2.7\times10^{-5}$ & 13 \\
\hline
McKenna et al., \textit{Optica} \textbf{7}, 1737--1745 (2020) & 6.80 & Loaded $3.2\times10^{2}$; Intrinsic $5.3\times10^{2}$ & Loaded $2.1\times10^{5}$; Intrinsic $3.3\times10^{5}$ & $6.6\times10^{-6}$ & 20 \\
\hline
Xu et al., \textit{Nat. Commun.} \textbf{12}, 4453 (2021) & 7.84 & Loaded $8.7\times10^{2}$; Intrinsic $1.3\times10^{3}$ & Loaded $5.1\times10^{5}$ & $1.02\times10^{-2}$ & 9.1 \\
\hline
Sahu et al., \textit{Nat. Commun.} \textbf{13}, 1276 (2022) & 8.80 & Loaded $6.4\times10^{2}$; Intrinsic $1.1\times10^{3}$ & Loaded $7.5\times10^{6}$ & $1.5\times10^{-1}$ & 24 \\
\hline
Xu et al., \textit{Phys. Rev. Applied} \textbf{18}, 064045 (2022) & 6.47 & Loaded $1.6\times10^{4}$; Intrinsic $1.7\times10^{4}$ & Loaded $8.1\times10^{5}$; Intrinsic $1.0\times10^{6}$ & $1.70\times10^{-3}$ & 0.40\textsuperscript{b} \\
\hline
Gaier et al., arXiv:2505.04585 (2025) & 123.2 & Loaded 6.0; Intrinsic 8.0 & Loaded $8.8\times10^{5}$ & $3.1\times10^{-7}$\textsuperscript{a} & 220\textsuperscript{b} \\
\hline
Gaier et al., arXiv:2505.04585 (2025) & 307.9 & Loaded 12.0; Intrinsic 16.0 & Loaded $8.8\times10^{5}$ & $4.5\times10^{-7}$\textsuperscript{a} & 220\textsuperscript{b} \\
\hline
Suresh et al., \textit{Opt. Express} \textbf{33}, 10302--10311 (2025) & 157 & N.A. & Loaded $1.3\times10^{8}$ & $2.0\times10^{-5}$ & 1.5\textsuperscript{b} \\
\hline
Warner et al., \textit{Nat. Phys.} \textbf{21}, 831--838 (2025) & 3.71 & Loaded $\sim3.7\times10^{2}$ (pulsed operation) & Loaded $2.4\times10^{6}$; Intrinsic $7.7\times10^{6}$ & $1.18\times10^{-2}$ & 30 \\
\hline
Multani et al., \textit{Nat. Commun.} \textbf{17}, 1166 (2026) & 105.29 & Loaded $6.4\times10^{1}$; Intrinsic $8.5\times10^{1}$ & Loaded $9.2\times10^{5}$ & $8.6\times10^{-7}$ & 210 \\
\hline
\textbf{This work} & \textbf{60.5} & \textbf{Loaded 17.8; Intrinsic 35.6} & \textbf{Loaded} $\mathbf{1.3\times10^{6}}$\textbf{; Intrinsic} $\mathbf{2.6\times10^{6}}$ & $\mathbf{1.0\times10^{-1}}$ & \textbf{155} \\
\hline
\end{tabular}
\par\vspace{0.3em}
\begin{minipage}{\linewidth}
\raggedright\scriptsize $^{a}$ Estimated from the reported EO-comb spectrum. $^{b}$ When unreported, the bandwidth is estimated from the narrower of the applicable loaded optical and microwave cavity linewidths, as determined by cavity electro-optic theory.
\end{minipage}
\caption{\textbf{Quality factors and performance of cavity electro-optic transducers.} Reported microwave and optical quality factors are compared with photon-number conversion efficiencies (\(\eta\)) and conversion bandwidths. Each quality factor is identified as intrinsic or loaded.}
\label{tab:SI_quality_factor_comparison}
\end{table}
\Fref{fig:SI_bandwidth_efficiency} compares the reported conversion efficiencies and bandwidths of these cavity electro-optic transducers / receivers.

\begin{figure}[htbp]
  \centering
  \includegraphics[width=0.58\linewidth]{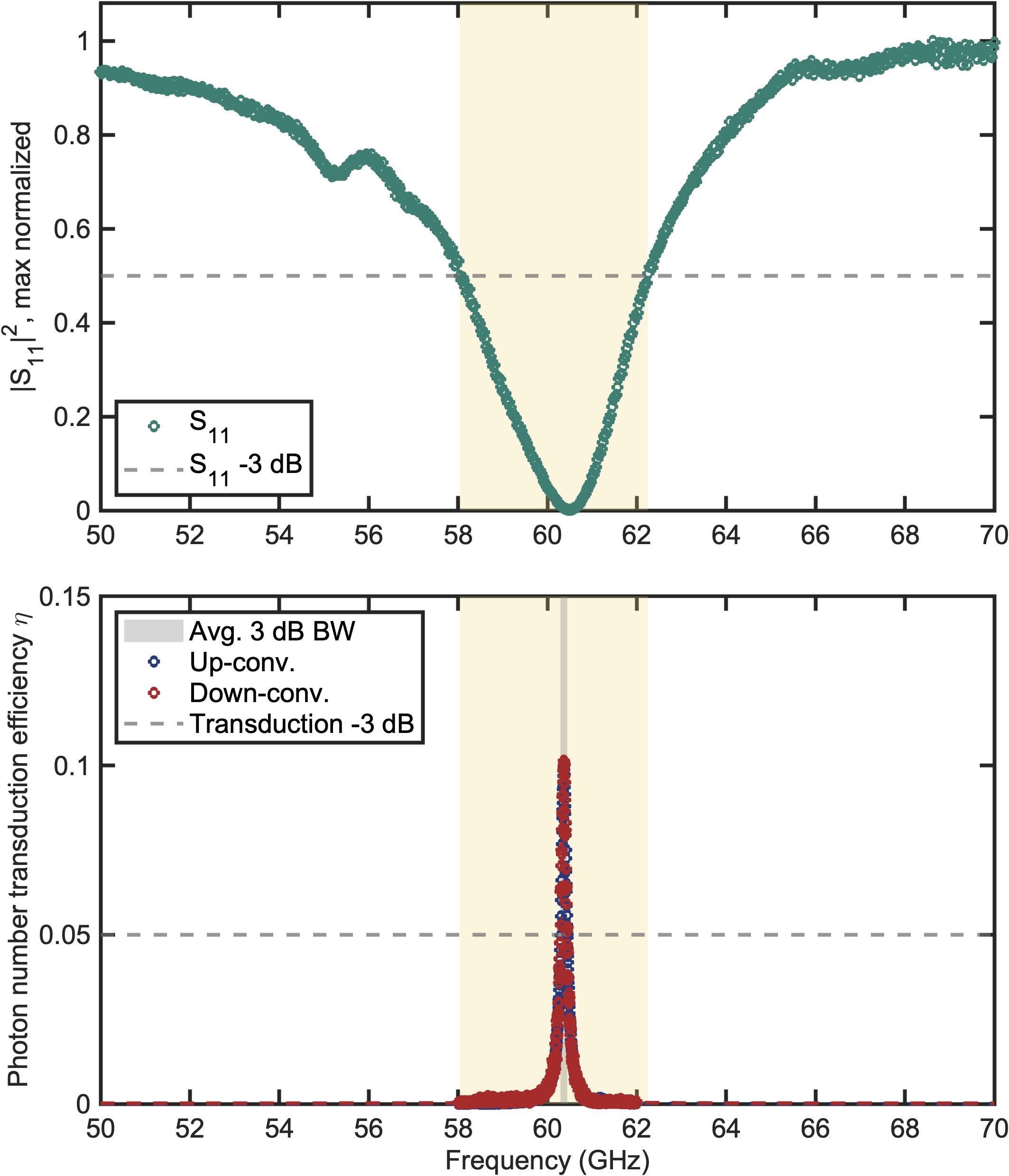}
  \caption{\textbf{Microwave and transduction bandwidths.} Top: measured normalized microwave reflection $|S_{11}|^2$ on a linear scale, with the yellow region indicating the microwave $3$-dB bandwidth. Bottom: measured photon-number conversion efficiency for upconversion (blue) and downconversion (red), with the gray region indicating the average transduction $3$-dB bandwidth of approximately $155~\mathrm{MHz}$. The bi-directional transduction responses lie within the broader microwave response.}
  \label{fig:SI_microwave_transduction_bandwidths}
\end{figure}

Since the total optical linewidth, $\kappa/2\pi\approx155~\mathrm{MHz}$, is much smaller than the total microwave linewidth, $\Gamma/2\pi\approx3.4~\mathrm{GHz}$, the transduction bandwidth of our device is limited by the optical linewidth. \Fref{fig:SI_microwave_transduction_bandwidths} compares the measured microwave response with the upconversion and downconversion responses. Both $S_{\mathrm{oe}}$ and $S_{\mathrm{eo}}$ lie within the broader microwave $3$-dB response.

\begin{figure}[htbp]
  \centering
  \includegraphics[width=1.0\linewidth]{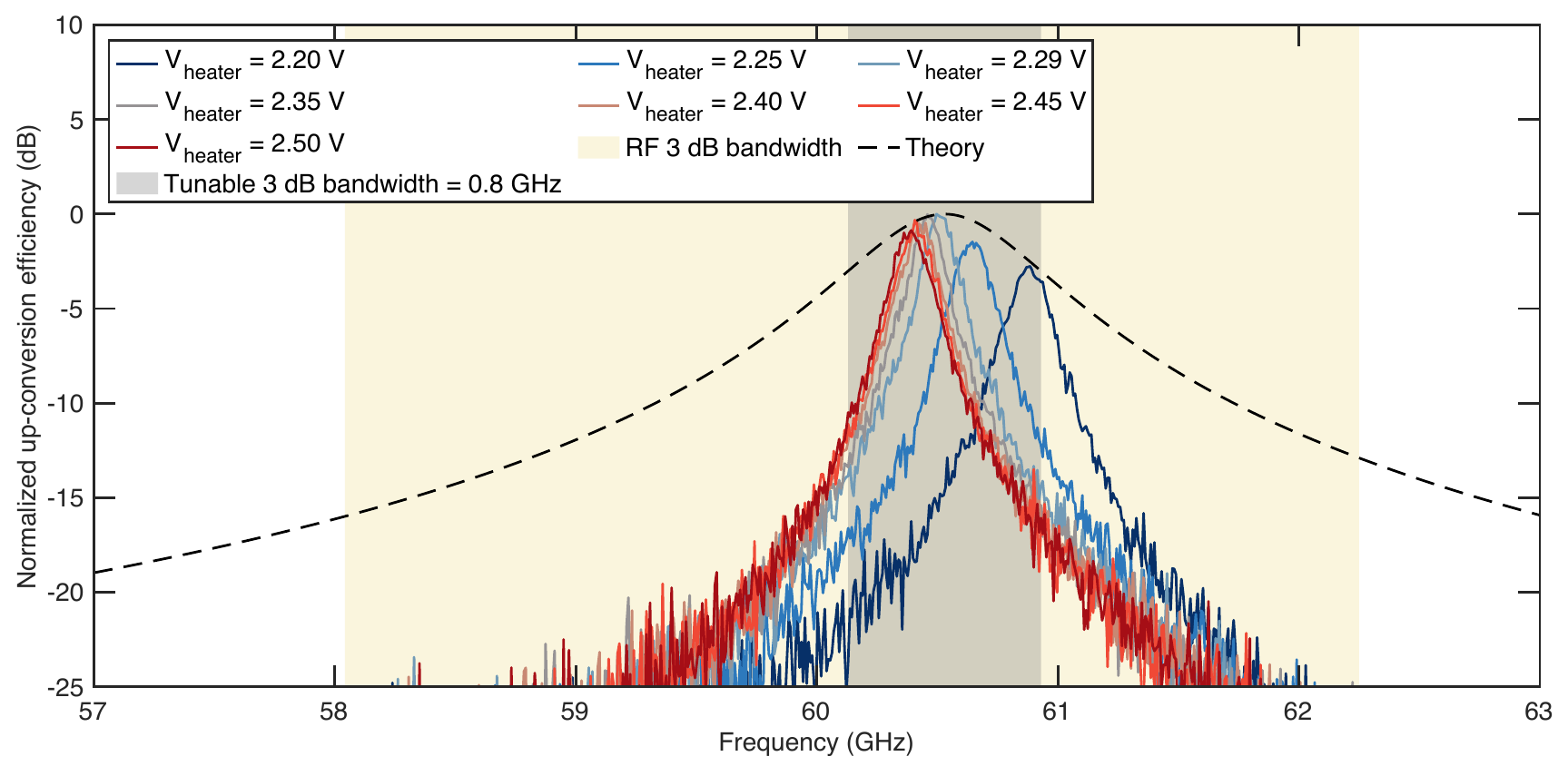}
  \caption{\textbf{Thermally tunable transduction response.} Measured normalized upconversion efficiency for heater voltages from $2.20$ to $2.50~\mathrm{V}$. Thermal tuning provides a combined $3$-dB frequency coverage of approximately $0.8~\mathrm{GHz}$ (gray region), while the measured responses remain within the broader microwave $3$-dB response (yellow region). The dashed curve shows the microwave-response envelope. The $0.8~\mathrm{GHz}$ coverage represents the demonstrated thermal tuning range rather than the instantaneous bandwidth at a single heater setting. The gaps in spectral coverage arise from the Vernier effect between the auxiliary and main rings, which have similar free spectral ranges (FSRs) in the current design, and can be overcome by using auxiliary rings with larger FSRs.}
  \label{fig:SI_heater_tunable_transduction}
\end{figure}

The coupled-ring structure used for side-mode suppression also provides a finite frequency-tuning range governed by the inter-ring coupling rate $J$. As shown in \fref{fig:SI_heater_tunable_transduction}, thermally / electro-optically tuning the relative detuning between the two rings shifts the transduction response over an approximately $0.8~\mathrm{GHz}$ frequency range. This tunability provides frequency reconfigurability for the photonic receiver (tunable bandwidth), while the instantaneous conversion bandwidth at an individual heater setting remains approximately $155~\mathrm{MHz}$.

\subsection{Receiver noise figure and bandwidth context}

\begin{figure}[htbp]
  \centering
  \includegraphics[width=1.0\linewidth]{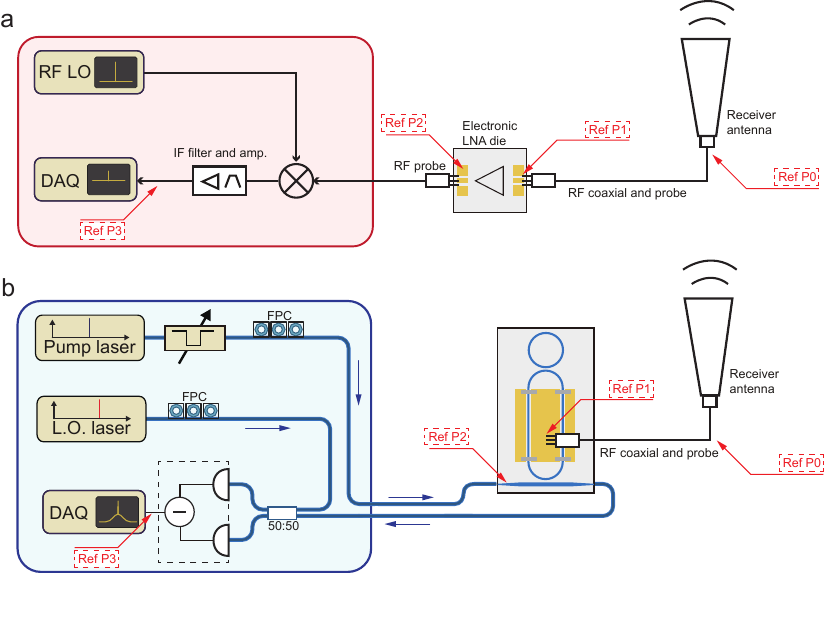}
  \caption{\textbf{Receiver reference planes.} Ref.~P0 is the antenna terminal, Ref.~P1 is the on-chip microwave input, Ref.~P2 is the on-chip device output (the output RF pads for an electronic LNA and the optical edge coupler for the photonic receiver), and Ref.~P3 is the recovered IF output after the electronic mixer chain or optical heterodyne detection. The photonic on-chip noise figure is defined from Ref.~P1 to Ref.~P2, whereas the full-system noise figure is defined from Ref.~P1 to Ref.~P3 and includes the off-chip optical heterodyne-detection and IF chain. Published research electronic LNA noise figures are usually referenced from Ref.~P1 to Ref.~P2 rather than to the IF output at Ref.~P3.}
  \label{fig:SI_receiver_reference_planes}
\end{figure}

\Fref{fig:SI_receiver_reference_planes} and Table~\ref{tab:SI_reference_plane_definitions} define the reference planes used for the noise-figure comparison.

\begin{table}[htbp]
\centering
\scriptsize
\setlength{\tabcolsep}{3pt}
\renewcommand{\arraystretch}{1.18}
\begin{tabular}{|>{\raggedright\arraybackslash}p{0.25\linewidth}|>{\centering\arraybackslash}p{0.08\linewidth}|>{\centering\arraybackslash}p{0.08\linewidth}|>{\raggedright\arraybackslash}p{0.41\linewidth}|}
\hline
\textbf{Metric} & \textbf{Input} & \textbf{Output} & \textbf{Included receiver section} \\
\hline
Electronic LNA device noise figure & Ref.~P1 & Ref.~P2 & LNA die between input and output CPW pads; the quantity commonly reported in research articles. \\
\hline
Photonic on-chip noise figure & Ref.~P1 & Ref.~P2 & Electro-optic transducer from the on-chip microwave input to optical output, assuming ideal heterodyne detection. \\
\hline
Photonic full-system noise figure & Ref.~P1 & Ref.~P3 & Electro-optic transducer, measured off-chip optical heterodyne detection, and recovered IF output. \\
\hline
End-to-end receiver noise figure & Ref.~P0 & Ref.~P3 & Adds the antenna and deployment-dependent RF feed between Ref.~P0 and Ref.~P1. \\
\hline
\end{tabular}
\caption{\textbf{Receiver noise-figure reference-plane definitions.} Input and output reference planes and receiver sections included in the electronic and photonic noise figures and transduction efficiencies used in this work.}
\label{tab:SI_reference_plane_definitions}
\end{table}

Published research electronic LNAs generally report the de-embedded device noise figure between Ref.~P1 and Ref.~P2, whereas connectorized commercial modules may additionally include package and connector losses. For the photonic receiver, the on-chip noise figure is likewise defined between Ref.~P1 and Ref.~P2, while the full-system noise figure is defined between Ref.~P1 and Ref.~P3 and includes the off-chip optical detection chain.

\begin{figure}[htbp]
  \centering
  \includegraphics[width=0.6\linewidth]{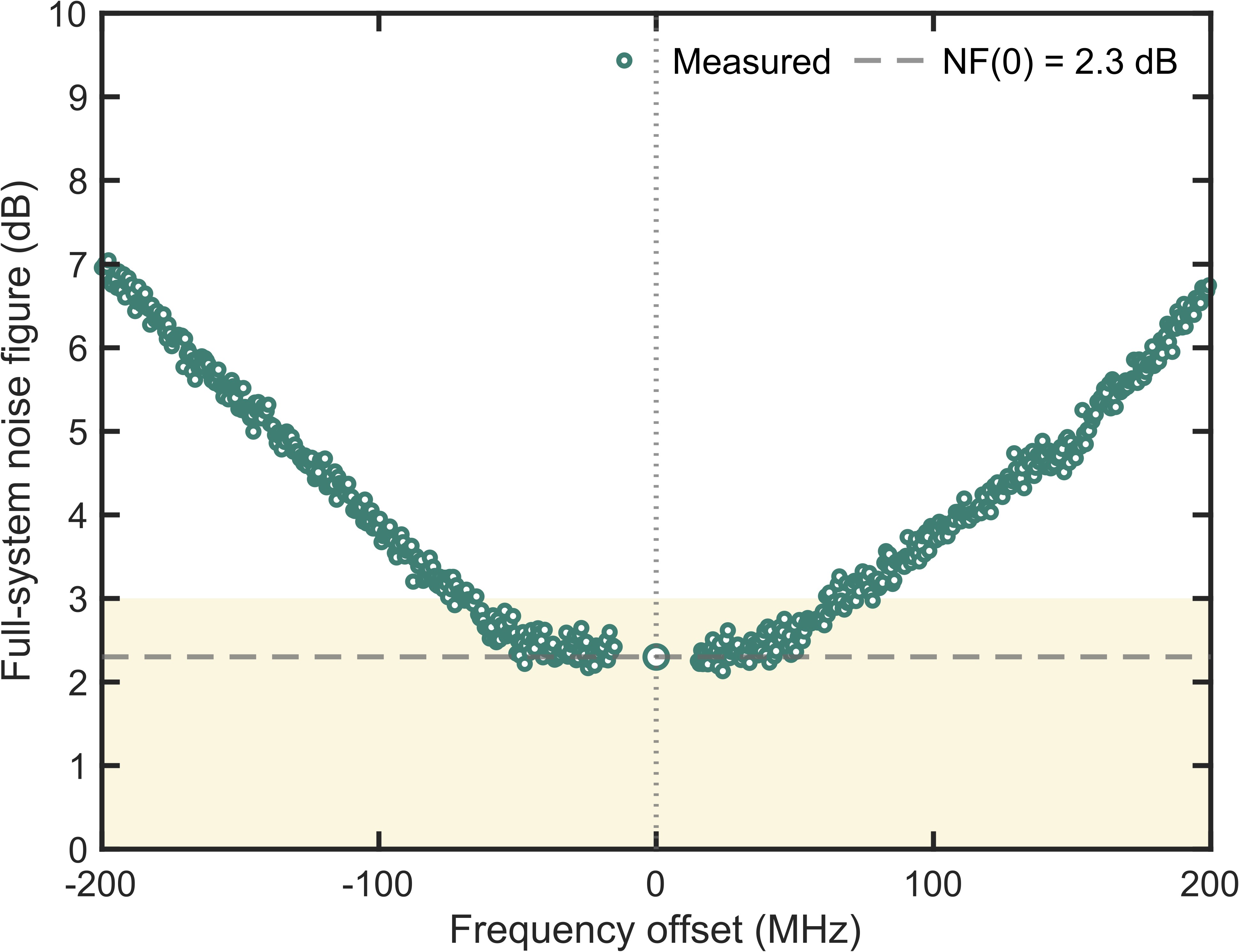}
  \caption{\textbf{Frequency-dependent full-system noise figure.} Full-system noise figure from the on-chip microwave input at Ref.~P1 to the detected IF output at Ref.~P3, with the reference planes defined in \fref{fig:SI_receiver_reference_planes}. The frequency axis shows the offset from the resonance centre, where the noise figure is $2.3\pm0.3~\mathrm{dB}$. The shaded region indicates $\mathrm{NF}<3~\mathrm{dB}$. At the lower and upper edges of the $155~\mathrm{MHz}$ $3$-dB conversion bandwidth, the noise figures are approximately $3.2~\mathrm{dB}$ and $3.0~\mathrm{dB}$, respectively.}
  \label{fig:SI_frequency_dependent_nf}
\end{figure}

We study the receiver noise figure as a function of frequency, as shown in \fref{fig:SI_frequency_dependent_nf}. The full-system noise figure is referred from the on-chip microwave input at Ref.~P1 to the detected IF output at Ref.~P3, following the reference-plane definitions in \fref{fig:SI_receiver_reference_planes}. It reaches $2.3\pm0.3~\mathrm{dB}$ at the resonance centre and is approximately $3.2~\mathrm{dB}$ and $3.0~\mathrm{dB}$ at the lower and upper edges of the $155~\mathrm{MHz}$ $3$-dB conversion bandwidth, respectively.

\begin{figure}[htbp]
		\centering
	\includegraphics[width=1.0\linewidth]{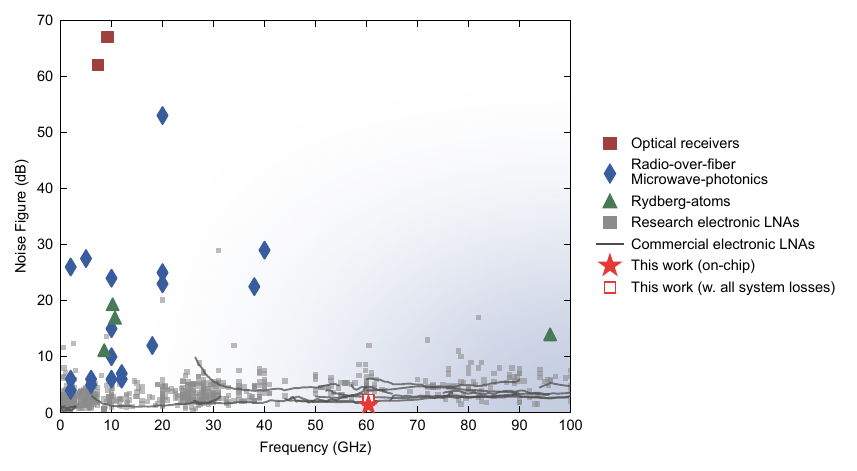}
		\caption{Comparison of noise figures for Rydberg atom-based microwave receivers, photonic receivers, microwave photonics systems, and state-of-the-art electronic low-noise amplifiers (LNAs).
		Gray squares represent electronic LNAs from peer-reviewed journals~\cite{wang2025lna,belostotskiLNA}. 
		Red squares denote optical microwave receiver schemes~\cite{hsuAlldielectricPhotonicassistedRadio2007,ilchenko2002sub}. 
        Blue rhombuses represent photonic-based microwave receivers / radio-over-fiber systems \cite{karim2007low,zhang2022low,xie2014high,ackerman2007signal,daulay2022ultrahigh,wei2025programmable,deng2025single,roussell2007gain,peng2021high,rady202220}
		Green triangles indicate Rydberg atom-based receivers~\cite{cai2022sensitivity,photonics9040250,wu2024enhancing,legaie2024millimeter}, with noise figures inferred from electric-field sensitivity as described in Ref.~\cite{santamaria2022comparison}.
        Commercial low noise amplifier data taken from Refs ~\cite{zx6033lndatasheet, zx60p103lndatasheet, minicircuits-wva-751103lnx-datasheet, minicircuits-wva-44603ln-datasheet,  lnf-lnr10_30a_sv-datasheet, lnf-lnr26_62a_sv-datasheet, lnf-lnr45_77wa_sv-datasheet, lnf-lnr55_96wa_sv-datasheet, lnf-lnr65_115wd_sv-datasheet, ums-cha2157-99f00-datasheet, ums-cha2159-datasheet, adi-hmc-alh382-datasheet, marki-amm-7203uc-datasheet, rpg-lna-datasheet, rpg-w-lna-75-110-40-3-datasheet, vdi-low-noise-amplifier-product-manual, mws-w-band-lna-75-110ghz-datasheet, eravant-sbl-1442041570-0505-e1-datasheet}}
\label{fig:SI_noise_figure_comparison}
\end{figure}
In \fref{fig:SI_noise_figure_comparison}, we compare our measured full-system receiver noise figure, including the optical-detection and IF-chain contributions, with published receiver and LNA noise figures.

We enhance the electro-optic response through a resonant cavity, which trades receiver bandwidth for response at resonance. Electronic LNAs have a different bandwidth dependence: reducing the amplifier's operating bandwidth does not intrinsically reduce its noise figure. For an electronic LNA, the noise figure is instead determined by the minimum noise figure of the first transistor and the source admittance seen at its input:
\begin{equation}
F=F_{\min}+\frac{R_N}{G_S}\left|Y_S-Y_{\mathrm{opt}}\right|^2,
\label{eq:SI_lna_noise_parameters}
\end{equation}
where $F_{\min}$ is the transistor minimum noise figure, $R_N$ is the equivalent noise resistance, $Y_S=G_S+jB_S$ is the source admittance, and $Y_{\mathrm{opt}}$ is the optimum source admittance. Narrowband impedance matching can reduce the mismatch term by bringing $Y_S$ closer to $Y_{\mathrm{opt}}$, but it cannot reduce the noise figure below $F_{\min}$. The resonant bandwidth trade-off of the cavity electro-optic receiver and the impedance-matching bandwidth of an electronic LNA arise from different mechanisms.

\subsection{Thermal-noise visibility across electro-optic transducers}

Resolving converted equilibrium thermal noise requires the thermal noise transduced to the measurement output to exceed the background (quantum shot noise):
\begin{equation}
\eta n_{\mathrm{th}}>N_{\mathrm{bg}},\qquad
n_{\mathrm{th}}=\frac{1}{\exp\!\left(\hbar\Omega/k_\mathrm{B}T\right)-1}
\simeq\frac{k_\mathrm{B}T}{\hbar\Omega}.
\label{eq:SI_thermal_visibility}
\end{equation}
The output-referred thermal signal is expressed as $\eta n_{\mathrm{th}}$, where $\eta$ is the total efficiency from the intrinsic microwave thermal bath to the measured optical output. The total output background is $N_{\mathrm{bg}}=N_\mathrm{q}+N_\mathrm{parasitic}$, with a phase-preserving heterodyne quantum-noise floor of $N_\mathrm{q}=1~\mathrm{photon\,s^{-1}\,Hz^{-1}}$ in the photon-number convention used here. In Table~\ref{tab:SI_thermal_visibility}, we use the reported photon conversion efficiency $\eta$ to calculate $\eta n_{\mathrm{th}}$ as the thermal noise-visibility metric.

\begin{table}[htbp]
\centering
\scriptsize
\setlength{\tabcolsep}{2pt}
\renewcommand{\arraystretch}{1.30}
\begin{tabular}{|>{\raggedright\arraybackslash}p{0.25\linewidth}|>{\centering\arraybackslash}p{0.07\linewidth}|>{\centering\arraybackslash}p{0.07\linewidth}|>{\centering\arraybackslash}p{0.12\linewidth}|>{\centering\arraybackslash}p{0.13\linewidth}|>{\centering\arraybackslash}p{0.09\linewidth}|>{\centering\arraybackslash}p{0.11\linewidth}|}
\hline
Reference & $T$ & $f_{\mathrm{mw}}$ & On-chip $\eta$ & Off-chip $\eta$ & $n_{\mathrm{th}}$ & $\eta n_{\mathrm{th}}$ \\
 & (K) & (GHz) & & & & \\
\hline
Fan et al., \textit{Sci. Adv.} \textbf{4}, eaar4994 (2018) & 2.0 & 8.31 & $2.05\%$ & -- & 4.53 & $9.3\times10^{-2}$ \\
\hline
Holzgrafe et al., \textit{Optica} \textbf{7}, 1714--1720 (2020) & 1.0 & 3.89 & $2.7\times10^{-5}$ & -- & 4.88 & $1.3\times10^{-4}$ \\
\hline
McKenna et al., \textit{Optica} \textbf{7}, 1737--1745 (2020) & 1.0 & 6.80 & $6.6\times10^{-6}$ & $3.9\times10^{-7}$ & 2.59 & $1.7\times10^{-5}$ \\
\hline
Xu et al., \textit{Nat. Commun.} \textbf{12}, 4453 (2021) & 1.9 & 7.84 & $1.02\%$ & -- & 4.57 & $4.7\times10^{-2}$ \\
\hline
Sahu et al., \textit{Nat. Commun.} \textbf{13}, 1276 (2022) & $\leq0.060^{a}$ & 8.80 & -- & $8.7\%$ (low noise); $15\%$ (increased noise) & $\leq8.8\times10^{-4}$ & $\leq1.3\times10^{-4}$ \\
\hline
Warner et al., \textit{Nat. Phys.} \textbf{21}, 831--838 (2025) & 0.038 & 3.71 & $1.18\%$ & -- & $9.3\times10^{-3}$ & $1.1\times10^{-4}$ \\
\hline
Multani et al., \textit{Nat. Commun.} \textbf{17}, 1166 (2026) & 4.89 & 105.29 & $0.82\times10^{-6}$ & -- & 0.552 & $4.5\times10^{-7}$ \\
\hline
\textbf{This work} & \textbf{290} & \textbf{60.5} & \textbf{10\%} & \textbf{\boldmath \(6.46\%\)} & \textbf{99.38} & \textbf{\boldmath \(9.9\times10^{0}\) (on-chip); \(6.4\times10^{0}\) (off-chip)} \\
\hline
\end{tabular}
\par\vspace{0.3em}
\begin{minipage}{\linewidth}
\raggedright\scriptsize $^{a}$The baseplate temperature reached at most $0.060~\mathrm{K}$.
\end{minipage}
\caption{\textbf{Thermal-noise visibility in electro-optic transducers.} Reported operating temperatures, microwave frequencies, transduction efficiencies, thermal occupations, and calculated thermal-visibility metrics are compared for representative devices.}
\label{tab:SI_thermal_visibility}
\end{table}

\section{Theory for cavity electro-optics in triply-resonant system}
\subsection{Hamiltonian approach for symmetric cavity electro-optics}

For a triply-resonant cavity electro-optic system, the interaction Hamiltonian can be expressed as $H_{eo}=\hbar g_{\mathrm{0}} a_{\mathrm{p}}a_{\mathrm{as}}^\dagger b + \hbar g_{\mathrm{0}} a_{\mathrm{p}} a_{\mathrm{s}}^\dagger b^\dagger + h.c.$, where $g_{\mathrm{0}}$ is the vacuum (single-photon) electro-optic coupling rate, $a_{\mathrm{p}}$, $a_{{\mathrm{as}}}$, $a_{\mathrm{s}}$ and $b$ are the optical pump, anti-Stokes, Stokes modes, and microwave mode respectively. This Hamiltonian can be linearized in different regimes, depending on whether the strong drive is applied to the microwave mode \( b \) or the optical pump mode $ a_\mathrm{p}$.

To derive the expression for $g_{\mathrm{0}}$, we need to compute the interaction energy arising from the electro‑optic effect. In this work, we focus on the linear electro‑optic (Pockels) effect, in which the induced second‑order polarization $P^{(2)}$ is proportional to the applied microwave field $E_m$:

\begin{equation}
    P^{(2)}_i=\chi^{(2)}_{ijk}E_{s,j}E_{m,k}
\end{equation}
where $\chi^{(2)}_{ijk}$ is the second-order dielectric susceptibility, $E_{s,j}$ and $E_{m,k}$ denote the components of the electric field along the $j$- and $k$-directions of the optical signal and the added microwave electrical field with the frequency of $\omega_s$ and $\Omega$.

Then under the rotating‑wave approximation, the interaction energy takes the form:
\begin{equation}
    U_\mathrm{int}=-\frac{\epsilon_{\mathrm{0}}}{4}\int_{LT}dV\chi^{(2)}_{ijk}E_{{\mathrm{p}},i}E_{{\mathrm{s}},j}^*E_{{\mathrm{m}},k} + c.c.
\end{equation}
Where $E_{{\mathrm{p}},i}$ denote the $i$-direction of the optical pump electrical field with the frequency of $\omega_{\mathrm{p}}$, and it obeys $\omega_{\mathrm{s}}=\omega_{\mathrm{p}}\pm\Omega$.

To derive the single photon coupling rate $g_{\mathrm{0}}$, the electrical fields need to be normalized with their respective energies:
\begin{equation}
    W_{\mathrm{p,s,m}}=\frac{\epsilon_0}2\int\epsilon_{\mathrm{r}}(\mathrm{p,s,m})|\vec{\textbf{E}}_\mathrm{p,s,m}|^2dV
\end{equation}
where $\epsilon_{\mathrm{r}}(\mathrm{p,s,m})$ is the space-dependent relative permittivity of the medium at the corresponding mode frequency. Then we can derive the expression of $g_0$ as:
\begin{equation}
    g_0 = \frac{\varepsilon_0}{4} \sqrt{ \frac{\hbar  \omega^2 \Omega}{W_{\mathrm{m}} W_{\mathrm{p}} W_{\mathrm{s}}} } \int_{\mathrm{LiTaO_3}} \chi^{(2)}_{ijk} E_{\mathrm{p},i} E^{*}_{\mathrm{s},j} E_{\mathrm{m},k} \, \mathrm{d}V\label{SI:eq:g0}
\end{equation}

In case of strong microwave drive, the interaction Hamiltonian can be written as:
\begin{equation}
    H_\mathrm{int}=\hbar g_\mathrm{0} \langle b\rangle a_\mathrm{p}a_\mathrm{as}^\dagger + \hbar g_\mathrm{0} \langle b\rangle a_\mathrm{p}a_\mathrm{s}^\dagger + h.c.
\end{equation}

Here, $g = \langle b \rangle g_\mathrm{0}$ is the effective EO coupling strength, proportional to the microwave drive amplitude $\langle b \rangle$. Under this Hamiltonian, a strong microwave drive enables cascaded energy transfer from the optical pump mode $a_{\mathrm{p}}$ to the Stokes and anti-Stokes sidebands ($a_{\mathrm{s}}$ and $a_{\mathrm{as}}$), as well as to higher-order sidebands, generating the EO comb~\cite{zhangUltrabroadbandIntegratedElectrooptic2025a}.

In case of strong optical drive, the interaction Hamiltonian can be written as:
\begin{equation}
    H_\mathrm{int}=\hbar g_\mathrm{0} \langle a_\mathrm{p}\rangle a_\mathrm{as}^\dagger b + \hbar g_\mathrm{0} \langle a_\mathrm{p}\rangle a_\mathrm{s}^\dagger b^\dagger + h.c.
\end{equation}

where $g=\langle a_\mathrm{p}\rangle g_\mathrm{0}$ is the effective EO coupling strength, proportional to the optical pump drive amplitude $\langle a_\mathrm{p}\rangle$. This Hamiltonian represents two basic cavity EO processes, one is the beam-splitter process with anti-Stokes sideband $H_\mathrm{BS}=\hbar g(a_\mathrm{as}b^\dagger+ a_\mathrm{as}^\dagger b)$, the other is the two mode squeezing process with Stokes sideband $H_\mathrm{squeeze} = \hbar g(a_\mathrm{s} b+a_\mathrm{s}^\dagger b^\dagger)$. The beam-splitter Hamiltonian shows that the strong optical pump can enhance the coherent microwave-optic conversion, while the squeezing Hamiltonian represents the enhanced pair generation.

We define the mode susceptibility for an optical mode $j$ ($j = s, as, p$ for Stokes, anti-Stokes, or pump mode) as
\begin{equation}
	\chi_j(\Delta)^{-1} = \frac{\kappa_j}{2} - i\Delta ,
\end{equation}
and for the microwave mode as
\begin{equation}
	\chi_\mathrm{e}(\Delta)^{-1} = \frac{\Gamma}{2} - i\Delta .
\end{equation}
Here, $\kappa_j$ is the total dissipation rate of the optical mode $j$, $\Gamma$ is the total dissipation rate of the microwave mode, and $\Delta$ is the detuning between the probe frequency and the mode.

Though in ideality, the Stokes and anti-Stokes modes are located on the lower or higher side of the optical pump by microwave resonant frequency $\Omega$ symmetrically. In practice, the optical cavity dispersion and phase mismatch between optical FSR and microwave resonant frequency will cause the detuning of Stokes and anti-Stokes sidebands with the respective modes $\delta_\mathrm{s}, \delta_\mathrm{as}$.

In this work, we concentrate on the case with strong optical pump. We then can define the intra-cavity and output matrix with the intra-cavity and output operators as:
\begin{equation}
    \begin{aligned}
        D &= [a_\mathrm{s}, a_\mathrm{s}^\dagger, a_\mathrm{as}, a_\mathrm{as}^\dagger, b, b^\dagger]^{-1} \\
    D_\mathrm{in} &= [a_\mathrm{s,in}, a_\mathrm{s,in}^\dagger, a_\mathrm{as,in}, a_\mathrm{as,in}^\dagger, b_\mathrm{in}, b_\mathrm{in}^\dagger]^{-1}
    \end{aligned}
\end{equation}

In the rotating frame of microwave resonant frequency, Stokes mode, and anti-Stokes mode, we can obtain the full expression of intra-cavity matrix from  the quantum Langevin equations:
\begin{equation}
    \begin{aligned}
        D = \textbf{M}\cdot\textbf{L}\cdot D_\mathrm{in}
    \end{aligned}
\end{equation}
where,
\begin{equation}
    \begin{aligned}
        \textbf{M}&= \begin{pmatrix}
            \chi_\mathrm{s}^{-1}(\delta_\mathrm{s}+\Delta) & 0 & 0 & 0 & 0 & ig\\
            0 & \chi_\mathrm{s}^{-1}(-\delta_\mathrm{s}+\Delta) & 0 & 0 & -ig & 0\\
            0 & 0 & \chi_\mathrm{as}^{-1}(\delta_\mathrm{as}+\Delta) & 0 & ig & 0\\
            0 & 0 & 0 & \chi_\mathrm{as}^{-1}(\delta_\mathrm{as}+\Delta) & 0 & -ig\\
            0 & ig & ig & 0 & \chi_\mathrm{e}^{-1}(\Delta) & 0\\
            -ig & 0 & 0 & -ig & 0 & \chi_\mathrm{e}^{-1}(\Delta)\\
        \end{pmatrix}^{-1}
        \\ \textbf{L}&=\mathrm{diag}(\sqrt{\kappa_\mathrm{s, ex}}, \sqrt{\kappa_\mathrm{s, ex}}, \sqrt{\kappa_\mathrm{as, ex}}, \sqrt{\kappa_\mathrm{as, ex}}, \sqrt{\Gamma_\mathrm{ex}}, \sqrt{\Gamma_\mathrm{ex}})
    \end{aligned}
\end{equation}
where $\kappa_{j, \mathrm{ex}}$ and $\Gamma_{ex}$ represent the external coupling rate for optical mode j and the microwave mode, and the intrinsic coupling rate is $\kappa_{j,0}$ and $\Gamma_\mathrm{0}$. By diagonalizing the above equations, we can obtain the effective susceptibility $\chi_{j,eff}$ for different mode j.

The output matrix can be obtained from the input-output theorem $a_{j,\mathrm{out}}=a_{j,\mathrm{in}}-\sqrt{\kappa_{j,\mathrm{ex}}}a_j$. In this work, we focus on the coherent response of this CEO system, where the amplitude reflection efficiency can be defined as $S_{jj}(\Delta+\omega_j)=1-\kappa_{\mathrm{ex}}\chi_{j,eff}(\Delta)$.

\subsection{Bidirectional transduction in the anti-Stokes branch}
As shown in the previous equations, for the simplest system, with only a single optical ring resonator, the Stokes and anti‐Stokes processes always occur simultaneously. In this work, we focus on achieving millimeter‑wave optical detection with reduced added noise; accordingly, the anti‐Stokes process, as the coherent transduction channel, is our preferred mechanism. We therefore employ coupled rings to suppress the Stokes process, making the anti‑Stokes process the dominant pathway. We introduce an auxiliary mode, resonant with and coupled to the Stokes mode, which induces mode splitting at the Stokes sideband and thereby suppresses the Stokes process. Therefore, the Hamiltonian we are interested in can be expressed as:
\begin{equation}
    H_\mathrm{int}=\hbar g(ab^\dagger+ a^\dagger b)
\end{equation}

Simplify from the previous full expression of the system, in case of Stokes process being ideally suppressed, we can obtain the quantum Langevin equations as:
\begin{equation}
    \begin{pmatrix}
        a\\
        b
    \end{pmatrix}=
    \begin{pmatrix}
        \chi_\mathrm{o}^{-1}(\delta+\Delta) & ig \\
        ig & \chi_\mathrm{e}^{-1}(\Delta)
    \end{pmatrix}^{-1}
    \begin{pmatrix}
        \sqrt{\kappa_\mathrm{ex}} & 0\\
        0 & \sqrt{\Gamma_\mathrm{ex}}
    \end{pmatrix}
    \begin{pmatrix}
        a_\mathrm{in}\\
        b_\mathrm{in}
    \end{pmatrix}
\end{equation}
where $\chi_\mathrm{o}^{-1}(\Delta)=\kappa/2-i\Delta$, $\chi_\mathrm{e}^{-1}(\Delta)=\Gamma/2-i\Delta$ are the inverse susceptibilities of the optical and microwave cavities, respectively. The rate $\kappa_\mathrm{ex}$ and $\kappa_\mathrm{0}$ ($\Gamma_\mathrm{ex}$ and $\Gamma_\mathrm{0}$) denote the external and intrinsic decay rates of the optical (microwave) cavity, with $\kappa=\kappa_\mathrm{0}+\kappa_\mathrm{ex},\Gamma=\Gamma_\mathrm{0}+\Gamma_\mathrm{ex}$. $\delta$ is the detuning between the anti-Stokes sideband and the mode.

From the input-output theorem, we have,

\begin{equation}
    \begin{pmatrix}
        a_\mathrm{out}\\b_\mathrm{out}
    \end{pmatrix}=
    \begin{pmatrix}
        a_\mathrm{in}\\b_\mathrm{in}
    \end{pmatrix}-
    \begin{pmatrix}
        \sqrt{\kappa_\mathrm{ex}} & 0\\
        0 & \sqrt{\Gamma_\mathrm{ex}}
    \end{pmatrix}
    \begin{pmatrix}
        a\\b
    \end{pmatrix}
\end{equation}

Then we can define the scattering matrix $\textbf{S}$ as:
\begin{equation}
    \begin{pmatrix}
        a_\mathrm{out}\\b_\mathrm{out}
    \end{pmatrix}=\textbf{S}
    \begin{pmatrix}
        a_\mathrm{in}\\b_\mathrm{in}
    \end{pmatrix}
\end{equation}
where the scattering matrix can be written as:
\begin{equation}
    \textbf{S}=\textbf{I}-\begin{pmatrix}
        \sqrt{\kappa_\mathrm{ex}} & 0\\
        0 & \sqrt{\Gamma_\mathrm{ex}}
    \end{pmatrix}
    \begin{pmatrix}
        \chi_\mathrm{o}^{-1}(\delta+\Delta) & ig \\
        ig & \chi_\mathrm{e}^{-1}(\Delta)
    \end{pmatrix}^{-1}
    \begin{pmatrix}
        \sqrt{\kappa_\mathrm{ex}} & 0\\
        0 & \sqrt{\Gamma_\mathrm{ex}}
    \end{pmatrix}
\end{equation}

In the ideal frequency matched situation, the detuning between anti-Stokes sideband and the mode is zero ($\delta=0$), then we can obtain the full expression for the scattering matrix as:
\begin{equation}
\textbf{S}=\begin{pmatrix}
    S_\mathrm{oo} & S_\mathrm{oe}\\
    S_\mathrm{eo} & S_\mathrm{ee}
\end{pmatrix}=
\begin{pmatrix}
1 - \displaystyle\frac{\bigl(\tfrac{\Gamma}{2} - i\Delta\bigr)\,\kappa_\mathrm{ex}}
      {g^2 - \tfrac{i\Gamma\Delta}{2} - \Delta^2 + \tfrac{\Gamma\kappa}{4} - \tfrac{i\Delta\kappa}{2}}
&
\displaystyle\frac{i\,g\,\sqrt{\Gamma_{\mathrm{ex}}\,\kappa_{\mathrm{ex}}}}
      {g^2 - \tfrac{i\Gamma\Delta}{2} - \Delta^2 + \tfrac{\Gamma\kappa}{4} - \tfrac{i\Delta\kappa}{2}}
\\[2ex]
\displaystyle\frac{i\,g\,\sqrt{\Gamma_{\mathrm{ex}}\,\kappa_{\mathrm{ex}}}}
      {g^2 - \tfrac{i\Gamma\Delta}{2} - \Delta^2 + \tfrac{\Gamma\kappa}{4} - \tfrac{i\Delta\kappa}{2}}
&
1 - \displaystyle\frac{\Gamma_{\mathrm{ex}}\bigl(-\,i\Delta + \tfrac{\kappa}{2}\bigr)}
      {g^2 - \tfrac{i\Gamma\Delta}{2} - \Delta^2 + \tfrac{\Gamma\kappa}{4} - \tfrac{i\Delta\kappa}{2}}
\end{pmatrix}
\end{equation}

As for the room temperature cavity electro-optics in our case, the microwave dissipation rate $\Gamma$ is much larger than the optical dissipation rate $\kappa$, i.e., $\Gamma \gg \kappa$, we are working in the reversed-dissipation regime. In this regime, the microwave mode can be adiabatically eliminated, and the effective transduction bandwidth is set by the optical linewidth.

From the symmetry of the scattering matrix, the transduction efficiency between microwave to optic and optic to microwave are the same. Therefore, the Bidirectional transduction efficiency at resonance $\Delta = 0 $ can be defined as:
\begin{equation}
    \eta = |S_\mathrm{oe}(\Delta=0)|^2= |S_\mathrm{eo}(\Delta=0)|^2=\frac{4\mathcal{C}}{(\mathcal{C}+1)^2}\frac{\gamma_\mathrm{e}}{\gamma_\mathrm{e}+1}\frac{\gamma_\mathrm{o}}{\gamma_\mathrm{o}+1}
\end{equation}
where $\mathcal{C}=\frac{4g^2}{\Gamma\kappa}$ is the electro-optic cooperativity, and $\gamma_{\mathrm{e}}={\Gamma_{\mathrm{ex}}}/{\Gamma_0}, \gamma_{\mathrm{o}}={\kappa_{\mathrm{ex}}}/{\kappa_{0}}$. In the weak coupling regime, where cooperativity is much smaller than unity, $\mathcal{C}\ll1$, the transduction efficiency exhibits a linear relationship with $\mathcal{C}$:
\begin{equation}
    \eta\approx 4\mathcal{C}\frac{\gamma_\mathrm{e}}{\gamma_\mathrm{e}+1}\frac{\gamma_\mathrm{o}}{\gamma_\mathrm{o}+1}
\end{equation}

When microwave resonator and optical resonator are all critical coupled, we have the $\gamma_\mathrm{e} = 1, \gamma_\mathrm{o}=1$, causing $\eta\approx \mathcal{C}$.

\subsection{Link-equivalent half-wave voltage}

We use $V_{\pi,\mathrm{eff}}$ as a link-equivalent metric for transduction efficiency because $V_\pi$ is the standard modulation-efficiency parameter in microwave photonics link-gain analysis, where the small-signal link gain scales as $G_\mathrm{link}\propto1/V_\pi^2$. We define $V_{\pi,\mathrm{eff}}$ by mapping the measured photon-number conversion efficiency onto an ideal lossless electro-optic modulator operated with the same on-chip optical pump power and microwave impedance:
\begin{equation}
V_{\pi,\mathrm{eff}}
=\frac{\pi}{2}
\sqrt{\frac{Z_0P_\mathrm{p}}{\eta}\frac{\Omega}{\omega}}.
\end{equation}
Using $P_\mathrm{p}=26.1~\mathrm{mW}$ and $\eta\simeq10\%$ gives $V_{\pi,\mathrm{eff}}\simeq100~\mathrm{mV}$. Relative to a representative $6~\mathrm{V}$ modulator, this corresponds to an approximately $36~\mathrm{dB}$ enhancement in idealized small-signal link gain when the other link parameters are held constant. The extracted $V_{\pi,\mathrm{eff}}$ incorporates the interaction length and the optical and microwave resonant enhancement through the measured conversion efficiency. It characterizes the resonant link-level performance at the resonance centre over the measured $155~\mathrm{MHz}$ conversion bandwidth.

\section{Receiver-added noise and calibration}
\label{sec:SI_receiver_noise}

Using the receiver reference planes defined in \fref{fig:SI_receiver_reference_planes} and Table~\ref{tab:SI_reference_plane_definitions}, this section derives the receiver-added noise and details the laboratory RF-source-to-Ref.~P1 calibration and Ref.~P2--Ref.~P3 optical-detection calibration.
This section also includes measurements that characterized the microwave thermal-noise contribution and demonstrate coupling-dependent suppression of the cavity-internal thermal-noise contribution. Unless stated otherwise, insertion losses are quoted as positive quantities.

\subsection{Noise-temperature definition}

The equivalent input-referred receiver-added noise temperature $T_e$ is related to the noise factor $F$ and noise figure $\mathrm{NF}$ through
\begin{equation}
F=10^{\mathrm{NF}/10}=1+\frac{T_e}{T_0},
\qquad
T_e=T_0\left(10^{\mathrm{NF}/10}-1\right).
\label{eq:SI_noise_temperature_definition}
\end{equation}
We use $T_0=290~\mathrm{K}$ and report $T_e$ as the receiver-added noise temperature. For an external source or antenna with noise temperature $T_\mathrm{ant}$, the total equivalent input noise temperature is $T_\mathrm{ant}+T_e$.

For the Raman-suppressed device at 60.5~GHz, the on-chip noise figure is $1.4~\mathrm{dB}$, corresponding to $T_{e,\mathrm{on-chip}}\simeq110~\mathrm{K}$. Including the measured $5.27~\mathrm{dB}$ optical detection-chain penalty, the calculated full-system noise figure is $2.58~\mathrm{dB}$, corresponding to $T_{e,\mathrm{sys}}^\mathrm{th}\simeq236~\mathrm{K}$. The directly measured full-system noise figure is $2.3\pm0.3~\mathrm{dB}$, corresponding to $T_{e,\mathrm{sys}}^\mathrm{meas}=202^{+35}_{-33}~\mathrm{K}$ and agreeing with the independent loss-budget estimate within the experimental uncertainty.

\subsection{Input-referred thermal and quantum noise}

We consider the coupled Langevin equations for the anti-Stokes optical mode $a_\mathrm{as}$ and microwave mode $b$, retaining the anti-Stokes process:
\begin{equation}
\begin{aligned}
\partial_t a_\mathrm{as} &= -\frac{\kappa_0+\kappa_\mathrm{ex}}{2}a_\mathrm{as}+igb
+\sqrt{\kappa_0}\,a_\mathrm{in,0}+\sqrt{\kappa_\mathrm{ex}}\,a_\mathrm{in},\\
\partial_t b &= -\frac{\Gamma_0+\Gamma_\mathrm{ex}}{2}b+iga_\mathrm{as}
+\sqrt{\Gamma_0}\,b_\mathrm{in,0}+\sqrt{\Gamma_\mathrm{ex}}\,b_\mathrm{in}.
\end{aligned}
\label{eq:SI_noise_langevin}
\end{equation}
Here, $a_\mathrm{in,0}$ and $b_\mathrm{in,0}$ are the noise operators associated with the intrinsic optical and microwave loss channels. The microwave internal bath has thermal occupation
\begin{equation}
\bar n_\mathrm{th}(T)=\frac{1}{\exp\!\left(\hbar\Omega/k_\mathrm{B}T\right)-1}
\simeq\frac{k_\mathrm{B}T}{\hbar\Omega},
\label{eq:SI_thermal_occupation}
\end{equation}
where the final expression is the Rayleigh--Jeans approximation, valid at 60.5~GHz and room temperature. The optical thermal occupation is negligible.

With $\gamma_\mathrm{e}=\Gamma_\mathrm{ex}/\Gamma_0$, $\gamma_\mathrm{o}=\kappa_\mathrm{ex}/\kappa_0$, and $\mathcal{C}=4g^2/(\Gamma\kappa)$, the on-resonance photon-number conversion efficiency from the external microwave port to the optical output is
\begin{equation}
\eta_\mathrm{on-chip}=\frac{4\mathcal{C}}{(1+\mathcal{C})^2}
\frac{\gamma_\mathrm{e}}{1+\gamma_\mathrm{e}}
\frac{\gamma_\mathrm{o}}{1+\gamma_\mathrm{o}}.
\label{eq:SI_external_port_efficiency}
\end{equation}
The corresponding conversion efficiency from the intrinsic microwave loss channel is
\begin{equation}
\eta_0=\frac{4\mathcal{C}}{(1+\mathcal{C})^2}
\frac{1}{1+\gamma_\mathrm{e}}
\frac{\gamma_\mathrm{o}}{1+\gamma_\mathrm{o}}.
\end{equation}
Referring the internal thermal noise back to the external microwave input therefore gives
\begin{equation}
\bar N_\mathrm{th,add}
=\frac{\eta_0}{\eta_\mathrm{on-chip}}\bar n_\mathrm{th}(T_\mathrm{phys})
=\frac{\Gamma_0}{\Gamma_\mathrm{ex}}\bar n_\mathrm{th}(T_\mathrm{phys}).
\label{eq:SI_thermal_added_noise}
\end{equation}
The receiver-added thermal noise in Eq.~\eqref{eq:SI_thermal_added_noise} originates from the cavity's internal dissipative bath. Microwave overcoupling reduces this input-referred internal thermal-noise contribution in proportion to $\Gamma_0/\Gamma_\mathrm{ex}$. Thermal noise entering through the external coupling port belongs to the input source and is not included in the receiver-added noise.

Ideal phase-preserving heterodyne detection has a one-photon output noise floor in the photon-number convention used here. Referring this floor to the microwave input gives
\begin{equation}
\bar N_\mathrm{q,on-chip}=\frac{1}{\eta_\mathrm{on-chip}}.
\end{equation}
The total on-chip input-referred receiver-added noise is consequently
\begin{equation}
\bar N_\mathrm{add,on-chip}
=\frac{1}{\eta_\mathrm{on-chip}}
+\frac{\Gamma_0}{\Gamma_\mathrm{ex}}\bar n_\mathrm{th}(T_\mathrm{phys}).
\label{eq:SI_onchip_added_noise}
\end{equation}

Passive optical losses and detection penalties between Ref.~P2 and Ref.~P3 reduce the net detected efficiency to
\begin{equation}
\eta_\mathrm{sys}=\eta_\mathrm{on-chip}10^{-L_\mathrm{sys}^{(\mathrm{opt})}/10}.
\label{eq:SI_system_efficiency}
\end{equation}
The full-system input-referred receiver-added noise is then
\begin{equation}
\bar N_\mathrm{add,sys}
=\underbrace{\frac{10^{L_\mathrm{sys}^{(\mathrm{opt})}/10}}{\eta_\mathrm{on-chip}}}_{\bar N_\mathrm{q,sys}}
+\underbrace{\frac{\Gamma_0}{\Gamma_\mathrm{ex}}\bar n_\mathrm{th}(T_\mathrm{phys})}_{\bar N_\mathrm{th,add}}.
\label{eq:SI_system_added_noise}
\end{equation}
At room temperature, the corresponding receiver-added noise temperature is
\begin{equation}
T_{e,\mathrm{sys}}
=\frac{\hbar\Omega}{\eta_\mathrm{sys}k_\mathrm{B}}
+\frac{\Gamma_0}{\Gamma_\mathrm{ex}}T_\mathrm{phys}.
\label{eq:SI_system_noise_temperature}
\end{equation}
For $T_\mathrm{phys}=T_0$, Eqs.~\eqref{eq:SI_noise_temperature_definition} and \eqref{eq:SI_system_noise_temperature} give
\begin{equation}
\mathrm{NF}_\mathrm{sys}=10\log_{10}\!\left[
1+\frac{\hbar\Omega}{\eta_\mathrm{sys}k_\mathrm{B}T_0}
+\frac{\Gamma_0}{\Gamma_\mathrm{ex}}
\right].
\label{eq:SI_system_noise_figure}
\end{equation}
Replacing $\eta_\mathrm{sys}$ by $\eta_\mathrm{on-chip}$ gives the corresponding on-chip noise figure.

\subsection{RF and optical calibration}

The calibrated microwave input chain from the laboratory RF-source output to Ref.~P1 is summarized in Table~\ref{tab:SI_rf_input_loss}. The same calibration and de-embedding convention is used for the transduction and receiver-noise measurements.

\begin{table}[htbp]
\centering
\small
\setlength{\tabcolsep}{5pt}
\renewcommand{\arraystretch}{1.18}
\begin{tabular}{|>{\raggedright\arraybackslash}p{0.43\linewidth}|>{\centering\arraybackslash}p{0.17\linewidth}|>{\centering\arraybackslash}p{0.22\linewidth}|}
\hline
\textbf{Microwave input-chain contribution} & \textbf{Symbol} & \textbf{Insertion loss at 60.5~GHz} \\
\hline
RF probe (ACP65-A-GSG-150) & $L_\mathrm{probe}^{(\mathrm{RF})}$ & $0.98~\mathrm{dB}$ \\
\hline
Nominal 70-dB attenuator chain & $L_\mathrm{att}^{(\mathrm{RF})}$ & $68.6\pm0.1~\mathrm{dB}$ \\
\hline
Coaxial cable and RF-source calibration offset & $L_\mathrm{cable+src}^{(\mathrm{RF})}$ & $13.9\pm0.2~\mathrm{dB}$ \\
\hline
Total laboratory RF-source-to-Ref.~P1 input-chain loss & $L_\mathrm{in}^{(\mathrm{RF})}$ & $83.5\pm0.3~\mathrm{dB}$ \\
\hline
\end{tabular}
\caption{\textbf{Calibrated microwave input-chain losses.} RF insertion losses from the laboratory RF-source output to Ref.~P1 at 60.5~GHz. The total loss is calibrated out of both the transduction and receiver-noise measurements.}
\label{tab:SI_rf_input_loss}
\end{table}

The total input-chain loss is
\begin{equation}
L_\mathrm{in}^{(\mathrm{RF})}
=L_\mathrm{probe}^{(\mathrm{RF})}+L_\mathrm{att}^{(\mathrm{RF})}+L_\mathrm{cable+src}^{(\mathrm{RF})}
=83.5\pm0.3~\mathrm{dB}.
\end{equation}
Accordingly, the calibrated on-chip microwave power is $P_\mathrm{P1}=P_\mathrm{src}-L_\mathrm{in}^{(\mathrm{RF})}$. The uncertainty of the cable/source term includes the observed change caused by cable bending.

\begin{figure}[htbp]
  \centering
  \includegraphics[width=0.75\linewidth]{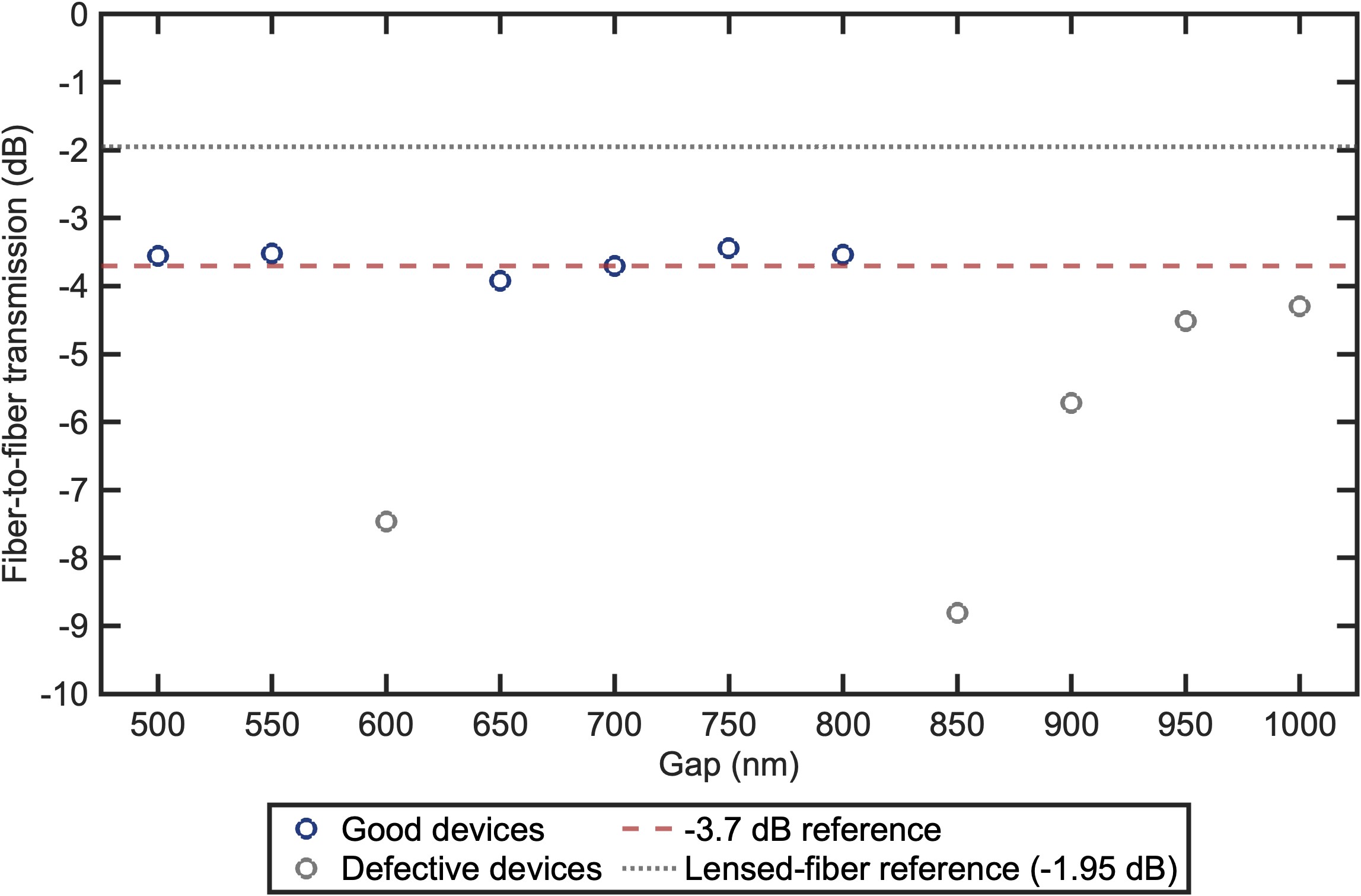}
  \caption{\textbf{Fiber-to-fiber coupling-loss calibration.} Measured transmission for devices with different coupling gaps between the bus waveguide and optical resonator. For devices without visible defects, the fiber-to-fiber loss is consistently $3.7~\mathrm{dB}$. The single-side insertion loss is therefore taken as half of the total. The measured $1.95~\mathrm{dB}$ loss of the lensed-fiber pair is not removed from the $3.7~\mathrm{dB}$.}
  \label{fig:SI_fiber_loss_calibration}
\end{figure}

After the electro-optic transducer, the converted optical signal experiences additional loss and detection penalties between Ref.~P2 and Ref.~P3. For the device used in the receiver-noise measurement, the measured fiber-to-fiber transmission loss is $3.7~\mathrm{dB}$, as shown in \fref{fig:SI_fiber_loss_calibration}. Because the input and output tapers are nominally identical and share the loss symmetrically, the inferred single-facet and lensed-fiber loss is $1.85~\mathrm{dB}$. The remaining optical loss before the balanced photodiode contributes $0.5~\mathrm{dB}$.

\begin{figure}[htbp]
  \centering
  \includegraphics[width=0.8\linewidth]{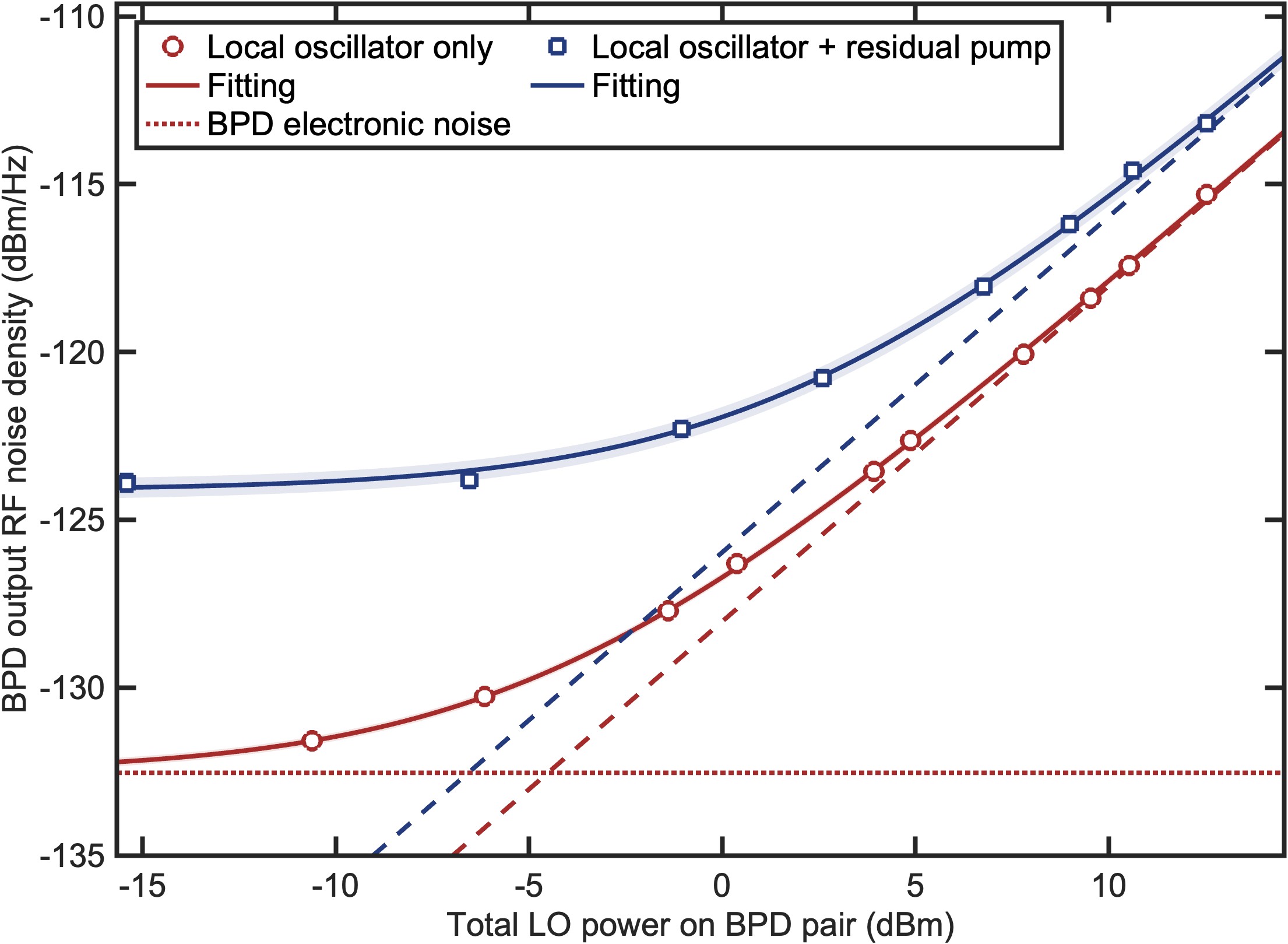}
  \caption{\textbf{Shot-noise calibration with residual pump and ASE leakage.} Measured balanced-photodiode RF-noise density versus total local-oscillator power incident on the photodiode pair. The solid curves are fits including the electronic and optical noise contributions; dashed lines show the corresponding shot-noise scaling.}
  \label{fig:SI_shot_noise_scaling}
\end{figure}

The balanced-photodiode quantum efficiency inferred from the shot-noise measurement is $\eta_\mathrm{BPD}=81\%$, equivalent to
\begin{equation}
L_\mathrm{BPD}^{(\mathrm{opt})}=-10\log_{10}(\eta_\mathrm{BPD})=0.92~\mathrm{dB}.
\end{equation}
Residual pump and ASE leakage adds a further $2.0~\mathrm{dB}$ detection-noise penalty. This is an equivalent signal-to-noise penalty rather than a passive insertion loss, and is folded into $L_\mathrm{sys}^{(\mathrm{opt})}$ only for full-system noise accounting. \Fref{fig:SI_shot_noise_scaling} verifies that the balanced detector follows the expected shot-noise scaling above its electronic-noise floor, both with the local oscillator alone and with residual pump and ASE light present.

\begin{table}[htbp]
\centering
\small
\setlength{\tabcolsep}{5pt}
\renewcommand{\arraystretch}{1.18}
\begin{tabular}{|>{\raggedright\arraybackslash}p{0.43\linewidth}|>{\centering\arraybackslash}p{0.17\linewidth}|>{\centering\arraybackslash}p{0.22\linewidth}|}
\hline
\textbf{Optical detection-chain contribution} & \textbf{Symbol} & \textbf{Equivalent loss or penalty} \\
\hline
Output-facet and lensed-fiber loss & $L_\mathrm{facet}^{(\mathrm{opt})}$ & $1.85~\mathrm{dB}$ \\
\hline
Optical path to the balanced photodiode & $L_\mathrm{path}^{(\mathrm{opt})}$ & $0.5~\mathrm{dB}$ \\
\hline
Balanced-photodiode quantum efficiency ($81\%$) & $L_\mathrm{BPD}^{(\mathrm{opt})}$ & $0.92~\mathrm{dB}$ \\
\hline
Residual pump and ASE leakage & $L_\mathrm{pump}^{(\mathrm{opt})}$ & $2.0~\mathrm{dB}$ \\
\hline
Total Ref.~P2--Ref.~P3 equivalent loss & $L_\mathrm{sys}^{(\mathrm{opt})}$ & $5.27~\mathrm{dB}$ \\
\hline
\end{tabular}
\caption{\textbf{Calibrated optical detection-chain losses.} Optical losses and equivalent detection penalties between Ref.~P2 and Ref.~P3. They are included in the full-system noise figure but are calibrated out of the on-chip noise figure.}
\label{tab:SI_optical_detection_loss}
\end{table}

\begin{figure}[htbp]
  \centering
  \includegraphics[width=0.80\linewidth]{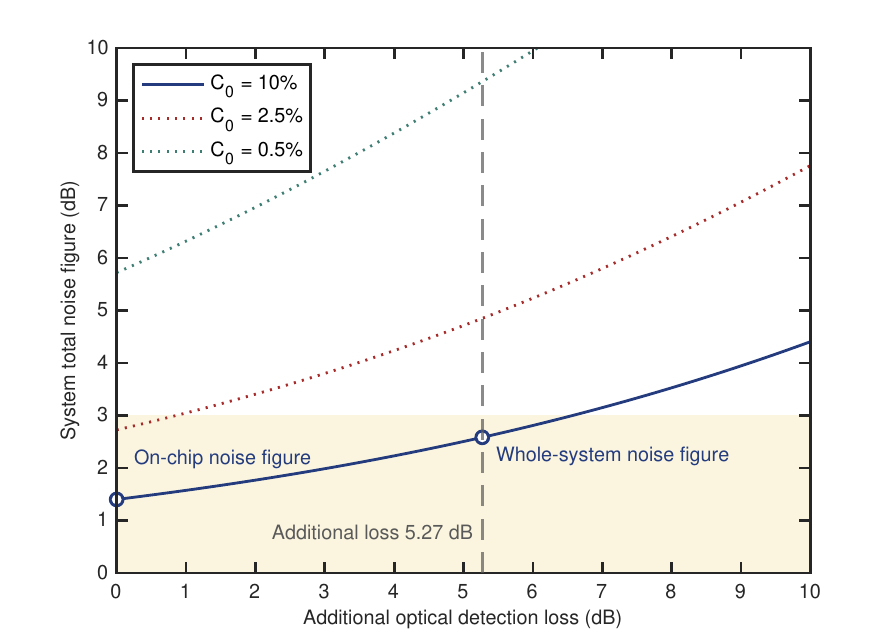}
  \caption{\textbf{Optical detection-loss contribution to the system noise figure.} Calculated full-system noise figure versus additional optical detection loss for different critical-coupling cooperativities $C_0$. For the present device with $C_0=10\%$, the measured $5.27~\mathrm{dB}$ penalty increases the calculated noise figure from $1.4~\mathrm{dB}$ on chip to $2.58~\mathrm{dB}$ for the full system. The yellow region indicates $\mathrm{NF}<3~\mathrm{dB}$, equivalent to $T_e<290~\mathrm{K}$.}
  \label{fig:SI_detection_loss_nf}
\end{figure}

The total optical detection-chain penalty is therefore
\begin{equation}
L_\mathrm{sys}^{(\mathrm{opt})}
=L_\mathrm{facet}^{(\mathrm{opt})}+L_\mathrm{path}^{(\mathrm{opt})}
+L_\mathrm{BPD}^{(\mathrm{opt})}+L_\mathrm{pump}^{(\mathrm{opt})}
=5.27~\mathrm{dB}.
\end{equation}
\Fref{fig:SI_detection_loss_nf} shows how this measured penalty changes the full-system noise figure. For the present device, the $5.27~\mathrm{dB}$ penalty increases the calculated noise figure from $1.4~\mathrm{dB}$ on chip to $2.58~\mathrm{dB}$ for the complete Ref.~P1--Ref.~P3 system.

Table~\ref{tab:SI_receiver_noise_parameters} summarizes the device and heterodyne experiment parameters. The optical pump power before the input facet is $40.0~\mathrm{mW}$. After the calibrated $1.85~\mathrm{dB}$ input-facet and lensed-fiber loss, the on-chip pump power is
\begin{equation}
P_\mathrm{p,on-chip}=40.0~\mathrm{mW}\times10^{-1.85/10}=26.1~\mathrm{mW}.
\end{equation}
For resonant pumping of the critically coupled optical mode, $\kappa_\mathrm{ex}=\kappa_0$ and $\kappa/2\pi=155~\mathrm{MHz}$. The corresponding intracavity pump photon number is
\begin{equation}
n_p=\frac{4\kappa_\mathrm{ex}}{\kappa^2}\frac{P_\mathrm{p,on-chip}}{\hbar\omega_p}
=\frac{2P_\mathrm{p,on-chip}}{\hbar\omega_p\kappa}
\simeq4.2\times10^8,
\end{equation}
where the pump wavelength is approximately $1550~\mathrm{nm}$.

\begin{table}[htbp]
\centering
\small
\setlength{\tabcolsep}{4pt}
\renewcommand{\arraystretch}{1.18}
\begin{tabular}{|>{\raggedright\arraybackslash}p{0.39\linewidth}|>{\centering\arraybackslash}p{0.20\linewidth}|>{\centering\arraybackslash}p{0.28\linewidth}|}
\hline
\textbf{Operating or calibration parameter} & \textbf{Symbol} & \textbf{Value or configuration} \\
\hline
Intrinsic mmWave-cavity loss rate & $\Gamma_0/2\pi$ & $1.7~\mathrm{GHz}$ \\
\hline
External mmWave coupling rate, bidirectional transduction measurement & $\Gamma_\mathrm{ex}/2\pi$ & $1.7~\mathrm{GHz}$ ($\Gamma_\mathrm{ex}/\Gamma_0=1$) \\
\hline
External mmWave coupling rate, receiver-noise measurement & $\Gamma_\mathrm{ex}/2\pi$ & $9.0~\mathrm{GHz}$ ($\Gamma_\mathrm{ex}/\Gamma_0=5.3$) \\
\hline
Pump power immediately before the input facet & $P_\mathrm{p,in}$ & $40.0~\mathrm{mW}$ \\
\hline
Input-facet and lensed-fiber pump-delivery loss & $L_\mathrm{p,in}^{(\mathrm{opt})}$ & $1.85~\mathrm{dB}$ \\
\hline
Calibrated on-chip pump power & $P_\mathrm{p,on-chip}$ & $26.1~\mathrm{mW}$ \\
\hline
Intracavity pump photon number & $n_p$ & $4.2\times10^8$ \\
\hline
Output-facet and lensed-fiber loss & $L_\mathrm{facet}^{(\mathrm{opt})}$ & $1.85~\mathrm{dB}$ \\
\hline
Pump-filter configuration & -- & Pre-chip FBG for input-ASE suppression; nearly critically coupled optical ring for on-chip pump rejection; no post-chip FBG \\
\hline
Residual pump and ASE leakage penalty above shot noise & $L_\mathrm{pump}^{(\mathrm{opt})}$ & $2.0~\mathrm{dB}$ \\
\hline
Total LO power incident on the two BPD ports & $P_\mathrm{LO}$ & $18~\mathrm{mW}$ \\
\hline
Balanced-photodiode quantum efficiency & $\eta_\mathrm{BPD}$ & $81\%$ \\
\hline
\end{tabular}
\caption{\textbf{Receiver-noise measurement parameters.} Device, optical-pump, and heterodyne operating parameters used for the receiver-noise measurement.}
\label{tab:SI_receiver_noise_parameters}
\end{table}

The optical pump-delivery loss is used to calibrate $P_\mathrm{p,on-chip}$ and $n_p$, while the output coupling loss, subsequent optical-path loss, balanced-photodiode quantum efficiency, and measured pump and ASE leakage penalty are included in $\mathrm{NF}_\mathrm{sys}$. The in-situ calibration tone provides the frequency-dependent response of the post-detection chain, whose electronic background remains below the optical shot-noise floor at the operating LO power.

\subsection{Experimental characterization of microwave thermal noise}

We characterize the microwave thermal noise through its dependence on optical detuning, the presence of the microwave resonator, and pump power.

\begin{figure}[htbp]
  \centering
  \includegraphics[width=0.6\linewidth]{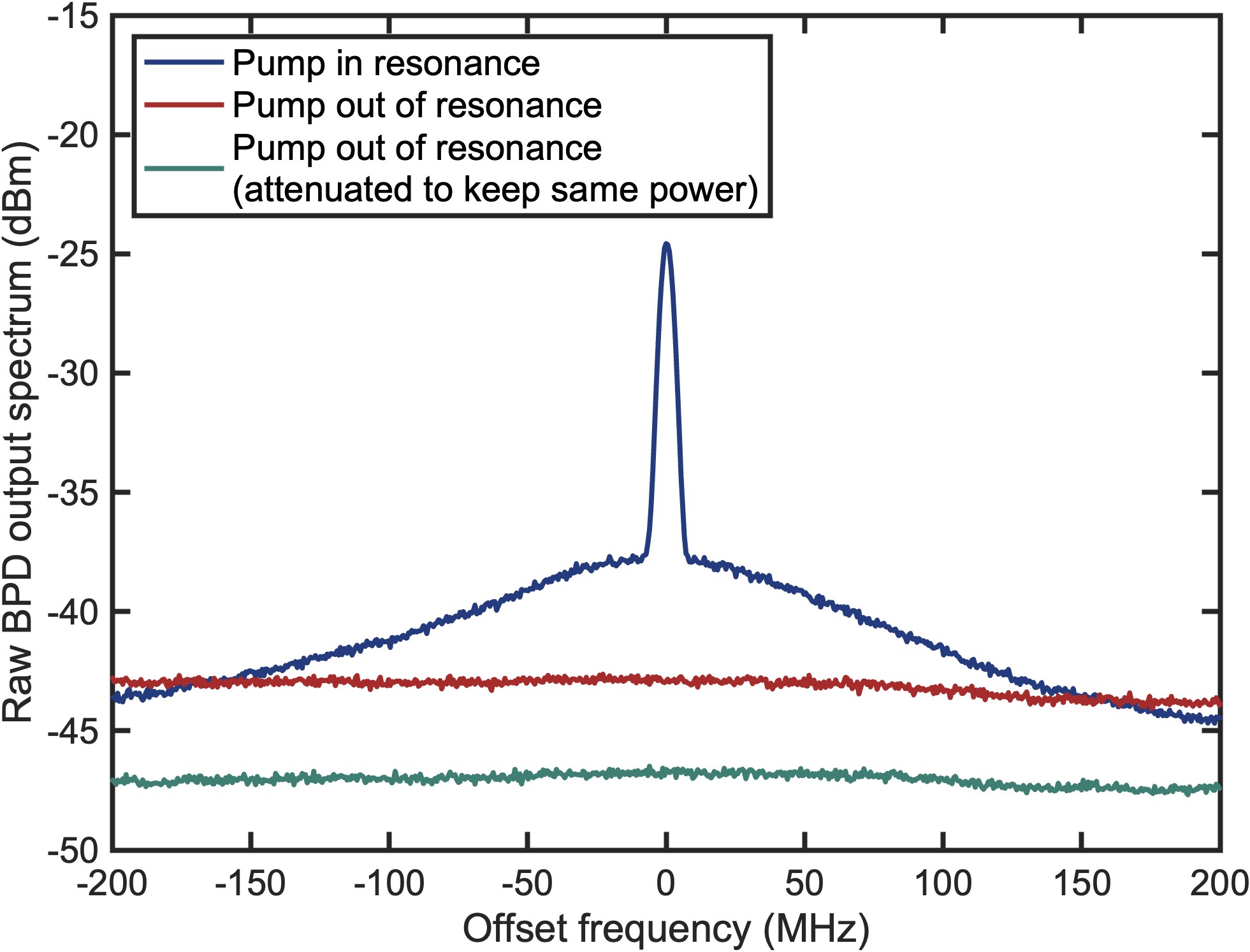}
  \caption{\textbf{Cavity noise versus optical-pump detuning.} Raw balanced-photodiode output spectra with the optical pump on resonance, off resonance, and off resonance at matched detected optical power. The narrow central peak is the calibration tone. The Lorentzian-shaped noise contribution appears only when the pump is resonant.}
  \label{fig:SI_pump_detuning_noise_control}
\end{figure}

We first measure noise with the optical pump on and off resonance, as shown in \fref{fig:SI_pump_detuning_noise_control}. To separate the effect of pump detuning from the accompanying change in detected optical power, we repeat the off-resonant measurement with the detected power matched to the resonant case. The Lorentzian-shaped noise contribution disappears when the pump is detuned and is observed only under resonant pumping.

\begin{figure}[htbp]
  \centering
  \includegraphics[width=1.0\linewidth]{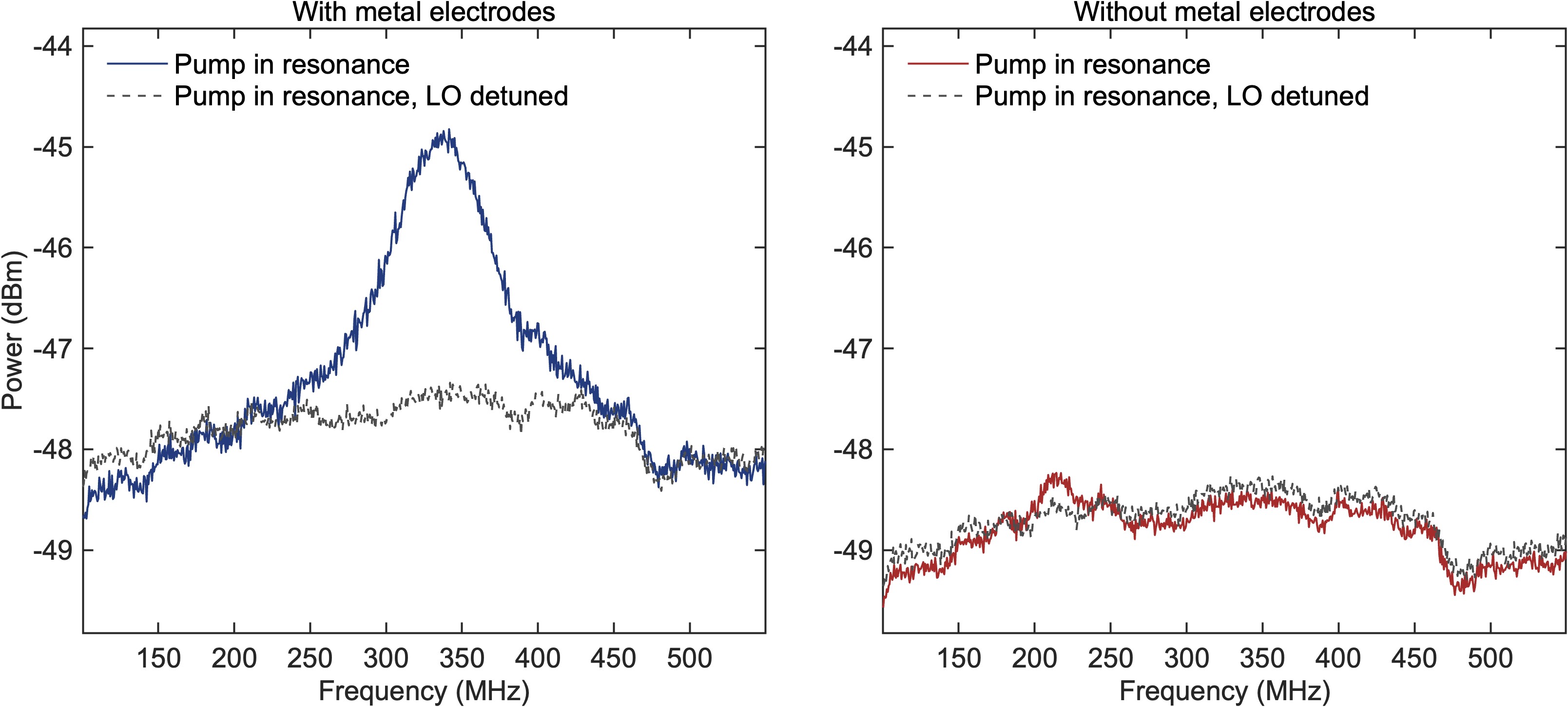}
  \caption{\textbf{Devices with and without a microwave resonator.} Raw heterodyne spectra for a device with metal electrodes forming the microwave resonator (left) and an otherwise similar device from which the metal layer was removed (right). Solid traces use a resonant optical pump; dashed traces detune the local oscillator from the converted sideband. The cavity-shaped feature is present only in the device containing the microwave resonator.}
  \label{fig:SI_with_without_microwave_resonator}
\end{figure}

We next compare resonantly pumped devices with and without the metal electrodes that form the microwave resonator, as shown in \fref{fig:SI_with_without_microwave_resonator}. The device with the microwave resonator exhibits a broad Lorentzian feature around the intermediate frequency corresponding to the $60.5~\mathrm{GHz}$ optical sideband. Detuning the local oscillator removes this feature. No corresponding feature is observed in the device without metal electrodes under the same resonant pumping condition. The dependence of the cavity-shaped noise on the presence of the metal resonator associates it with the microwave mode.

\begin{figure}[htbp]
  \centering
  \includegraphics[width=0.65\linewidth]{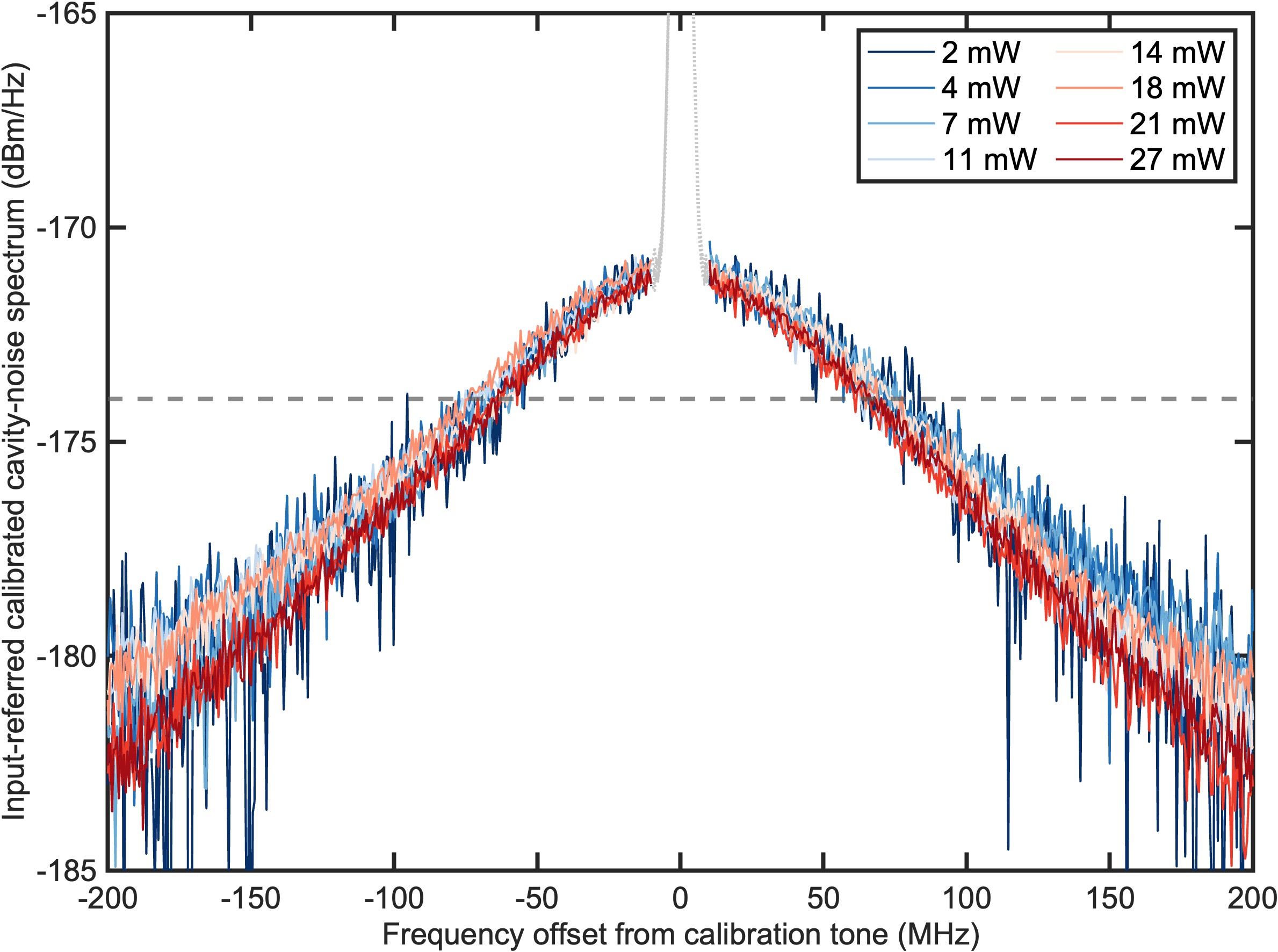}
  \caption{\textbf{Input-referred cavity-noise spectra versus pump power.} Calibrated spectra at critical microwave coupling for on-chip pump powers from $2$ to $27~\mathrm{mW}$. The central calibration tone is excluded from the noise analysis. The dashed horizontal line marks the ideal room-temperature thermal-noise level at critical coupling.}
  \label{fig:SI_noise_power_sweep}
\end{figure}

We then measure the input-referred spectra at critical microwave coupling for calibrated on-chip pump powers from $2$ to $27~\mathrm{mW}$, as shown in \fref{fig:SI_noise_power_sweep}. After calibration by the known microwave tone, the Lorentzian cavity-noise envelopes collapse to nearly the same level over the measured $\pm200~\mathrm{MHz}$ offset-frequency range. The input-referred noise is therefore independent of the optical transduction gain within the experimental uncertainty, as expected for equilibrium microwave thermal noise.

\begin{figure}[htbp]
  \centering
  \includegraphics[width=0.6\linewidth]{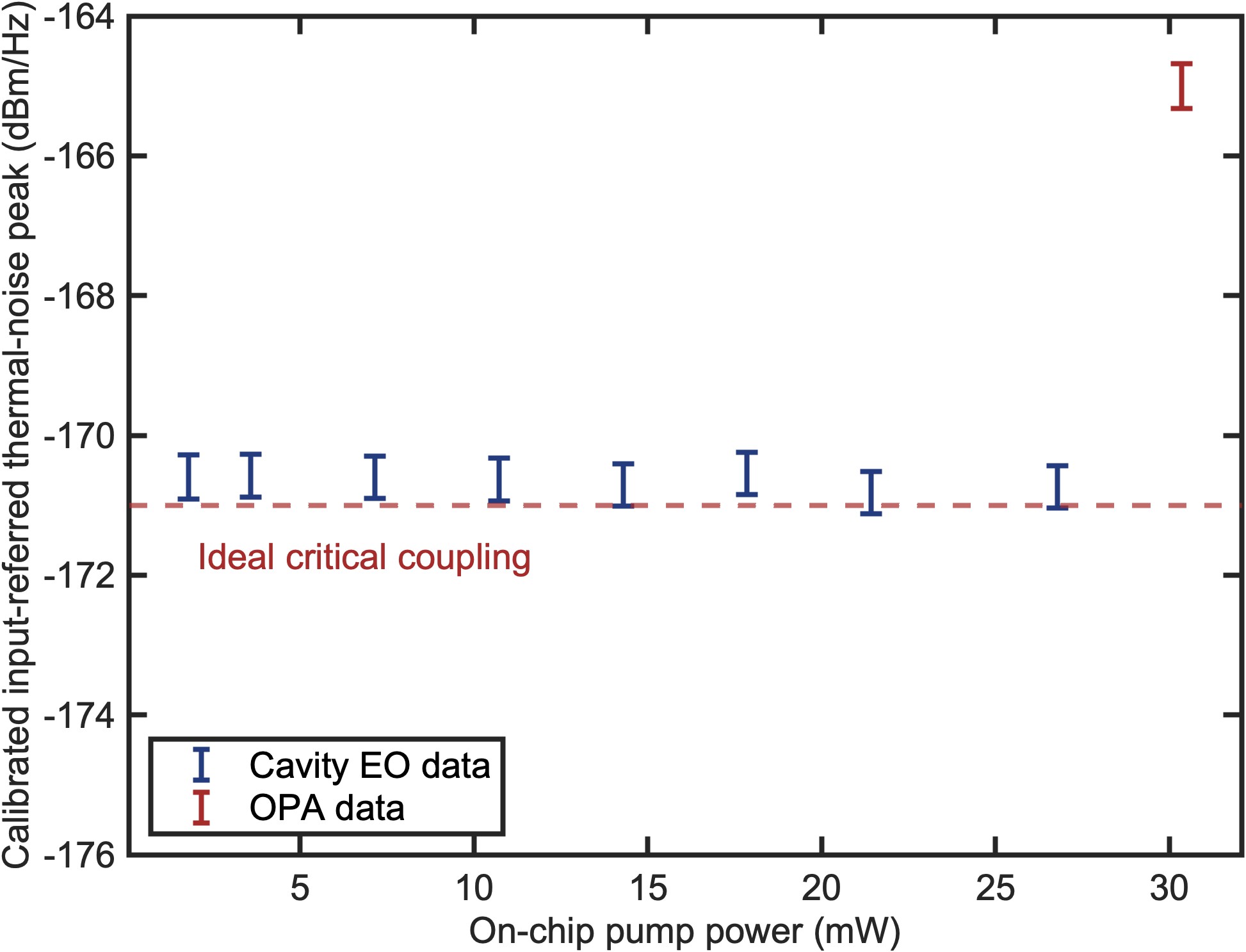}
  \caption{\textbf{Pump-power dependence of the input-referred thermal-noise peak.} Lorentzian peak levels extracted at critical microwave coupling. The blue points from $2$ to $27~\mathrm{mW}$ remain consistent with the ideal critical-coupling thermal-noise level, whereas the $30~\mathrm{mW}$ point rises after the onset of optical parametric instability. Error bars are obtained from the Lorentzian fits.}
  \label{fig:SI_noise_peak_power_dependence}
\end{figure}

\begin{figure}[htbp]
  \centering
  \includegraphics[width=1.0\linewidth]{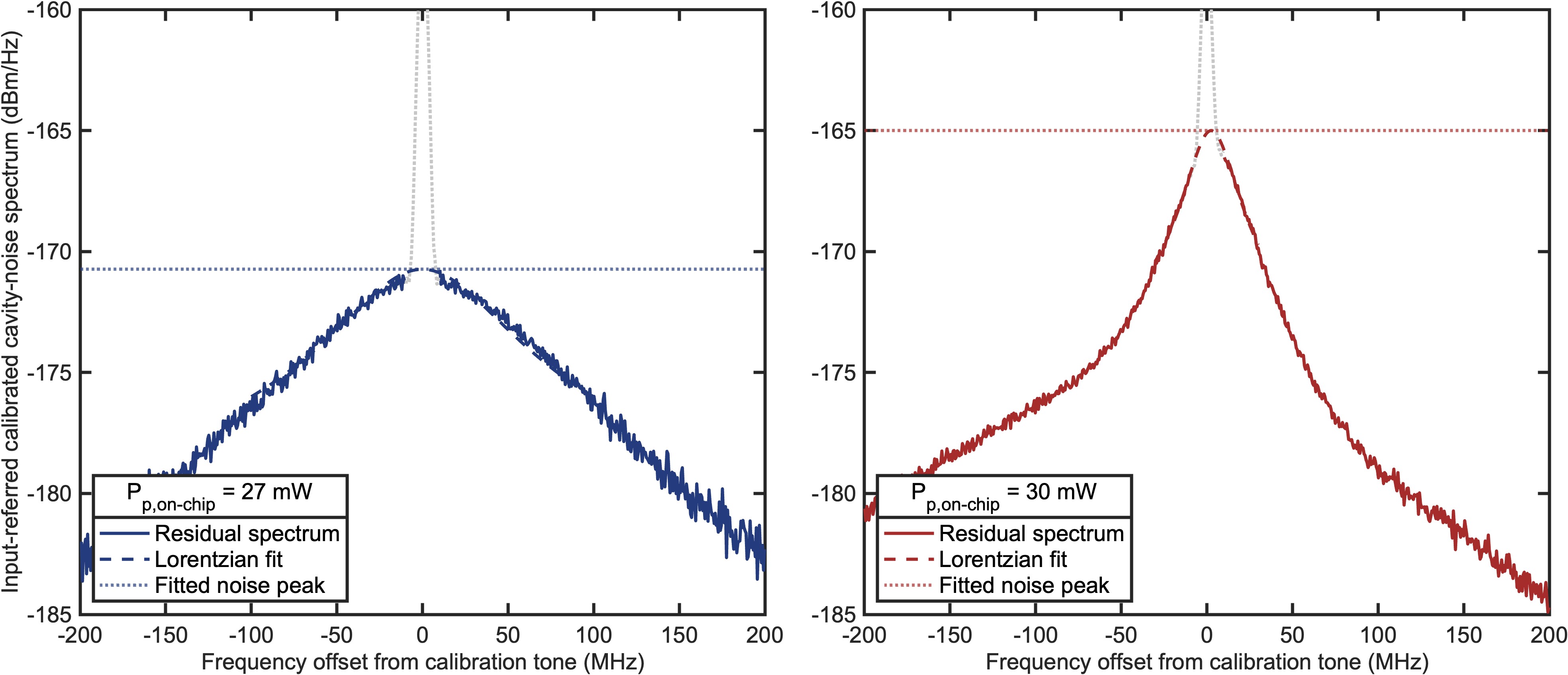}
  \caption{\textbf{Cavity-noise spectra below and above the optical-parametric-instability threshold.} Residual calibrated spectra and Lorentzian fits at $27~\mathrm{mW}$ (left) and $30~\mathrm{mW}$ (right) on-chip pump power. The fitted input-referred noise peak increases sharply at $30~\mathrm{mW}$.}
  \label{fig:SI_noise_parametric_threshold}
\end{figure}

\Fref{fig:SI_noise_peak_power_dependence} summarizes the Lorentzian peak extracted from the same power sweep. The fitted peak remains consistent with the critical-coupling thermal-noise prediction from $2$ to $27~\mathrm{mW}$. At $30~\mathrm{mW}$, however, the fitted noise rises sharply. The corresponding spectra in \fref{fig:SI_noise_parametric_threshold} show that this increase coincides with the onset of optical parametric instability. The receiver measurements reported here use pump powers below the optical parametric threshold.

Together, the optical-detuning and device-comparison measurements associate the Lorentzian noise with resonant electro-optic transduction of the microwave mode. The pump-power measurements further show that the input-referred noise remains at the equilibrium level below the optical-parametric threshold. We next examine its dependence on the microwave external coupling rate.

\subsection{Coupling-dependent suppression of internal thermal added noise}

We write the microwave coupling ratio as $r=\Gamma_\mathrm{ex}/\Gamma_0$. In the low-cooperativity coupling-dependence model, for fixed pump power and optical coupling, the on-chip conversion efficiency scales relative to its value at critical microwave coupling as
\begin{equation}
\eta_\mathrm{on-chip}(r)=\eta_\mathrm{crit}\frac{4r}{(1+r)^2}.
\label{eq:SI_efficiency_vs_coupling}
\end{equation}
Combining Eqs.~\eqref{eq:SI_system_noise_temperature} and \eqref{eq:SI_efficiency_vs_coupling} gives the on-chip noise temperature
\begin{equation}
T_{e,\mathrm{on-chip}}(r)
=\frac{(1+r)^2}{4r}\frac{\hbar\Omega}{\eta_\mathrm{crit}k_\mathrm{B}}
+\frac{T_\mathrm{phys}}{r}.
\label{eq:SI_noise_temperature_vs_coupling}
\end{equation}
The first term is minimized at critical coupling, whereas the internal thermal contribution decreases monotonically with overcoupling. Defining $T_{q0}=\hbar\Omega/(\eta_\mathrm{crit}k_\mathrm{B})$, the coupling ratio that minimizes Eq.~\eqref{eq:SI_noise_temperature_vs_coupling} is
\begin{equation}
r_\mathrm{opt}=\sqrt{1+\frac{4T_\mathrm{phys}}{T_{q0}}}.
\label{eq:SI_optimum_coupling}
\end{equation}
Low conversion efficiency gives $T_{q0}\gg T_\mathrm{phys}$ and $r_\mathrm{opt}\simeq1$. As the conversion efficiency increases, an overcoupled solution becomes optimal because the reduction in internal thermal noise outweighs the quantum-noise penalty associated with overcoupling.

\begin{figure}[htbp]
  \centering
  \includegraphics[width=1.0\linewidth]{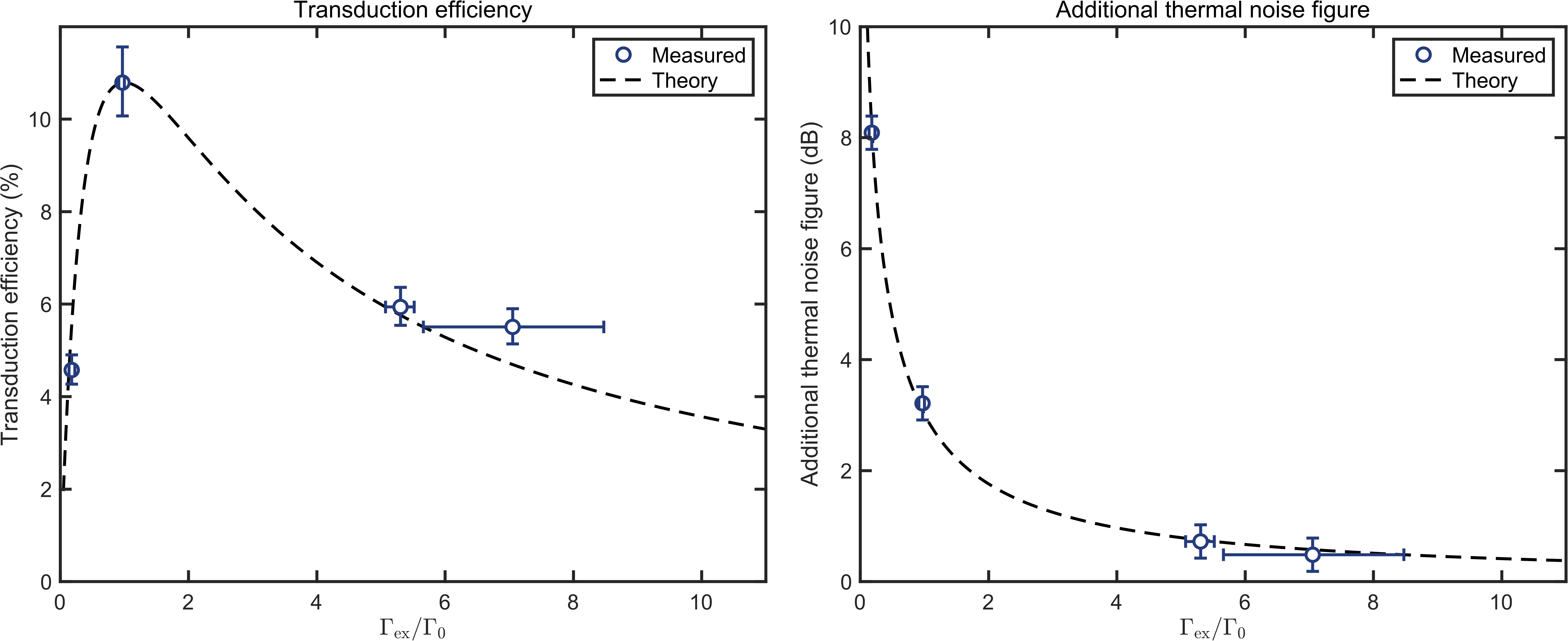}
  \caption{\textbf{Conversion-efficiency and thermal-noise trade-off.} Measured on-chip transduction efficiency (left) and additional internal thermal-noise figure (right) versus $\Gamma_\mathrm{ex}/\Gamma_0$, together with the theoretical curves.}
  \label{fig:SI_coupling_tradeoff}
\end{figure}

We test this theoretical prediction by varying the microwave external coupling rate through adjustments to the probe landing position while keeping the optical pump and detection chain unchanged. \Fref{fig:SI_coupling_tradeoff} compares the measured conversion efficiency and internal thermal-noise contribution with the predictions of Eqs.~\eqref{eq:SI_efficiency_vs_coupling} and \eqref{eq:SI_thermal_added_noise}. Overcoupling suppresses internal thermal noise while reducing the conversion efficiency away from critical coupling, thereby increasing the optical quantum-noise contribution as the transduction efficiency decreases.

\begin{figure}[htbp]
  \centering
  \includegraphics[width=1.0\linewidth]{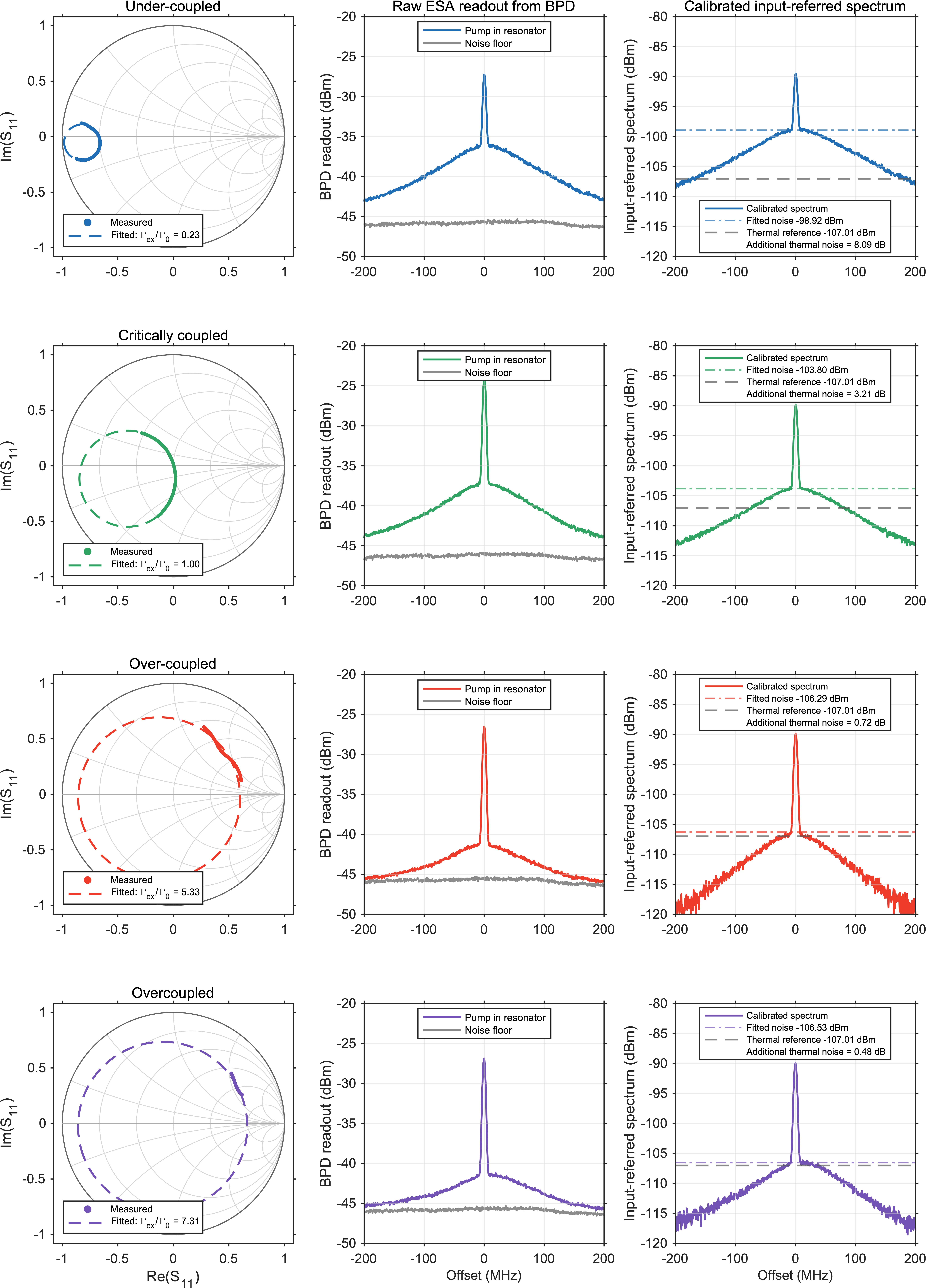}
  \caption{\textbf{Dependence of thermal noise on microwave coupling.} Rows show undercoupled, critically coupled, and two overcoupled operating points. Columns show the fitted $S_{11}$ response, raw balanced-photodiode spectrum, and calibrated input-referred spectrum. The extracted internal thermal-noise contribution decreases from $8.09$ to $0.48~\mathrm{dB}$ as $\Gamma_\mathrm{ex}/\Gamma_0$ increases.}
  \label{fig:SI_coupling_raw_noise}
\end{figure}

The fitted $S_{11}$ responses, raw balanced-photodiode spectra, and calibrated input-referred spectra underlying this comparison are shown in \fref{fig:SI_coupling_raw_noise} for undercoupled, critically coupled, and overcoupled operating points. The extracted internal thermal-noise contribution decreases from $8.09$ to $0.48~\mathrm{dB}$ as $\Gamma_\mathrm{ex}/\Gamma_0$ increases.

\begin{figure}[htbp]
  \centering
  \includegraphics[width=0.6\linewidth]{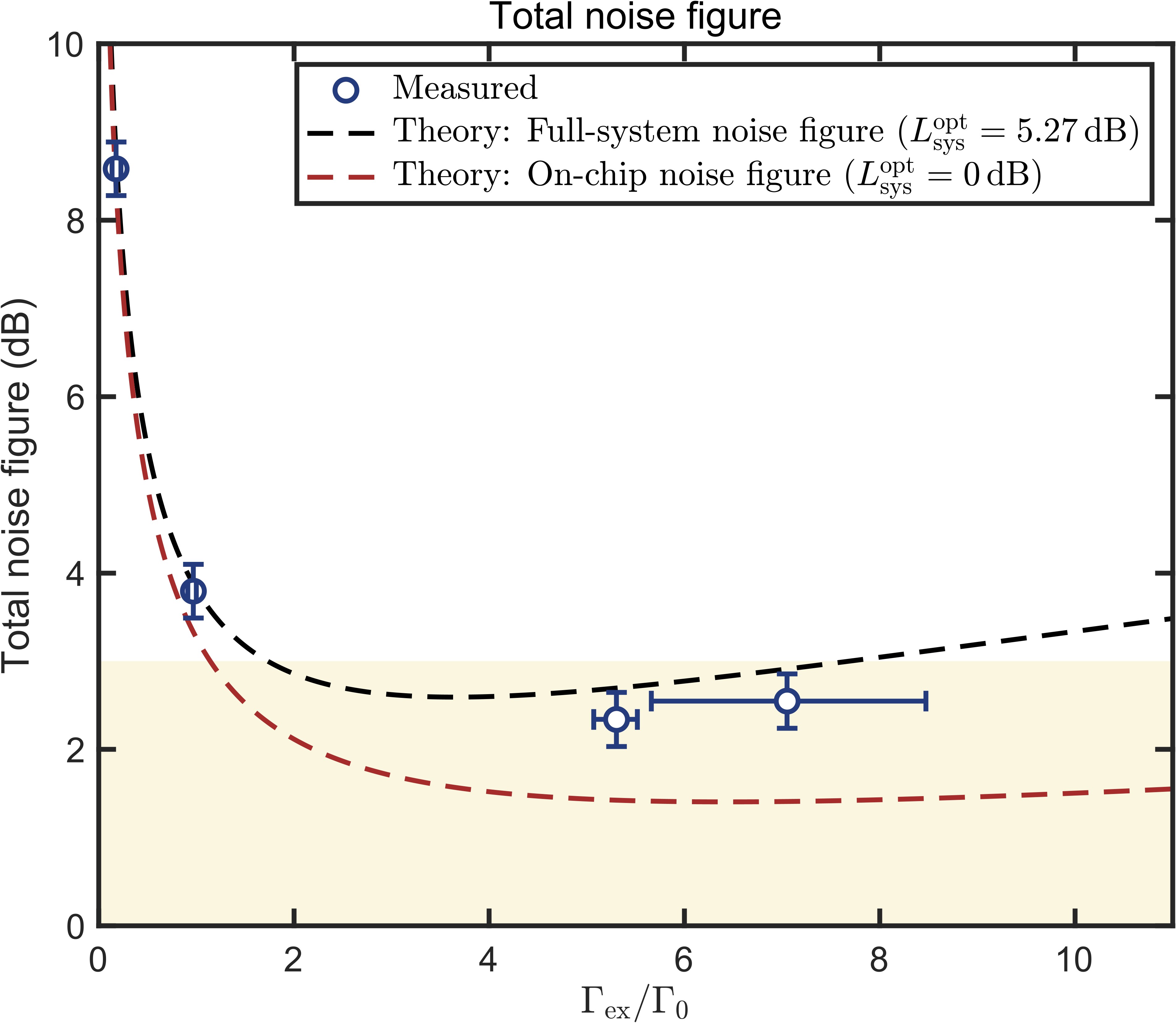}
  \caption{\textbf{Full-system noise figure versus microwave coupling.} Measured total noise figures are compared with full-system and on-chip theory as functions of $\Gamma_\mathrm{ex}/\Gamma_0$. The shaded region marks noise figures below $3~\mathrm{dB}$.}
  \label{fig:SI_total_nf_coupling}
\end{figure}

\begin{figure}[htbp]
  \centering
  \includegraphics[width=0.6\linewidth]{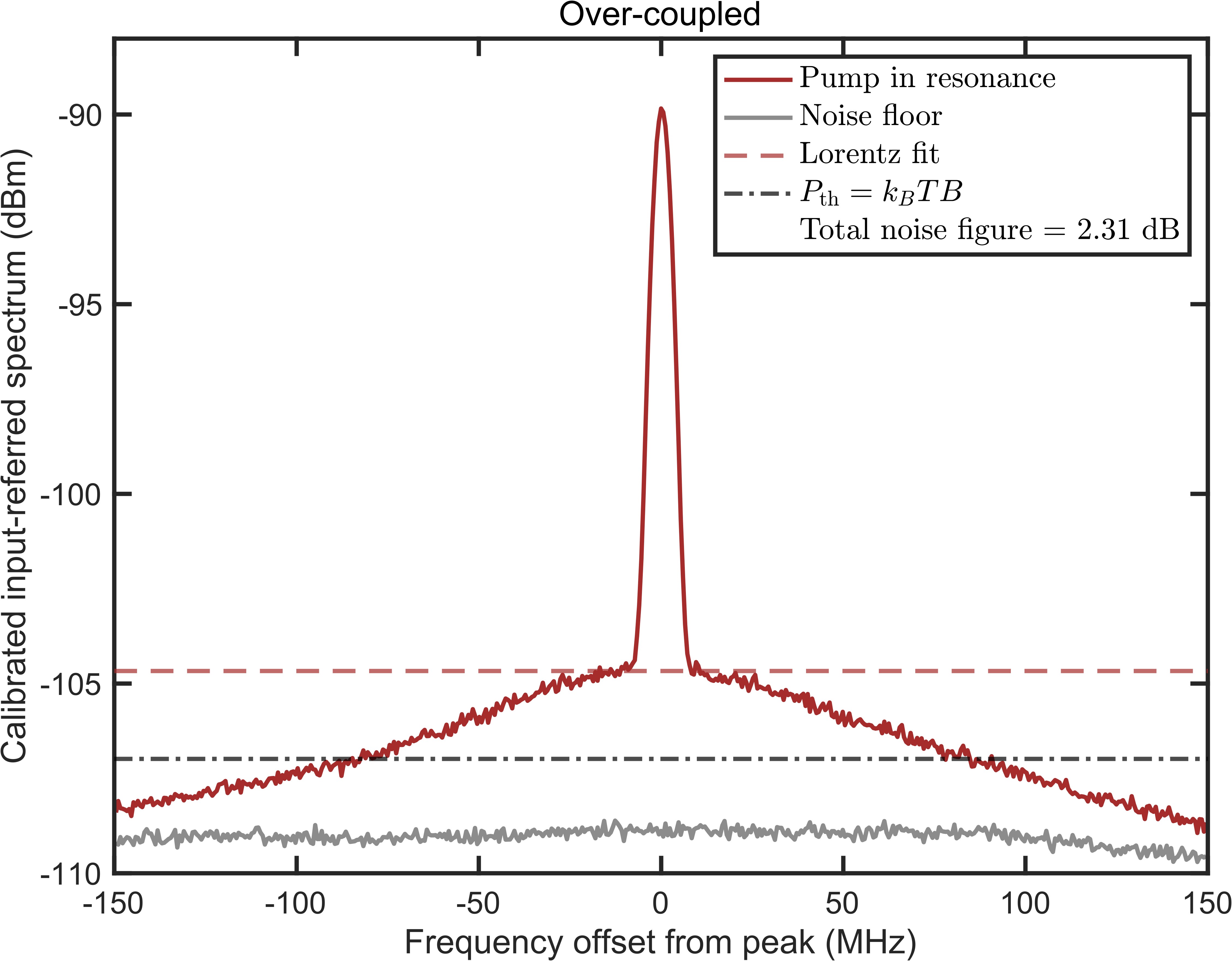}
  \caption{\textbf{Calibrated noise spectrum in the overcoupled regime.} Input-referred spectrum and Lorentzian fit for an overcoupled receiver, together with the measurement floor and room-temperature thermal reference. The extracted full-system noise figure is $2.31~\mathrm{dB}$ from the on-chip microwave input at Ref.~P1 to the detected IF output at Ref.~P3, including all optical detection penalties.}
  \label{fig:SI_overcoupled_spectrum}
\end{figure}

The resulting measured full-system noise figures are shown in \fref{fig:SI_total_nf_coupling}. The full-system noise figure falls below $3~\mathrm{dB}$ in the overcoupled regime and agrees with the model that includes the measured optical detection-chain penalty. A representative calibrated overcoupled spectrum is shown in \fref{fig:SI_overcoupled_spectrum}.

\section{Receiver linearity, high-power response, and survivability}

At low input powers, the electro-optic receiver operates linearly, as characterized by two-tone intermodulation and the input third-order intercept point (IIP3). Increasing the RF drive produces pump depletion and electro-optic comb formation, leading to compression and eventually saturation; the onset of compression is characterized by the input 1-dB compression point, $IP_{1\mathrm{dB}}$. In this section, we further measure the response at high RF power and examine voltage breakdown and overcurrent failure using destructive DC tests.

\subsection{Two-tone linearity, IIP3, and SFDR}

\begin{figure}[htbp]
  \centering
  \includegraphics[width=0.62\linewidth]{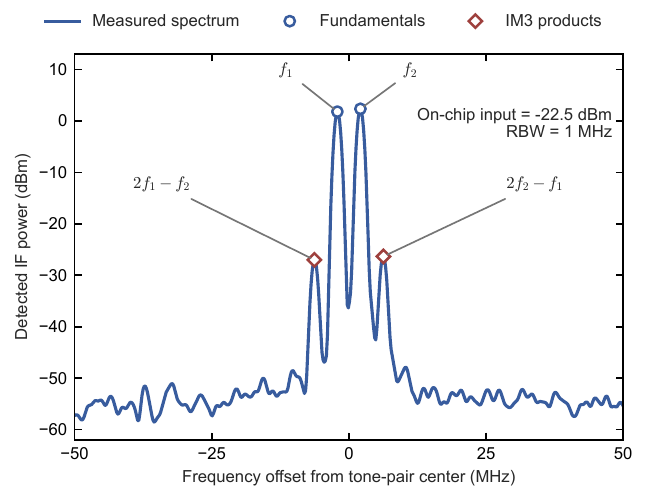}
  \caption{\textbf{Representative two-tone spectrum.} Detected IF spectrum for a calibrated on-chip RF input power of $-22.5~\mathrm{dBm}$. The labeled peaks are the two fundamentals, $f_1$ and $f_2$, and the third-order intermodulation products at $2f_1-f_2$ and $2f_2-f_1$. The resolution bandwidth is $1~\mathrm{MHz}$.}
  \label{fig:SI_iip3_spectrum}
\end{figure}

\begin{figure}[htbp]
  \centering
  \includegraphics[width=0.98\linewidth]{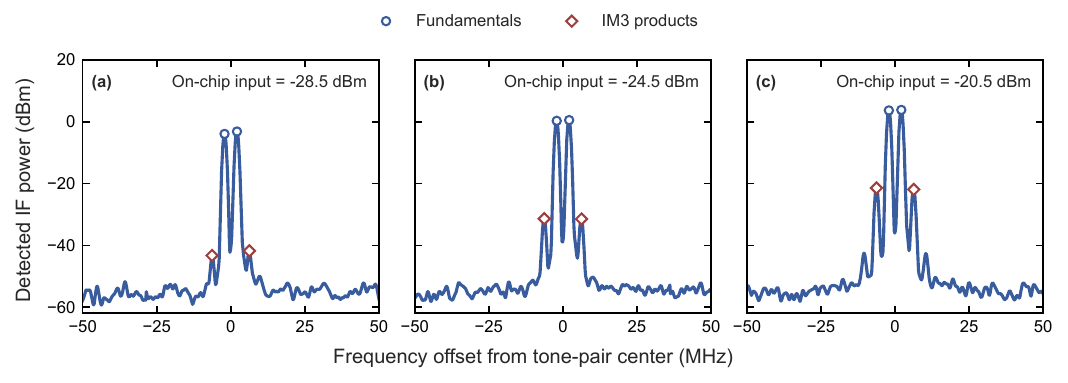}
  \caption{\textbf{Power-dependent evolution of the two-tone spectrum.} Panels (a)--(c) show the detected IF spectra at calibrated on-chip input powers of $-28.5$, $-24.5$, and $-20.5~\mathrm{dBm}$, respectively, without vertical offsets. Blue circles mark the fundamentals, and red diamonds mark the IM3 products. The IM3 products rise more rapidly than the fundamentals as the input power increases.}
  \label{fig:SI_iip3_spectral_evolution}
\end{figure}

\begin{figure}[htbp]
  \centering
  \includegraphics[width=0.52\linewidth]{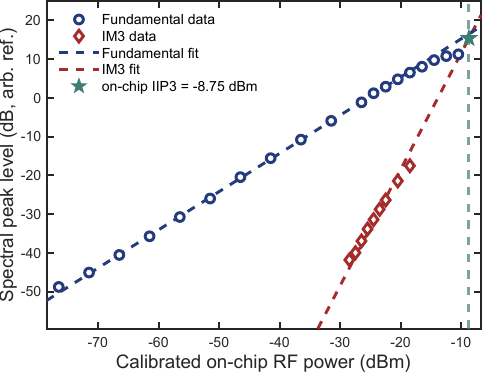}
  \caption{\textbf{On-chip two-tone linearity.} Measured fundamental tones and third-order intermodulation products as functions of the calibrated on-chip RF input power. Linear fits give an on-chip input third-order intercept point of $-8.75\pm0.8~\mathrm{dBm}$.}
  \label{fig:SI_iip3}
\end{figure}

We characterize third-order nonlinearity using two calibrated microwave input tones and record the resulting fundamental and third-order intermodulation (IM3) products at the detected intermediate-frequency output. A representative spectrum at an on-chip RF input power of $-22.5~\mathrm{dBm}$ is shown in \fref{fig:SI_iip3_spectrum}; both fundamentals and both IM3 products are resolved using the same resolution bandwidth and lie within the transducer's 3-dB bandwidth. \Fref{fig:SI_iip3_spectral_evolution} compares spectra at three input levels. From $-28.5$ to $-20.5~\mathrm{dBm}$, the IM3 products rise more rapidly than the fundamentals. \Fref{fig:SI_iip3} shows their measured spectral peak levels as functions of the calibrated on-chip RF input power. Linear fits to the fundamental and IM3 data give an on-chip input third-order intercept point of
\begin{equation}
\mathrm{IIP3}=-8.75\pm0.8~\mathrm{dBm}.
\end{equation}
The corresponding third-order spurious-free dynamic range is determined by this intercept point and the input-referred noise spectral density $N_\mathrm{in}$ at the same operating point,
\begin{equation}
\mathrm{SFDR}_3=\frac{2}{3}\left(\mathrm{IIP3}-N_\mathrm{in}\right),
\label{eq:SI_sfdr_definition}
\end{equation}
where IIP3 and $N_\mathrm{in}$ are expressed in dBm and dBm/Hz, respectively, giving $\mathrm{SFDR}_3$ in $\mathrm{dB\,Hz}^{2/3}$. We combine the measured on-chip IIP3 with the full-system input-referred noise spectral density, $N_\mathrm{in}=-171.7~\mathrm{dBm\,Hz^{-1}}$, measured at the same operating point and referred to the same on-chip microwave input, to obtain $\mathrm{SFDR}_3=108.6\pm0.6~\mathrm{dB\,Hz}^{2/3}$.

\FloatBarrier

\subsection{1-dB compression and pump depletion}

\begin{figure}[htbp]
  \centering
  \includegraphics[width=0.58\linewidth]{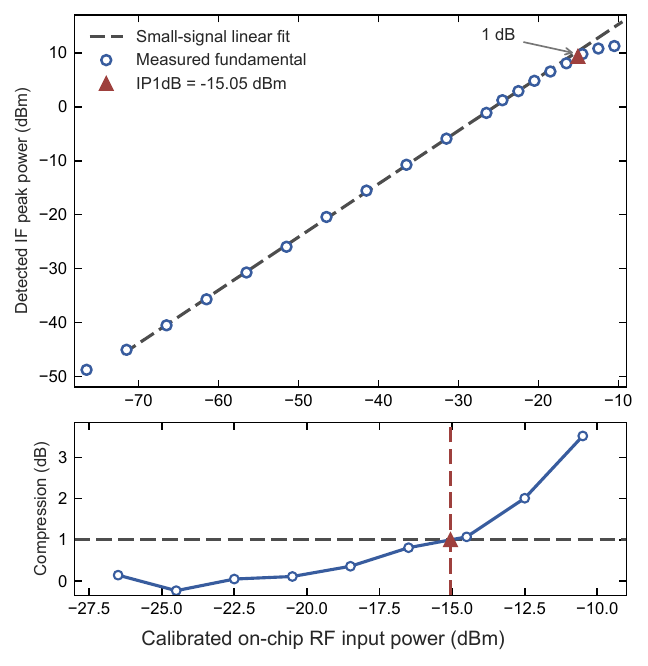}
  \caption{\textbf{Measured input 1-dB compression point.} Top, detected IF peak power as a function of calibrated on-chip RF input power, together with the small-signal linear fit. Bottom, compression relative to the linear fit. The $1~\mathrm{dB}$ crossing gives $IP_{1\mathrm{dB}}=-15.05~\mathrm{dBm}$.}
  \label{fig:SI_measured_IP1dB}
\end{figure}

When the input millimeter-wave power becomes sufficiently large, the transduction response compresses. The input 1-dB compression point, $IP_{1\mathrm{dB}}$, is the calibrated on-chip input power at which the measured transduction efficiency is $1~\mathrm{dB}$ below its small-signal linear extrapolation. The measured receiver has $IP_{1\mathrm{dB}}=-15.05~\mathrm{dBm}$, corresponding to the onset of appreciable pump depletion in the present operating configuration. The measured single-tone power sweep and its compression relative to the small-signal fit are shown in \fref{fig:SI_measured_IP1dB}.

\FloatBarrier

To describe this behavior, we analyze the steady-state response of the coupled-mode equations governing the frequency-multiplexed cavity modes. The time-domain dynamics of the mode amplitudes $a_\mu$ are

\begin{equation}
\frac{d a_\mu}{dt} = \left(-\frac{\kappa}{2} + i \Delta_\mu \right) a_\mu + i g (a_{\mu - 1} + a_{\mu + 1}) + \sqrt{\kappa_\mathrm{ex}} a_\mathrm{in} \delta_{\mu, 0}.
\end{equation}
where $g = g_0\langle b \rangle = g_0 \sqrt{\bar{n}_{\mathrm{mmW}}}$~\cite{zhangUltrabroadbandIntegratedElectrooptic2025a}.
In the steady state ($d a_\mu/dt = 0$) and assuming all detunings are the same, i.e., $\Delta_\mu = \Delta_L$, we introduce the Fourier transform
\begin{equation}
A_k = \frac{1}{\sqrt{N}} \sum_\mu a_\mu e^{\frac{2\pi i}{N} \mu k}, \quad
a_\mu = \frac{1}{\sqrt{N}} \sum_k A_k e^{-\frac{2\pi i}{N} \mu k},
\end{equation}
to diagonalize the system. Substituting into the steady-state equations, we obtain
\begin{equation}
\left(-\frac{\kappa}{2} + i \Delta_L + 2 i g \cos\left(\frac{2\pi k}{N}\right)\right) A_k + \frac{\sqrt{\kappa_\mathrm{ex}}}{\sqrt{N}} a_\mathrm{in} = 0,
\end{equation}
which yields the solution
\begin{equation}
A_k = \frac{-\sqrt{\kappa_\mathrm{ex}} a_\mathrm{in} / \sqrt{N}}{\left(\frac{\kappa}{2} - i \Delta_L\right) - 2 i g \cos\left(\frac{2\pi k}{N}\right)}.
\end{equation}

Transforming back to real space, we find
\begin{align}
a_\mu &= \sqrt{\kappa_\mathrm{ex}} \sum_k \frac{1}{\left(\kappa/2 - i \Delta_L\right) - 2 i g \cos\left(\frac{2\pi k}{N}\right)} e^{- \frac{2\pi i}{N} \mu k} \frac{1}{N} a_\mathrm{in} \nonumber \\
&\rightarrow \sqrt{\kappa_\mathrm{ex}} \int_{-\pi}^{\pi} \frac{du}{2\pi} \frac{e^{-i \mu u}}{(\kappa/2 - i \Delta_L) - 2 i g \cos u} a_\mathrm{in}, \quad (N \to \infty).
\end{align}

Using the substitution $z = e^{i u}$, we recast the integral as a contour integral around the unit circle:
\begin{equation}
a_\mu = \sqrt{\kappa_\mathrm{ex}} \oint_{|z|=1} \frac{dz}{2\pi i} \frac{z^{\mu}}{(\kappa/2 - i \Delta_L) - i g (z + 1/z)} a_\mathrm{in}.
\end{equation}

Evaluating this using the residue theorem, the solution is
\begin{equation}
a_\mu = \frac{i \sqrt{\kappa_\mathrm{ex}} z_0^\mu}{\sqrt{(\Delta_L - i \kappa/2)^2 - 4g^2}} a_\mathrm{in},
\end{equation}
where $z_0$ is the pole inside the unit circle:
\begin{equation}
z_0 = \frac{(\Delta_L - i \kappa/2) - \sqrt{(\Delta_L - i \kappa/2)^2 - 4g^2}}{2 i g}.
\end{equation}
The output fields are defined as
\begin{equation}
a_{0,\mathrm{out}} = a_0 - \sqrt{\kappa_\mathrm{ex}} a_\mathrm{in}, \quad a_{1,\mathrm{out}} = \sqrt{\kappa_\mathrm{ex}} a_1,
\end{equation}
allowing us to compute the transduction efficiency as a function of $g$.
This formulation enables accurate modeling of saturation and nonlinearities under strong input, and allows precise determination of the IP1dB point, beyond which the power transfer deviates from the small-signal regime.

\begin{figure}[htbp]
    \centering
    \includegraphics[width=0.8\linewidth]{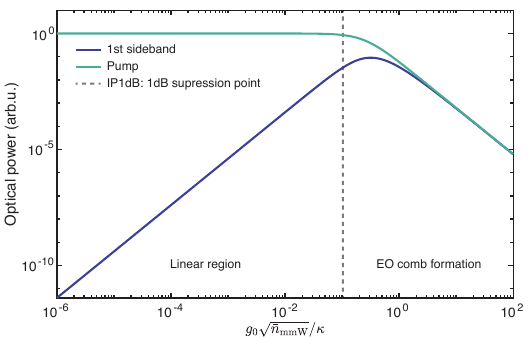}
    \caption{\textbf{Simulated first-sideband compression and pump depletion.} The input 1-dB compression point is defined as the input level at which the transduction efficiency is $1~\mathrm{dB}$ below its linear prediction. Beyond this point, energy transfer to multiple sidebands depletes the intracavity pump and compresses the first-sideband response.}
\label{fig:SI_simulated_IP1dB}
\end{figure}

\Fref{fig:SI_simulated_IP1dB} shows the simulated first-sideband power and intracavity optical pump power as functions of the normalized coupling rate \( g_0\sqrt{\bar{n}_{\mathrm{mmW}}}/\kappa \), assuming critical optical coupling, \( \kappa_{\mathrm{ex}} = \kappa/2 \). The first-sideband power initially grows linearly and then compresses as efficient transfer into multiple electro-optic sidebands depletes the pump. In the normalized simulation, the 1-dB compression point occurs near \( g_0\sqrt{\bar{n}_{\mathrm{mmW}}}/\kappa \approx 0.1 \).

\subsection{Electro-optic comb formation beyond compression}

\begin{figure}[htbp]
  \centering
  \includegraphics[width=0.65\linewidth]{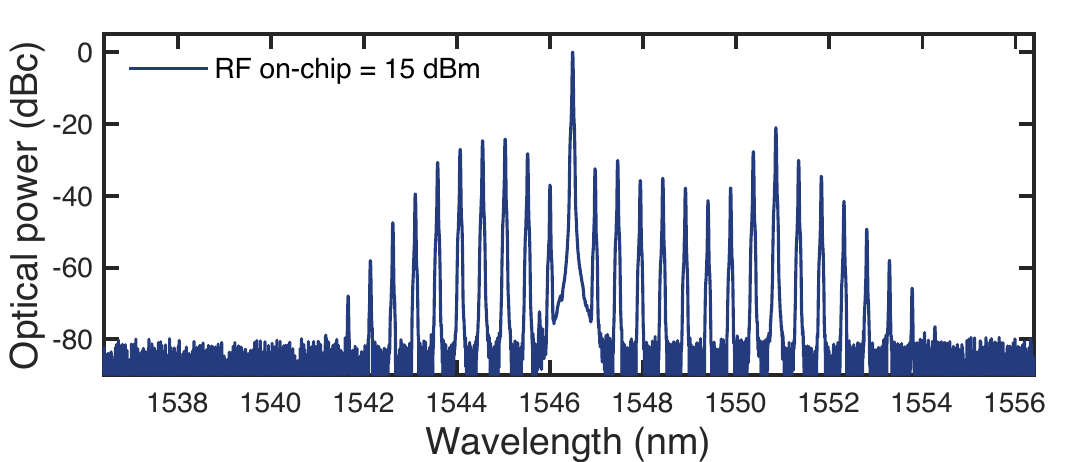}
  \caption{\textbf{Electro-optic comb under large RF drive.} Measured optical spectrum under a calibrated on-chip RF drive of $15~\mathrm{dBm}$. Multiple electro-optic sidebands form around the optical pump in the nonlinear operating regime.}
  \label{fig:SI_eo_comb_high_drive}
\end{figure}

We increased the calibrated on-chip RF power to $15~\mathrm{dBm}$, above $IP_{1\mathrm{dB}}$, and measured the resulting optical spectrum. \Fref{fig:SI_eo_comb_high_drive} shows an electro-optic comb comprising multiple sideband orders around the optical pump. The redistribution of intracavity optical power among these sidebands depletes the pump and compresses the first-order sideband relative to its small-signal linear extrapolation. Because the measured pump line contains a substantial uncoupled component that does not enter the resonator, the apparent pump-to-first-sideband power ratio differs from the ratios between adjacent generated sidebands.

\subsection{High-power RF survivability}

\begin{figure}[htbp]
  \centering
  \includegraphics[width=1.0\linewidth]{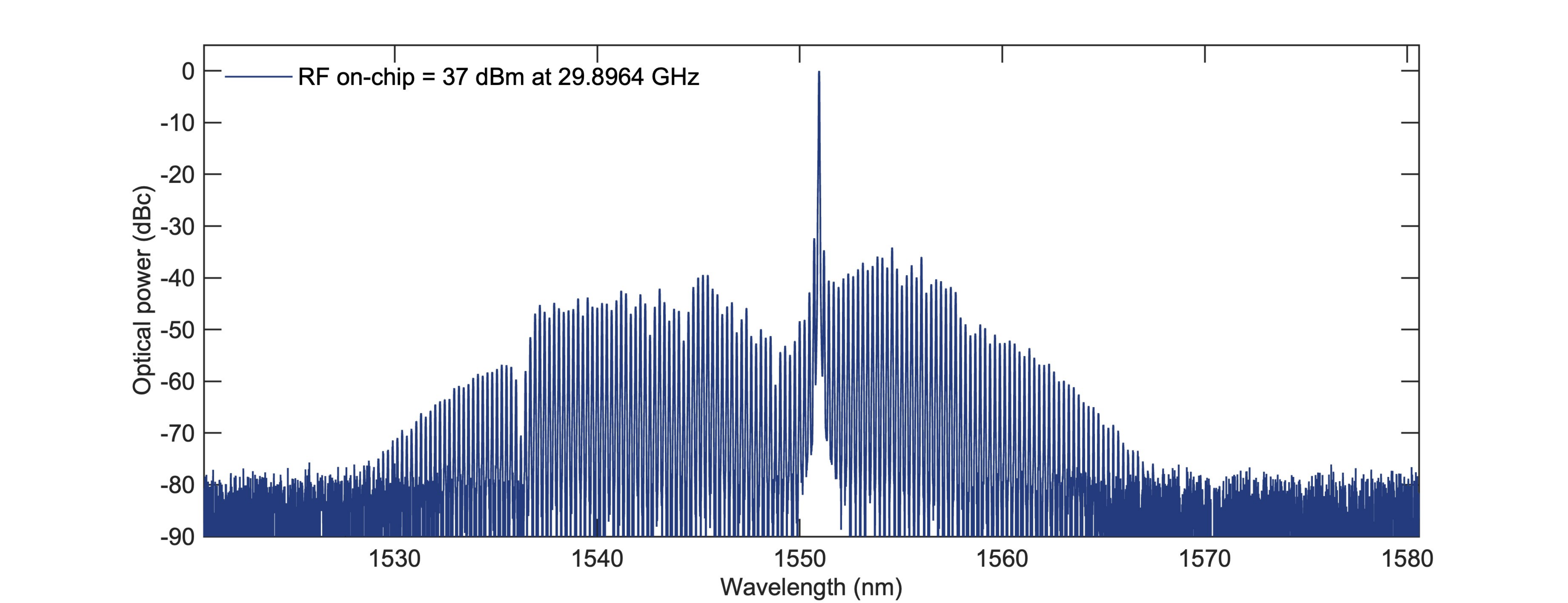}
  \caption{\textbf{High-power RF survivability under continuous-wave drive.} Optical spectrum measured under a calibrated on-chip RF power of $+37~\mathrm{dBm}$ ($\sim5~\mathrm{W}$) at $29.8964~\mathrm{GHz}$. The test structure uses the same electrode and metal-bridge architecture as the $60.5~\mathrm{GHz}$ receiver. The spectral envelope of the broadband electro-optic comb is shaped by the Raman-suppression structure. Subsequent measurements showed no observable irreversible degradation after the $5~\mathrm{min}$ continuous-wave exposure.}
  \label{fig:SI_high_power_comb_37dBm}
\end{figure}

We characterized the high-power RF survivability using a high-power amplifier (TLPA26.5G40G-47-40-HS) operating from $26.5$ to $40~\mathrm{GHz}$, with a maximum output power of $43~\mathrm{dBm}$. Because this frequency range does not cover the $60.5~\mathrm{GHz}$ receiver, we used a test structure resonant at $29.9~\mathrm{GHz}$ with the same electrode and metal-bridge architecture. After calibrating the RF link loss with a power meter and a vector network analyzer (VNA), we applied a continuous-wave on-chip RF power of $+37~\mathrm{dBm}$ ($\sim5~\mathrm{W}$) for $5~\mathrm{min}$. Under this drive, the measured optical spectrum exhibited a broadband electro-optic comb (\fref{fig:SI_high_power_comb_37dBm}). Subsequent measurements showed no observable irreversible degradation. The available RF source limited the maximum applied power; accordingly, $+37~\mathrm{dBm}$ is the highest tested continuous-wave level and provides a lower bound on the RF damage threshold.

\subsection{Voltage-breakdown and overcurrent failure mechanisms}

We performed separate destructive quasi-DC voltage and current tests to identify failure modes in the electrode structure. The voltage test shows that overvoltage breakdown occurs across a narrow electrode gap, whereas the current test shows that overcurrent burnout occurs at the metal bridges.

\begin{figure}[htbp]
  \centering
  \includegraphics[width=1.0\linewidth]{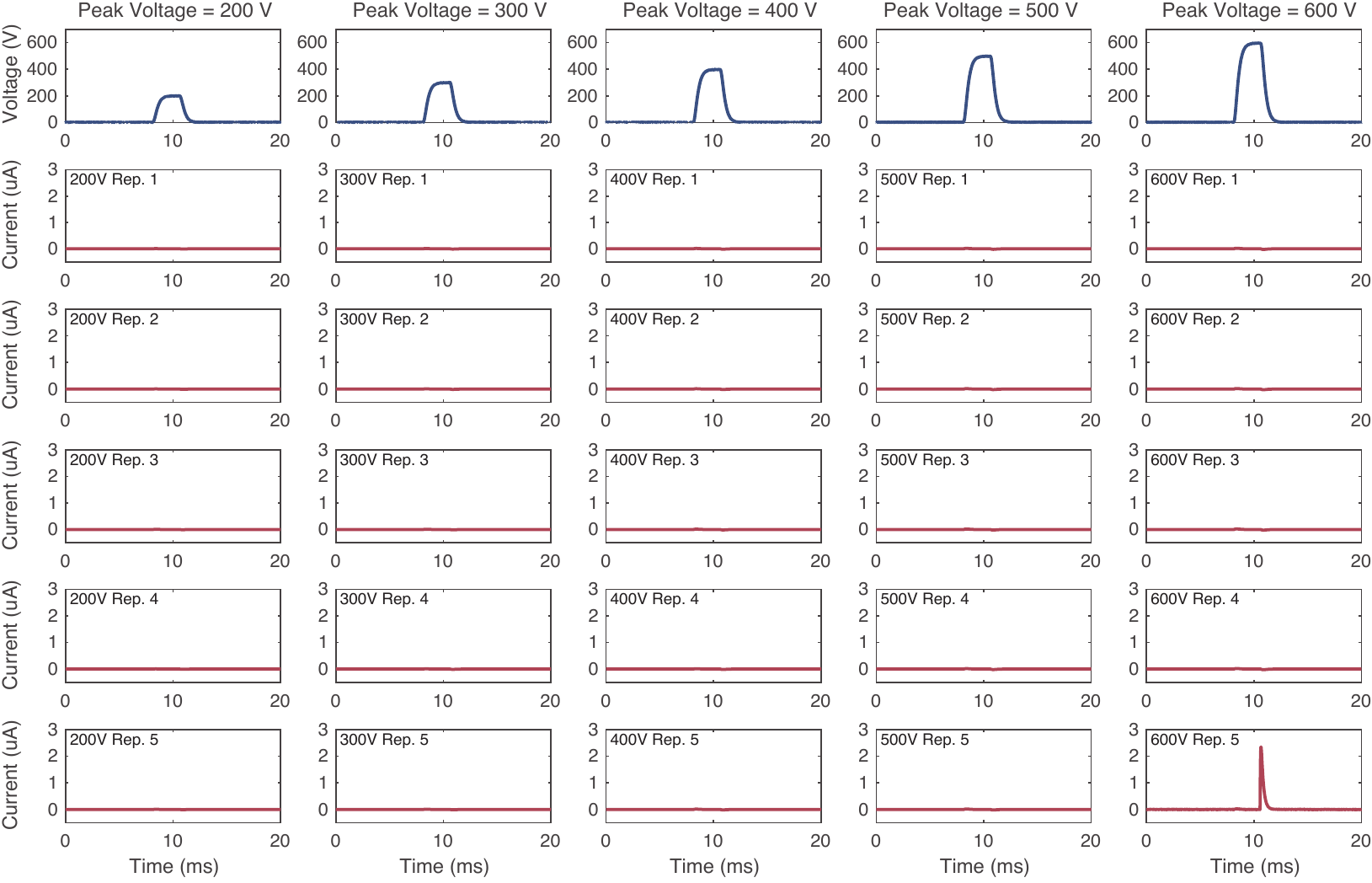}
  \caption{\textbf{Repeated DC-breakdown measurements.} Applied voltage pulses and measured current traces for five repetitions at each peak voltage from $200$ to $600~\mathrm{V}$. The tested device had its air bridges removed. The current spike in the fifth $600~\mathrm{V}$ trial marks the first breakdown event.}
  \label{fig:SI_dc_breakdown_trials}
\end{figure}

\begin{figure}[htbp]
  \centering
  \includegraphics[width=1.0\linewidth]{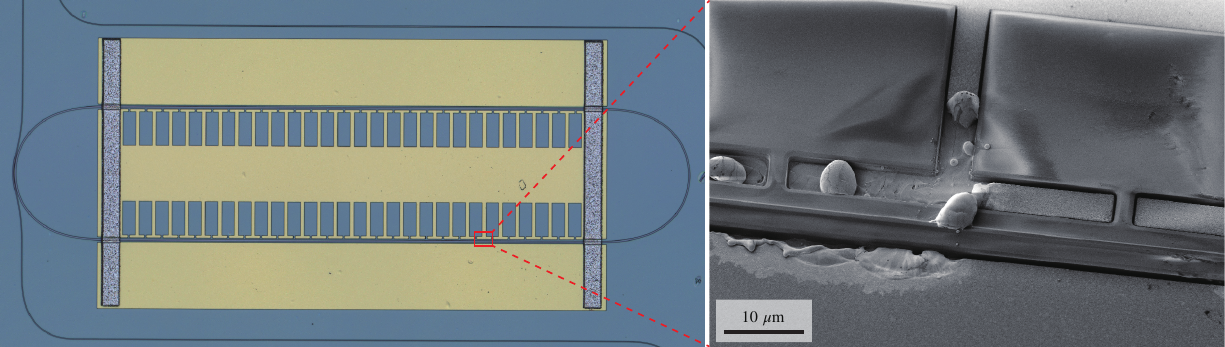}
  \caption{\textbf{Localized DC-breakdown site.} Optical-microscope overview of the relevant device region and SEM close-up after the $600~\mathrm{V}$ breakdown event. The damage is localized to a narrow electrode gap, where the discharge produces local melting and metal reflow.}
  \label{fig:SI_dc_breakdown_sem}
\end{figure}

For the voltage-breakdown test, the air bridges were removed to prevent a DC short through the bridge network. Voltage pulses from $200$ to $600~\mathrm{V}$ were applied in $100~\mathrm{V}$ increments. Each voltage was held for $2.5~\mathrm{ms}$ and repeated five times while the current was monitored. No measurable breakdown occurred from $200$ to $500~\mathrm{V}$ or during the first four $600~\mathrm{V}$ trials. The current spike in the fifth $600~\mathrm{V}$ trial marks the first breakdown event, as shown in \fref{fig:SI_dc_breakdown_trials}. Post-mortem imaging in \fref{fig:SI_dc_breakdown_sem} shows localized melting and metal reflow across a narrow electrode gap.

\begin{figure}[htbp]
  \centering
  \includegraphics[width=0.65\linewidth]{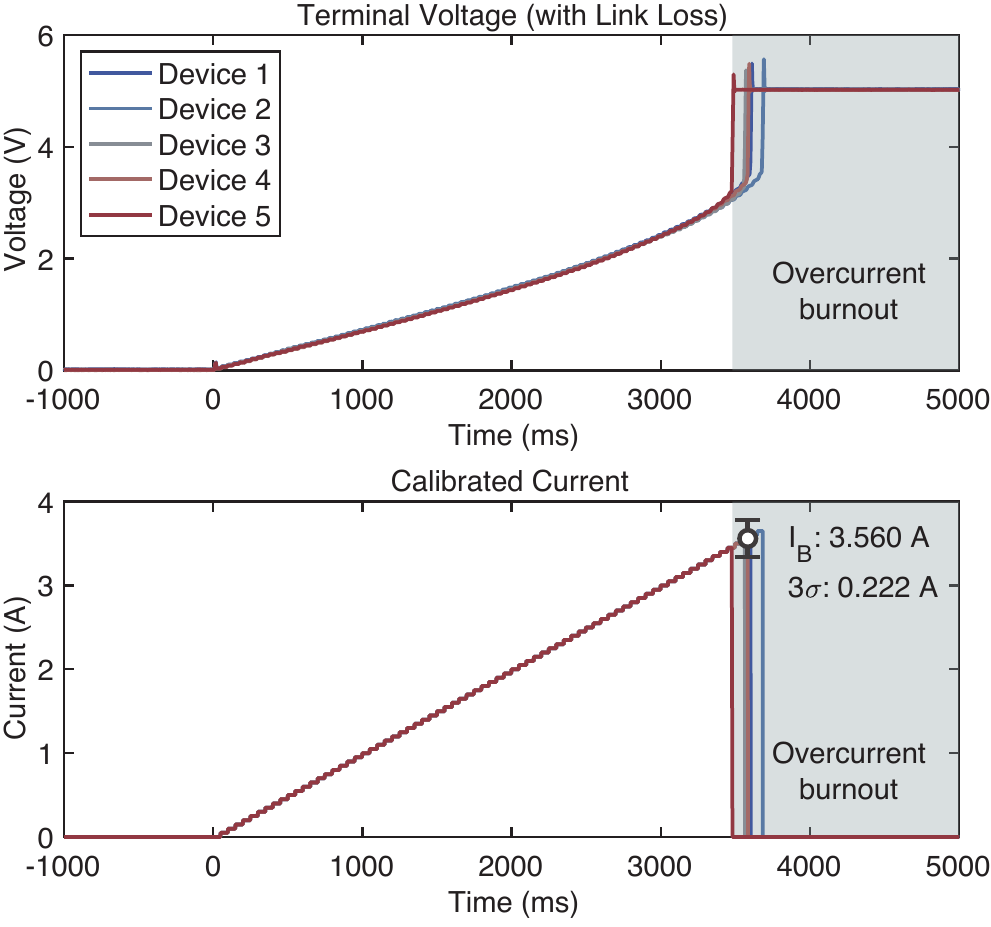}
  \caption{\textbf{Overcurrent-burnout threshold.} Terminal-voltage traces, including link loss and other series resistances (top), and calibrated current traces (bottom) for five devices during DC current ramps. The mean burnout current is $I_\mathrm{B}=3.560~\mathrm{A}$, with a $3\sigma$ spread of $0.222~\mathrm{A}$.}
  \label{fig:SI_overcurrent_trace}
\end{figure}

\begin{figure}[htbp]
  \centering
  \includegraphics[width=1.0\linewidth]{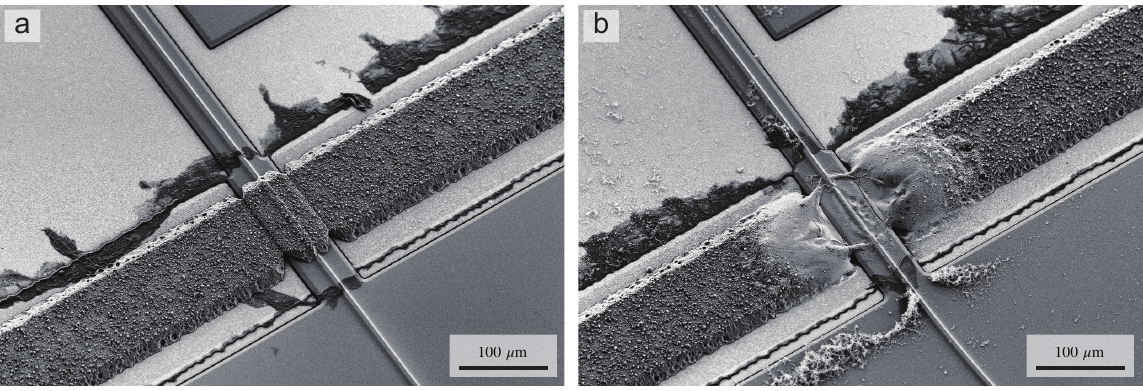}
  \caption{\textbf{Air-bridge failure under overcurrent stress.} Representative SEM images after a DC burnout event. The four air bridges at the CPW crossings show localized melting and metal reflow.}
  \label{fig:SI_overcurrent_sem}
\end{figure}

For the overcurrent test, five devices were driven by a programmable current source in $50~\mathrm{mA}$ steps with a $50~\mathrm{ms}$ dwell time at each step until burnout. The monitored terminal voltage includes link loss and other series resistances and is therefore not de-embedded. The mean burnout current was $I_\mathrm{B}=3.560~\mathrm{A}$, with a $3\sigma$ spread of $0.222~\mathrm{A}$ (\fref{fig:SI_overcurrent_trace}). Post-mortem SEM images show localized melting and metal reflow at the four air bridges, identifying them as the current-limiting failure points (\fref{fig:SI_overcurrent_sem}).

\subsection{Comparison with electronic frontend input-power limits}

We summarize representative manufacturer-specified input-power limits for millimeter-wave LNAs and RF limiters in Table~\ref{tab:SI_electronic_survivability}. The continuous-wave ratings span $-9$ to $+30~\mathrm{dBm}$ for the listed LNAs and $+28$ to $+30~\mathrm{dBm}$ for the listed RF limiters. In comparison, the electro-optic test structure remained operational at the maximum available continuous-wave on-chip power of $+37~\mathrm{dBm}$.

\begin{table}[htbp]
\centering
\scriptsize
\setlength{\tabcolsep}{2pt}
\renewcommand{\arraystretch}{1.08}
\begin{tabular}{|>{\raggedright\arraybackslash}p{0.29\linewidth}|>{\raggedright\arraybackslash}p{0.17\linewidth}|>{\raggedright\arraybackslash}p{0.21\linewidth}|>{\raggedright\arraybackslash}p{0.23\linewidth}|}
\hline
\textbf{Example frontend} & \textbf{Band} & \textbf{Type} & \textbf{Maximum input power} \\
\hline
MACOM MAAL-011238 & $27$--$31.5~\mathrm{GHz}$ & GaAs pHEMT LNA & CW: $+18~\mathrm{dBm}$ \\
\hline
MACOM MAAL-011240 & $27$--$31.5~\mathrm{GHz}$ & GaAs pHEMT LNA & CW: $+20~\mathrm{dBm}$ \\
\hline
ADI/Hittite HMC-ALH364 & $24$--$32~\mathrm{GHz}$ & GaAs HEMT LNA & CW: $-9~\mathrm{dBm}$ \\
\hline
MACOM MAAL-011251-DIE & $24$--$34~\mathrm{GHz}$ & GaN LNA die & CW: $+30~\mathrm{dBm}$ \\
\hline
ADI/Hittite HMC566LP4E & $28$--$36~\mathrm{GHz}$ & GaAs pHEMT LNA & CW: $+5~\mathrm{dBm}$ \\
\hline
ADI HMC-ALH445 & $18$--$40~\mathrm{GHz}$ & GaAs HEMT LNA & CW: $+10~\mathrm{dBm}$ \\
\hline
ADI HMC-ALH244 & $24$--$40~\mathrm{GHz}$ & GaAs HEMT LNA & CW: $+6~\mathrm{dBm}$ \\
\hline
ADI HMC1040LP3CE & $24$--$43.5~\mathrm{GHz}$ & GaAs pHEMT LNA & CW: $+5~\mathrm{dBm}$ \\
\hline
AT Microwave AT-LNA-3350-2328 & $33$--$50~\mathrm{GHz}$ & LNA module & CW: $+10~\mathrm{dBm}$ \\
\hline
ADI ADL8106 & $20$--$54~\mathrm{GHz}$ & GaAs pHEMT LNA & CW: $+17~\mathrm{dBm}$ \\
\hline
Marki AMM-7203CH/UC & $30$--$60~\mathrm{GHz}$ & Broadband LNA module & CW: $+20~\mathrm{dBm}$ \\
\hline
Low Noise Factory LNF-LNR26\_62A\_SV & $26$--$62~\mathrm{GHz}$ & InP LNA module & CW: $0~\mathrm{dBm}$ \\
\hline
ADI HMC-ALH382 & $57$--$65~\mathrm{GHz}$ & GaAs HEMT LNA & CW: $-5~\mathrm{dBm}$ \\
\hline
MI-Wave 955V-50/68/35/18/383 & $50$--$68~\mathrm{GHz}$ & V-band LNA module & CW: $0~\mathrm{dBm}$ \\
\hline
Gotmic gANZ0031C & $52$--$72~\mathrm{GHz}$ & V-band GaAs LNA & CW: $+15~\mathrm{dBm}$ \\
\hline
AT Microwave AT-LNA-5075-3825 & $50$--$75~\mathrm{GHz}$ & V-band LNA module & CW: $+10~\mathrm{dBm}$ \\
\hline
Low Noise Factory LNF-LNR45\_77WA\_SV & $45$--$77~\mathrm{GHz}$ & InP LNA module & CW: $0~\mathrm{dBm}$ \\
\hline
ADI ADL7003 & $50$--$95~\mathrm{GHz}$ & GaAs pHEMT LNA & CW: $+15~\mathrm{dBm}$ \\
\hline
Qorvo TGL2203 & $30$--$38~\mathrm{GHz}$ & RF limiter (IL $<1~\mathrm{dB}$) & CW: $+30~\mathrm{dBm}$ \\
\hline
Pasternack PE80L2002 & $18$--$40~\mathrm{GHz}$ & RF limiter (max. IL $\sim4~\mathrm{dB}$) & CW: $+30~\mathrm{dBm}$; pulse: $+43~\mathrm{dBm}$ \\
\hline
Keysight N9355F & $0.01$--$50~\mathrm{GHz}$ & RF limiter & CW: $+28~\mathrm{dBm}$ \\
\hline
Keysight N9355G & $0.01$--$54~\mathrm{GHz}$ & RF limiter & CW: $+30~\mathrm{dBm}$ \\
\hline
Marki HLM-67ABH & DC--$67~\mathrm{GHz}$ & RF limiter (typ. IL $\sim1.35~\mathrm{dB}$) & CW: $+30~\mathrm{dBm}$ \\
\hline
\end{tabular}
\caption{\textbf{Millimeter-wave frontend input-power comparison.} Representative electronic LNAs are listed first, followed by RF limiters. The electro-optic frontend showed no irreversible degradation under a calibrated $+37~\mathrm{dBm}$ on-chip RF drive. The listed electronic values are manufacturer ratings; CW denotes continuous wave and IL denotes insertion loss.}
\label{tab:SI_electronic_survivability}
\end{table}

\section{Microwave resonator design and single-photon coupling rate \texorpdfstring{$g_0$}{g0}}
\subsection{Microwave resonator design and coupling}




We designed a coplanar waveguide (CPW) microwave resonator embedding the optical racetrack to maximize transduction efficiency. Since the optical resonator is electrically large at the microwave frequency $f_{\mathrm{mw}}=\Omega/(2\pi)$, a distributed structure is required. The optimal transmission line resonator employing a CPW geometry is terminated by short-circuits on either end (as opposed to reactive or open-circuit terminations) \cite{zhangUltrabroadbandIntegratedElectrooptic2025a}. We also employ periodic structures along the transmission line (T-cells) to reduce the phase velocity and decrease the microwave loss \cite{zhangUltrabroadbandIntegratedElectrooptic2025a, kharel_breaking_2021}. The structures can be interpreted as sub-wavelength slots loaded in series, thus increasing the distributed inductance of the line, and reducing the attenuation constant for a fixed distributed resistance. 

The cross-section of the transmission line was designed to confine the microwave fields inside the optical waveguides as much as possible without affecting the optical quality factor. The microwave electrodes are separated $\mathrm{2.05~\mu m}$ from the $\mathrm{2.4~\mu m}$-wide ridge waveguide, thus increasing the electric field in the waveguide per Volt on the electrodes. The spacing between the straight sections of the optical racetrack, and therefore the lateral dimension of the transmission line, is chosen to allow for a standard Ground-Signal-Ground, 150~$\mathrm{\mu m}$ pitch (GSG-150) microwave probe to land on the resonator.

\begin{figure}[htbp]
    \centering
    \includegraphics[width=0.9\linewidth]{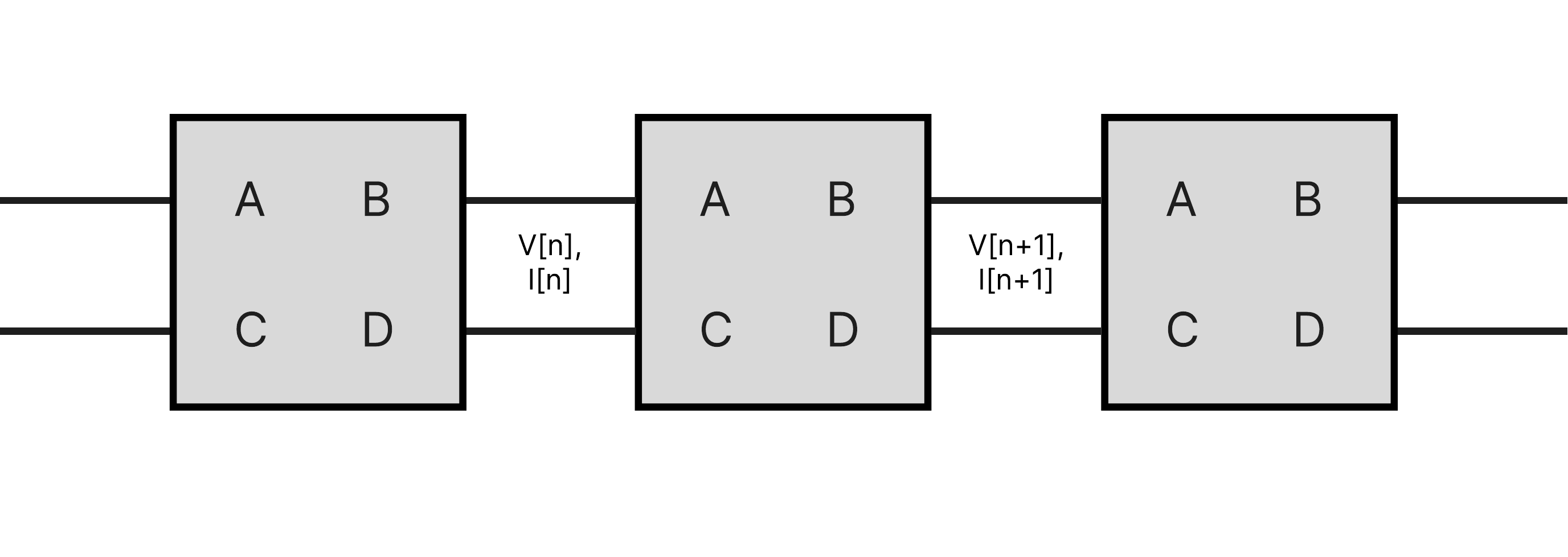}
    \caption{Transfer matrices for a unit cell cascaded infinitely in each direction to analyze uni-directional traveling waves in the structure.}
    \label{SIUnitCells}
\end{figure}

In order to avoid large parameter sweeps in finite-element-method (FEM) electromagnetic simulations involving the entire structure, we employed a semi-analytical approach to generate the microwave resonator layout. Given the cross-section of the transmission-line resonator and the geometry of the T-cells that optimize loss, the length of the resonator can be calculated from the phase constant of the periodic structure. 
%
%
This can be obtained using a method similar to \cite{itoh_periodic_2004}. First, consider an infinite cascade of unit cells and characterize each unit cell by the transfer (ABCD) matrix, as in \fref{SIUnitCells}. For each voltage/current column vector and node $n+1$, the voltage and current at the previous node are

\begin{equation}
    \begin{bmatrix}
        V\left[n+1\right] \\
        I\left[n+1\right]
    \end{bmatrix} = 
    \begin{bmatrix}
        A & B\\
        C & D
    \end{bmatrix}
    \begin{bmatrix}
        V\left[n\right] \\
        I\left[n\right]
    \end{bmatrix}.
    \label{eq:ABCD}
\end{equation}

Solutions to the Helmholtz equation tell us that for any wave traveling in one direction, the E-field and B-field (voltage and current, respectively) must be related by a homogeneous complex constant, the wave impedance. So for a wave traveling in the periodic structure, the voltage and current will accumulate a common phase, $\gamma\Delta z$, from one node to another, 

\begin{equation}
    \begin{bmatrix}
        V\left[n\right] \\
        I\left[n\right]
    \end{bmatrix} = 
    e^{\gamma\Delta z}
    \begin{bmatrix}
        V\left[n+1\right] \\
        I\left[n+1\right]
    \end{bmatrix} = e^{\gamma\Delta z}
    \begin{bmatrix}
        A & B\\
        C & D
    \end{bmatrix}
    \begin{bmatrix}
        V\left[n\right] \\
        I\left[n\right]
    \end{bmatrix}. 
    \label{eq:phaseDeley}
\end{equation}
Equation \eqref{eq:phaseDeley} states the eigen-problem of the transfer matrix, where the eigenvalue is the phase constant of the traveling wave in radians per unit cell. The eigenvectors indicate the allowed voltage (current) states at each node for a travelling wave which can be used to calculate the characteristic impedance, $Z_{0l}$. 

\begin{figure}[htbp]
    \centering
    \includegraphics[width=0.6\linewidth]{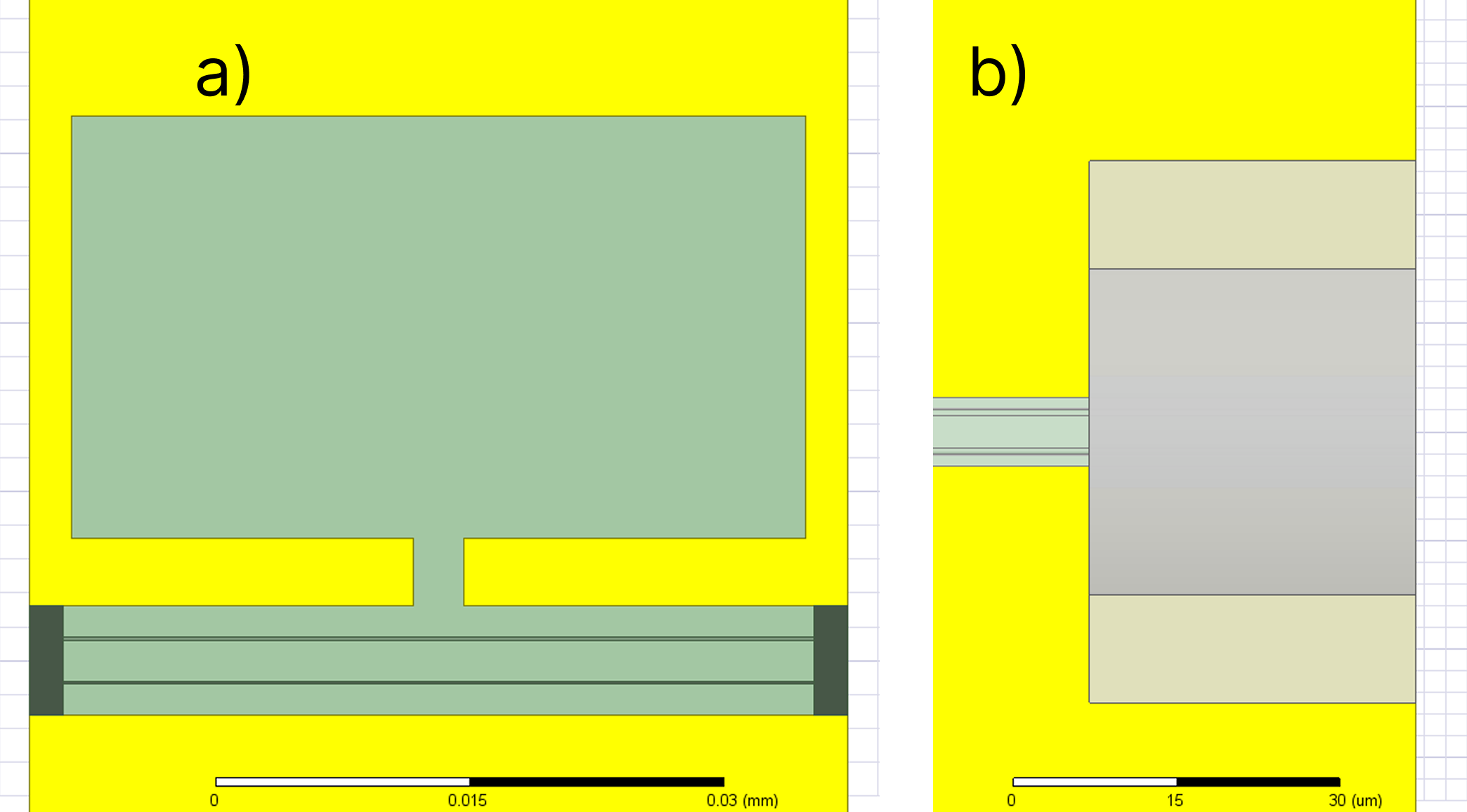}
    \caption{(a) Rendering of one unit cell from the FEM solver. The optical layer is shown as green and the electrode layer is shown as yellow. The S-parameters are referenced to the terminal ports in dark green at either end of the waveguide gap. (b) Rendering of the airbridge short-circuit sections. Both images show a portion of only half of the physical structure. A symmetry boundary on the symmetry plane was used to reduce simulation time.}
    \label{SI_HFSSUnitCell}
\end{figure}

We used a commercial FEM solver to calculate the 2-port scattering parameters for a single unit cell, shown in \fref{SI_HFSSUnitCell} (a), and the 1-port reflection coefficient for the airbridges, shown in \fref{SI_HFSSUnitCell} (b). Since each structure is symmetric about the center, we simulated only half of the models and defined an H-Field symmetry plane that bisects the center conductor of the CPW. Using the methods described above, we calculated the complex characteristic impedance and propagation constant of a line consisting on unit cells.
Given the phase progression per unit cell and the reflected phase from the airbridge short-circuits, we calculated the number of unit cells required to place the fundamental half-wave resonance near $60~\mathrm{GHz}$. Once we had an initial estimate for the resonator length, we modeled the complete structure in HFSS and slightly scaled its length to match the target resonance near $60.5~\mathrm{GHz}$.
In our experience, only one iteration was required to achieve the desired resonance frequency. 

\begin{figure}[htbp]
    \centering
    \includegraphics[width=0.55\linewidth]{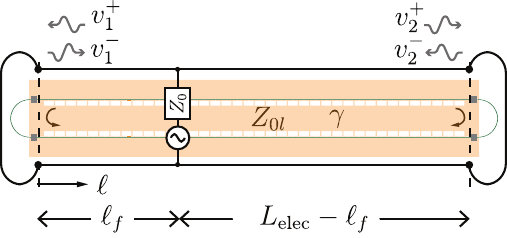}
    \caption{Transmission line model of resonator terminated in short circuits and fed somewhere in between.}
    \label{fig:TLmodel}
\end{figure}

Typically, transmission line resonators are fed from one end or the other where an impedance transformer allows the designer to achieve the desired coupling. In our case, we can land the microwave probe at any point along the resonator. Since the center of the resonator would have an input impedance approaching an open-circuit and the ends are shorted by the air bridges, some landing position between the two must have a 50 $\Omega$ input impedance (critical coupling) as long as the structure remains in resonance \cite{zhangUltrabroadbandIntegratedElectrooptic2025a}. 
If we consider a half-wavelength CPW resonator short-circuited on either end with characteristic impedance $Z_{0}$ and intrinsic quality factor $Q_{m,0}$, the achievable input resistance ranges from $R_\mathrm{in}=0~\Omega$ to $R_\mathrm{in}=2 \pi Q_{m,0}Z_{0}$ \cite{zhangUltrabroadbandIntegratedElectrooptic2025a}. This means that for any practical value of $Q_{m,0}$, critical coupling can be achieved even if the transmission line impedance, $Z_{0l}$, is far from $50~\Omega$. A simple transmission line model (\fref{fig:TLmodel}) is used to obtain the input impedance seen at the feed (with impedance $Z_0$) located at the position $\ell_f$ on the resonator of length $L_\mathrm{elec}$ as the reciprocal of the sum of the admittances seen on either side of the feed, i.e.,

\begin{equation}
     Z_{\mathrm{in}}={Z_{0l}}\left[\frac{1+e^{-2\gamma\ell_{f}}}{1-e^{-2\gamma\ell_{f}}}+\frac{1+e^{-2\gamma\left(L_{\mathrm{elec}}-\ell_{f}\right)}}{1-e^{-2\gamma\left(L_{\mathrm{elec}}-\ell_{f}\right)}}\right]^{-1}.\label{eq:SI_Zin}
\end{equation}
Knowing the input impedance, we postulate incident waves $v_{1,2}^+$ and reflected waves $v_{1,2}^- = -v_{1,2}^+$ at either short circuit reference plane. Solving for the incident wave amplitudes for a 1~V source, we obtain

\begin{equation}
    v_{1}^{+}=\frac{Z_{\mathrm{in}}}{Z_{\mathrm{in}}+Z_{0}}\left(e^{\gamma\ell_{f}}-e^{-\gamma\ell_{f}}\right)^{-1},
\end{equation}
and,
\begin{equation}
    v_{2}^{+}=\frac{Z_{\mathrm{in}}}{Z_{\mathrm{in}}+Z_{0}}\left[e^{\gamma\left(L_{\mathrm{elec}}-\ell_{f}\right)}-e^{-\gamma\left(L_{\mathrm{elec}}-\ell_{f}\right)}\right]^{-1},
\end{equation}
which can then be used to calculate the field (or voltage) distribution along the line as

\begin{equation}
        v(\ell)=\begin{cases}
            v_{1}^{+}\left(e^{\gamma \ell}-e^{-\gamma \ell}\right) & \ell\leq\ell_{f}\\
            v_{2}^{+}\left[e^{\gamma\left(L_{\mathrm{elec}}-\ell\right)}-e^{-\gamma\left(L_{\mathrm{elec}}-\ell\right)}\right] & \ell\geq\ell_{f}
        \end{cases}\label{eq:SI_Vdstr}.
\end{equation}

\begin{figure}[htbp]
    \centering
    \includegraphics[width=1.0\linewidth]{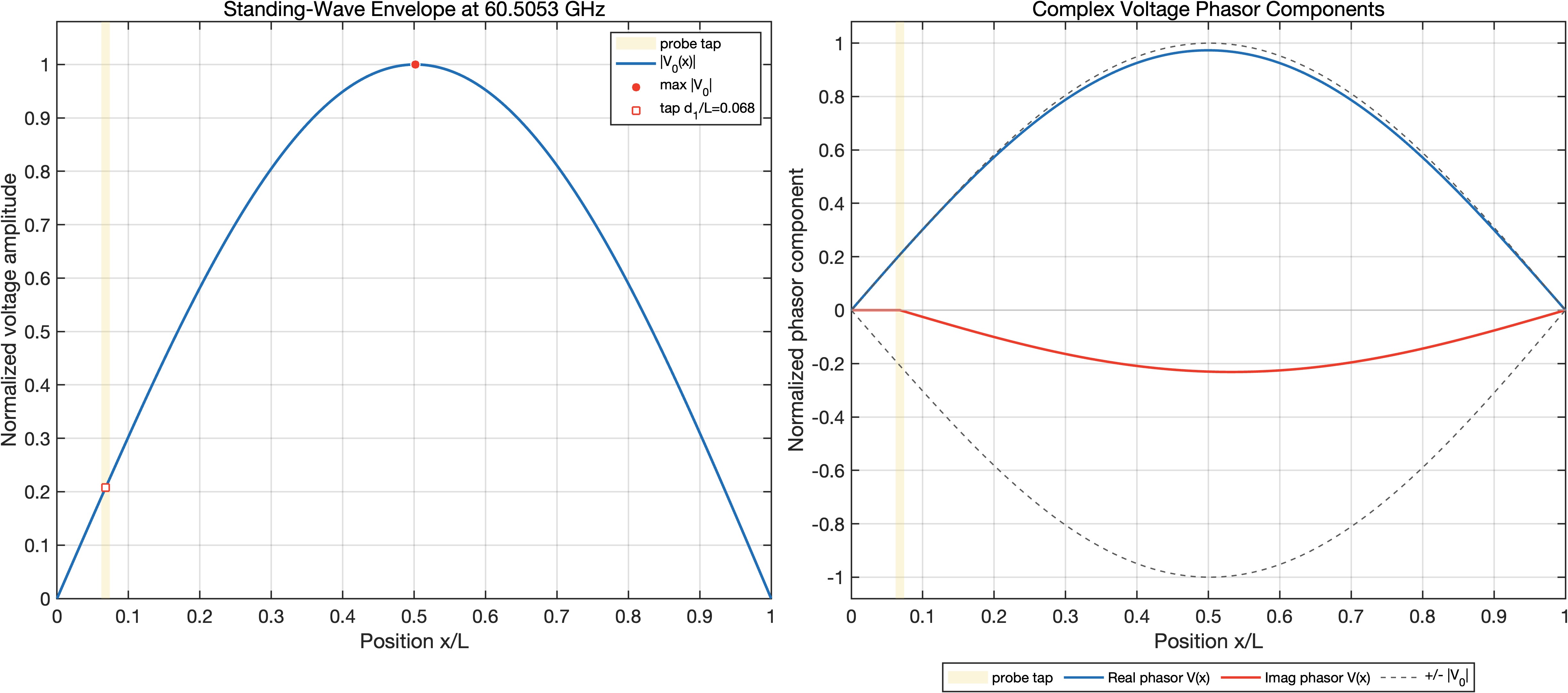}
    \caption{\textbf{Calculated voltage distribution of the $60~\mathrm{GHz}$ fundamental mode.} Left: normalized standing-wave envelope at $60.5053~\mathrm{GHz}$ for the selected probe-tap position. The voltage vanishes at the short-circuit terminations and has a single central antinode. Right: real and imaginary components of the normalized complex voltage phasor. The nearly uniform phase supports constructive electro-optic overlap along the interaction region.}
    \label{fig:SI_feedpos}
\end{figure}

The probe landing position controls the impedance presented to the source and therefore the external coupling rate of the fundamental mode. \Fref{fig:SI_feedpos} shows the calculated standing-wave voltage distribution at $60.5053~\mathrm{GHz}$ for the selected probe-tap position. The voltage vanishes at the two short-circuit terminations and has a single antinode near the centre of the resonator, as expected for the fundamental half-wave mode. The phase remains nearly uniform along the interaction region, apart from the small quadrature component produced by microwave loss, which enables constructive overlap with the optical fields (Sec.~\ref{sec:PMg0}).

\begin{figure}[htbp]
    \centering
    \includegraphics[width=0.6\linewidth]{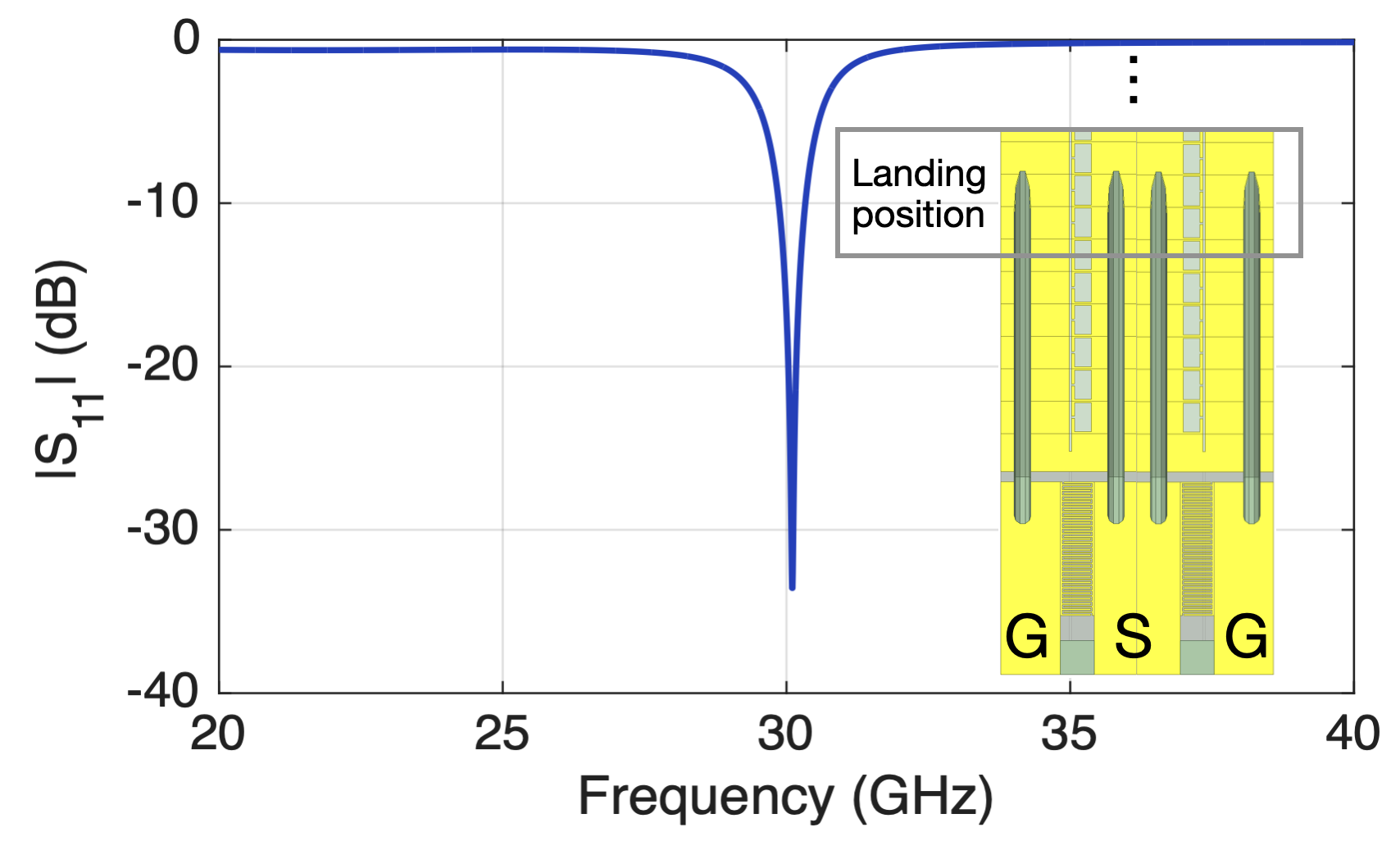}
    \caption{\textbf{Simulated fixed-feed coupling through a compensated wire-bond transition.} Electromagnetic simulation of a representative $30~\mathrm{GHz}$ microwave resonator connected at the selected feed position through $1$-mil-diameter wire bonds. The inset shows the modeled CPW feed pads, wire bonds, and on-chip capacitive structures. In this design, the compensation network offsets the wire-bond inductive reactance and produces near-critical coupling.}
    \label{fig:SI_packaged_fixed_feed}
\end{figure}

In the current demonstration reported in the manuscript, the microwave external coupling rate $\Gamma_\mathrm{ex}$ is tuned by changing the microwave probe landing position. In a practical implementation, the external coupling can instead be set by connecting to the corresponding feed position through wire bonds together with an on-chip capacitive compensation network, as shown by the HFSS simulation in \fref{fig:SI_packaged_fixed_feed}. In the representative $30~\mathrm{GHz}$ design, the compensation network offsets the inductive reactance of the $1$-mil-diameter wire bonds and produces near-critical coupling.

\subsection{High-frequency microwave-resonator scaling}\label{sec:SI_high_frequency_resonator}

To evaluate the feasibility of extending the present architecture toward $150$--$200~\mathrm{GHz}$, we performed full-wave simulations of CPW transmission lines on silicon and quartz substrates using the same CPW geometry and electrode parameters as the present $60.5~\mathrm{GHz}$ device. The attenuation constant was extracted from the simulated two-port scattering parameters.

\begin{figure}[htbp]
    \centering
    \includegraphics[width=0.78\linewidth]{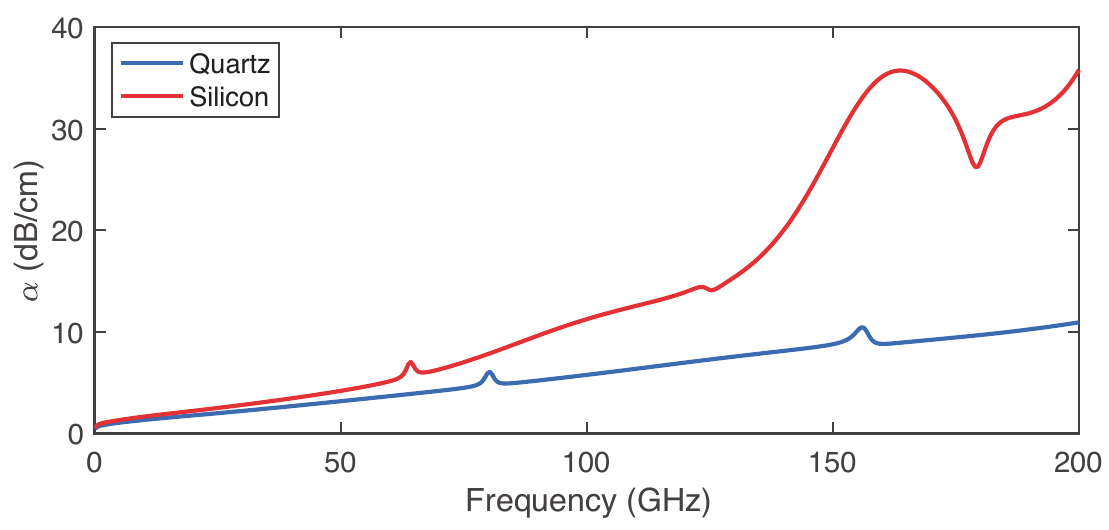}
    \caption{\textbf{Simulated microwave attenuation on quartz and silicon substrates.} The attenuation constant $\alpha$, extracted from full-wave simulated two-port scattering parameters using the CPW geometry and electrode parameters of the present device, is plotted as a function of frequency. The narrow peaks are numerical artifacts arising from imperfect port de-embedding in the finite-length simulation. At high frequencies, the loss on silicon increases rapidly because of microwave-mode leakage into the high-index substrate, whereas quartz suppresses this leakage and retains a slow, approximately linear loss increase up to $200~\mathrm{GHz}$.}
    \label{fig:SI_high_frequency_substrate_attenuation}
\end{figure}

The reduced substrate leakage on quartz motivates a frequency-specific resonator design for the $183~\mathrm{GHz}$ water-vapor line. The simulated design has a fundamental mode near $93~\mathrm{GHz}$ and a second-order mode near $183~\mathrm{GHz}$. The second-order mode has an ohmic-loss-dominated intrinsic quality factor of approximately $44$, with radiation loss below $10\%$. The separation between the straight racetrack arms is reduced to $140~\mathrm{\mu m}$ to balance electrode coverage and microwave mode volume.

\begin{figure}[htbp]
    \centering
    \includegraphics[width=0.78\linewidth]{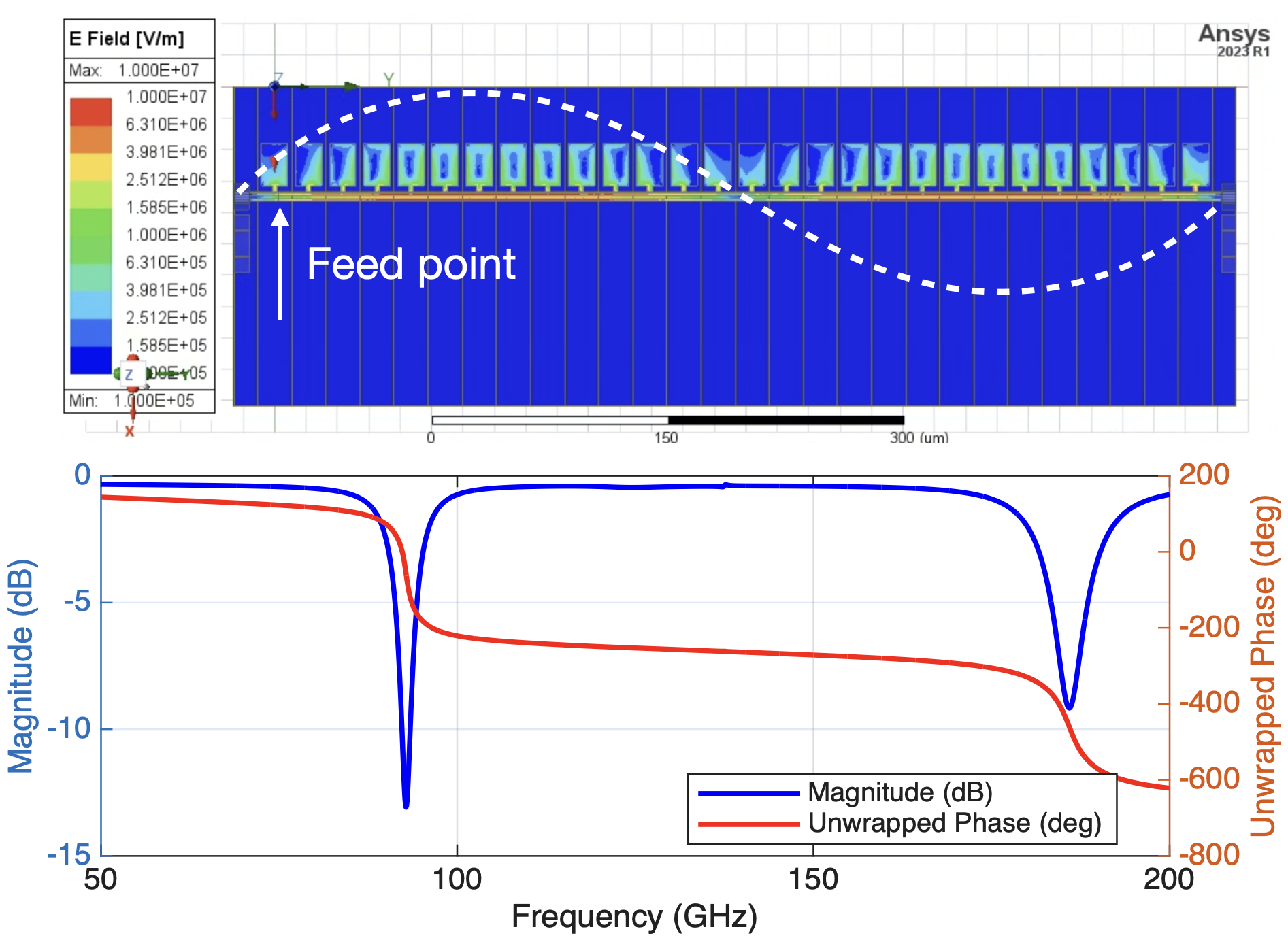}
    \caption{\textbf{Quartz microwave-resonator design for the $183~\mathrm{GHz}$ water-vapor line.} Top: simulated electric-field distribution of the second-order microwave mode, with the feed position indicated. Bottom: simulated reflection-coefficient magnitude and unwrapped phase, showing the fundamental resonance near $93~\mathrm{GHz}$ and the second-order resonance near $183~\mathrm{GHz}$. The quartz substrate increases the characteristic impedance and suppresses radiation loss.}
    \label{fig:SI_quartz_183GHz_resonator}
\end{figure}
\FloatBarrier


\subsection{Phase matching and quantitative determination of \texorpdfstring{$g_0$}{g0}}\label{sec:PMg0}

The vacuum coupling rate $g_0$ is obtained from the 3D overlap integral of the microwave, pump, and sideband fields in Eq.~\eqref{SI:eq:g0}. For the revised device, the optical free spectral range and the fundamental microwave resonance are both approximately $60~\mathrm{GHz}$, corresponding to $m=1$ operation. The symmetry of the racetrack and CPW resonators reduces the longitudinal part of the 3D overlap to one straight racetrack section, while the transverse microwave-field confinement is captured by the equivalent modulation efficiency $(V_\pi L)_\mathrm{MZM}$~\cite{zhangUltrabroadbandIntegratedElectrooptic2025a,santamaria-botello_electro-optic_2025}.

\begin{figure}[htbp]
    \centering
    \includegraphics[width=1.0\linewidth]{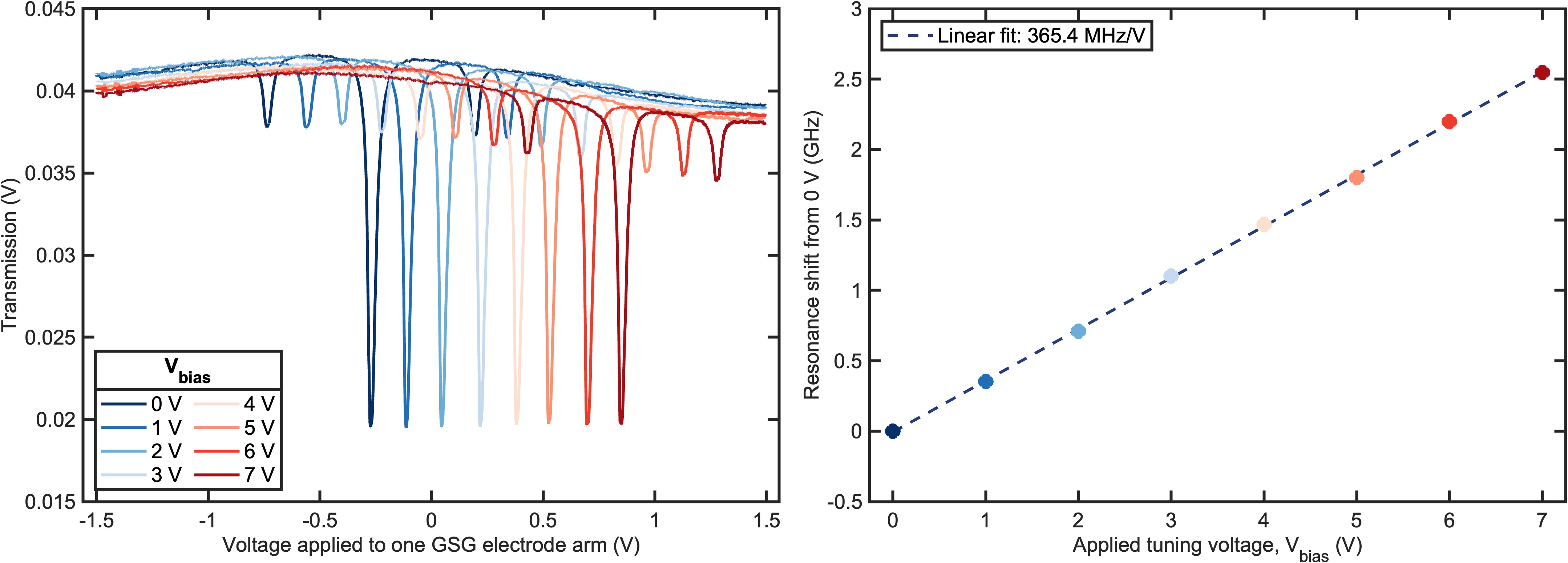}
    \caption{\textbf{DC electro-optic tuning calibration.} Left: measured optical-transmission traces as the DC bias applied to one GSG electrode arm is increased from $0$ to $7~\mathrm{V}$. Right: extracted resonance shift relative to $0~\mathrm{V}$. A linear fit gives $|\mathrm{d}f_\mathrm{opt}/\mathrm{d}V|=365.4~\mathrm{MHz/V}$, from which $(V_\pi L)_\mathrm{MZM}$ is determined using Eq.~\eqref{eq:SI_vpiL_from_dc_tuning}.}
    \label{fig:SI_dc_tuning_g0}
\end{figure}

We determine $(V_\pi L)_\mathrm{MZM}$ from the DC electro-optic tuning measurement in \fref{fig:SI_dc_tuning_g0}. The optical resonance shifts linearly with the voltage applied to one GSG electrode arm, giving $|\mathrm{d}f_\mathrm{opt}/\mathrm{d}V|=365.4~\mathrm{MHz/V}$. The corresponding equivalent push--pull modulation efficiency is
\begin{equation}
V_\pi=\frac{\mathrm{FSR}_\mathrm{opt}}{2\left|\mathrm{d}f_\mathrm{opt}/\mathrm{d}V\right|},
\qquad
(V_\pi L)_\mathrm{MZM}=\frac{V_\pi L_\mathrm{elec}}{2}
=\frac{\mathrm{FSR}_\mathrm{opt}L_\mathrm{elec}}
{4\left|\mathrm{d}f_\mathrm{opt}/\mathrm{d}V\right|}.
\label{eq:SI_vpiL_from_dc_tuning}
\end{equation}

Retaining the distinct optical phase index $n_\mathrm{opt}$ and microwave effective index $n_\mathrm{mw}$, the expression used for the numerical prediction is
\begin{equation}
g_0=
\frac{\pi}{2(V_\pi L)_\mathrm{MZM}}
\sqrt{\frac{\hbar\Omega Z_{0l}c^3}{2n_\mathrm{opt}^2n_\mathrm{mw}}}
\sqrt{\frac{\mathcal{G}_1}{L}},
\label{eq:SI_g0_quantitative_prediction}
\end{equation}
where $\mathcal{G}_1$ is the dimensionless longitudinal-overlap factor of the fundamental mode. In terms of the normalized coordinate $u$ along one straight racetrack section,
\begin{equation}
\mathcal{G}_1=
\frac{\ell_s}{2L}
\left[
\int_0^1
\frac{e_1(u)g_1(u)}
{\sqrt{\int_0^1|e_1(u)|^2\,\mathrm{d}u}}
\,\mathrm{d}u
\right]^2,
\label{eq:SI_g0_longitudinal_overlap}
\end{equation}
with $e_1(u)$ the fundamental microwave-field distribution and $g_1(u)=2\cos[2\pi(\ell_s/L)(u-1/2)]$ the corresponding phase-matching kernel. Thus an even microwave-field distribution about the centre of the resonator contributes constructively to the overlap. This calculation gives $g_0/(2\pi)=6.4789~\mathrm{kHz}$.

\begin{figure}[htbp]
    \centering
    \includegraphics[width=1.0\linewidth]{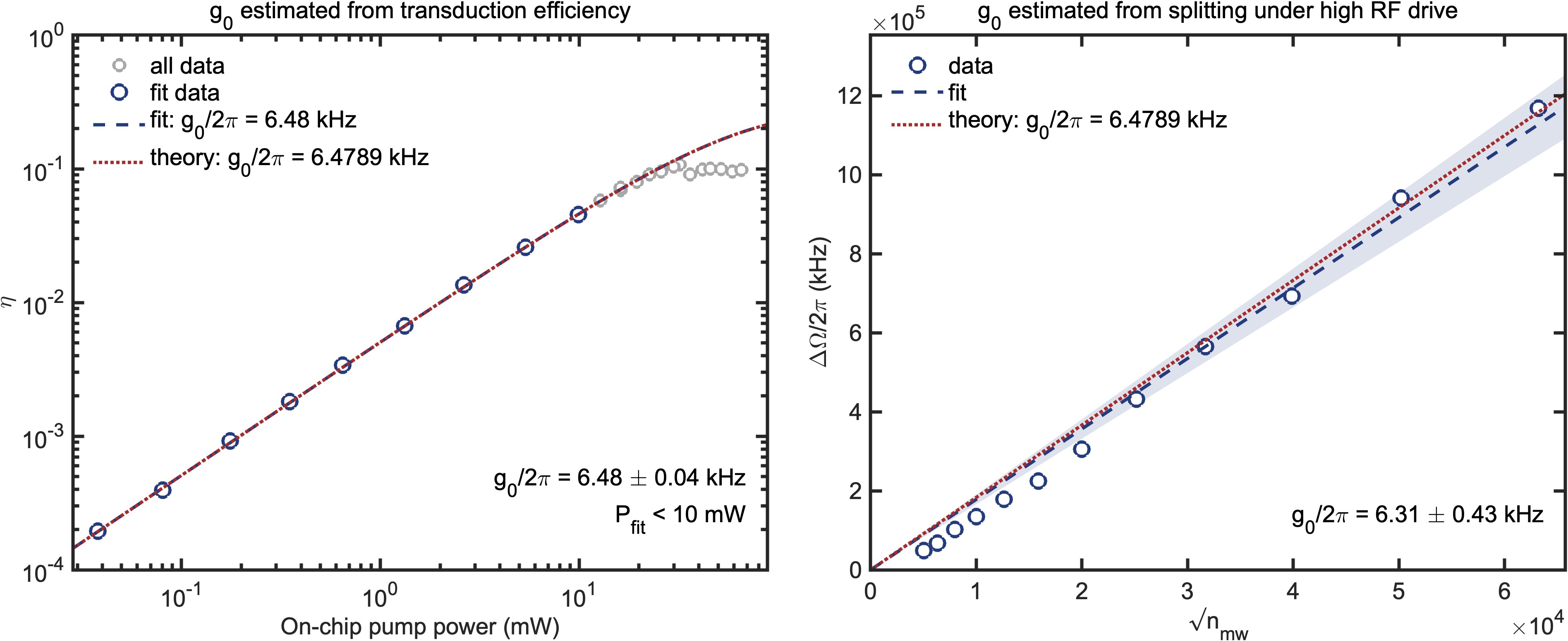}
    \caption{\textbf{Independent measurements of the vacuum electro-optic coupling rate.} Left: photon-number conversion efficiency versus calibrated on-chip pump power. Fitting the small-signal data below $10~\mathrm{mW}$ gives $g_0/(2\pi)=6.48\pm0.04~\mathrm{kHz}$. Right: RF-induced optical-mode splitting versus the square root of the intracavity microwave photon number. The fitted splitting gives $g_0/(2\pi)=6.31\pm0.43~\mathrm{kHz}$. In both panels, the red dotted curve is the theoretical prediction $g_0/(2\pi)=6.4789~\mathrm{kHz}$; uncertainties are $3\sigma$.}
    \label{fig:SI_g0_two_methods}
\end{figure}

We independently determine $g_0$ using two measurements, as summarized in \fref{fig:SI_g0_two_methods}. First, in the small-signal regime the photon-number conversion efficiency is fitted using
\begin{equation}
\eta=\frac{\mathcal{C}}{(1+\mathcal{C})^2},
\qquad
\mathcal{C}=\frac{4g_0^2\bar n_\mathrm{p}}{\Gamma\kappa}.
\label{eq:SI_g0_efficiency_fit}
\end{equation}
Restricting the fit to on-chip pump powers below $10~\mathrm{mW}$, before the measured efficiency saturates, gives $g_0/(2\pi)=6.48\pm0.04~\mathrm{kHz}$. Second, under strong microwave drive with both neighboring optical sidebands present, the two outer RF-dressed optical modes are separated by
\begin{equation}
\frac{\Delta\Omega}{2\pi}
=2\sqrt{2}\,\frac{g_0}{2\pi}\sqrt{\bar n_\mathrm{mw}}.
\label{eq:SI_g0_splitting_fit}
\end{equation}
A zero-intercept fit gives $g_0/(2\pi)=6.31\pm0.43~\mathrm{kHz}$. The quoted uncertainties are $3\sigma$ fit uncertainties. Both measurements agree with each other and with the theoretical prediction.

\subsection{Fabrication tolerance of the electro-optic interaction}

\begin{figure}[htbp]
    \centering
    \includegraphics[width=0.50\linewidth]{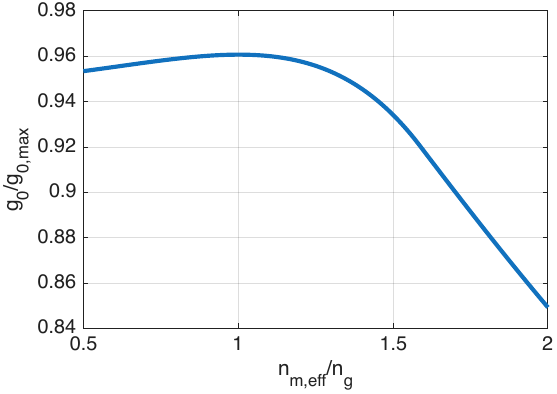}
    \caption{\textbf{Microwave-index dependence of the electro-optic coupling rate.} Calculated $g_0$, normalized to its ideal maximum, as a function of the ratio between the microwave effective index and the optical group index. The simulated indices of the device give $n_\mathrm{m,eff}/n_g\approx1.62$ and $g_0/g_{0,\mathrm{max}}\approx0.92$.}
    \label{fig:SI_g0_index_tolerance}
\end{figure}

The longitudinal contribution to $g_0$ is governed by the overlap factor $\mathcal{G}_1$ in Eq.~\eqref{eq:SI_g0_longitudinal_overlap}. For the standing-wave resonator, $0\leq\mathcal{G}_1\leq1/2$, with the upper limit corresponding to matching between the microwave effective phase index and the optical group index. \Fref{fig:SI_g0_index_tolerance} shows the calculated coupling rate as a function of the index ratio. For the simulated microwave effective index $n_\mathrm{m,eff}\approx3.57$ and optical group index $n_g\approx2.2$, $\mathcal{G}_1\approx0.42$ and $g_0/g_{0,\mathrm{max}}\approx0.92$, corresponding to an approximately $8\%$ reduction from the ideal maximum.

\begin{figure}[htbp]
    \centering
    \includegraphics[width=0.9\linewidth]{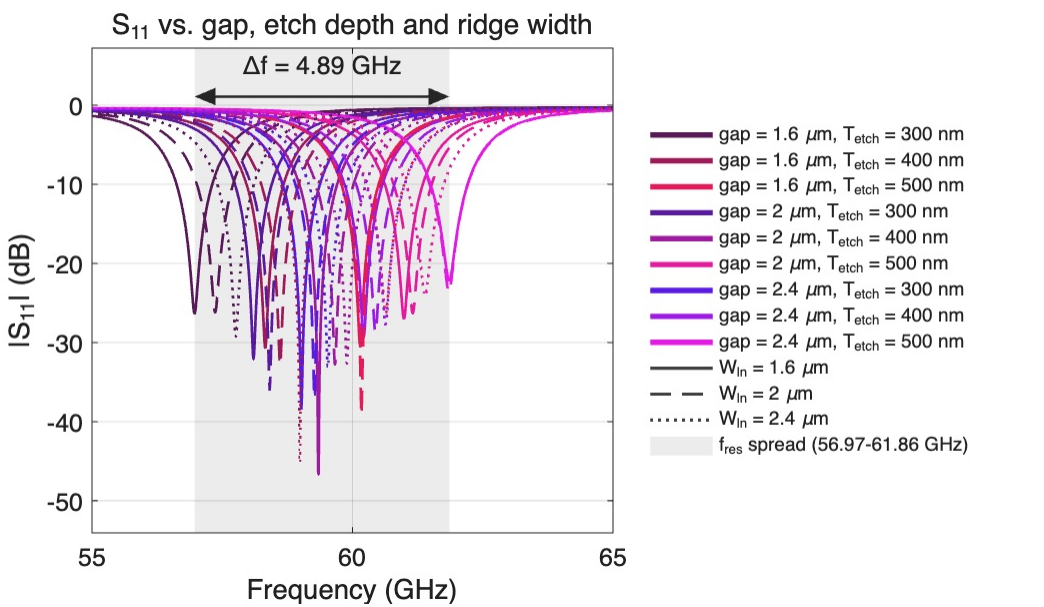}
    \caption{\textbf{Microwave-resonance sensitivity to fabrication geometry.} Calculated reflection coefficient for variations in the electrode-to-ridge gap, LiTaO$_3$ etch depth, and ridge width. The shaded region marks the resulting resonance-frequency range from $56.97$ to $61.86~\mathrm{GHz}$.}
    \label{fig:SI_microwave_geometry_tolerance}
\end{figure}

Geometric fabrication variations change $n_\mathrm{m,eff}$ and therefore shift the microwave resonance frequency. \Fref{fig:SI_microwave_geometry_tolerance} shows the calculated reflection response while varying the electrode-to-ridge gap, LiTaO$_3$ etch depth, and ridge width over the indicated ranges. The fundamental resonance remains between $56.97$ and $61.86~\mathrm{GHz}$, corresponding to a $4.89~\mathrm{GHz}$ spread and a maximum relative shift below $6\%$. This frequency shift can be compensated during design by adjusting the resonator length to align the fundamental microwave resonance with the optical free spectral range. After this frequency compensation, the residual variation in $g_0$ associated with the longitudinal phase-matching factor is approximately $0.04\%$ over the simulated range.

\begin{figure}[htbp]
    \centering
    \includegraphics[width=1.0\linewidth]{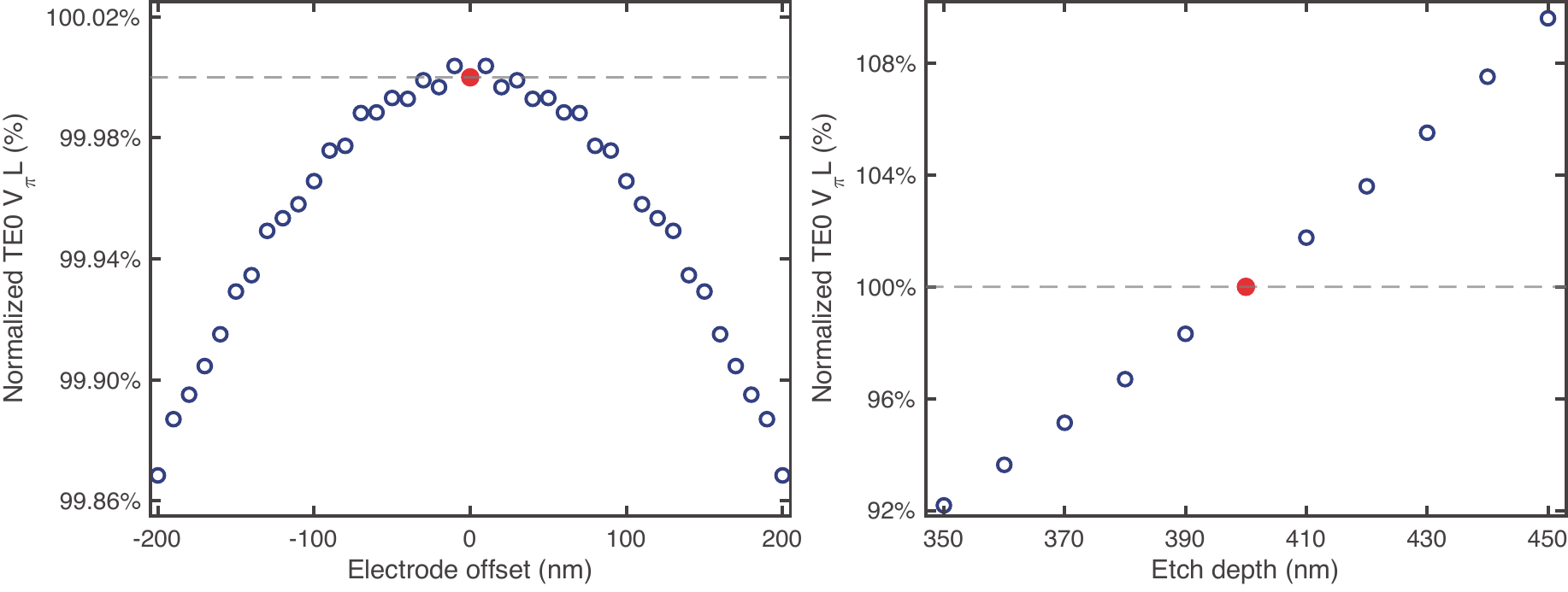}
    \caption{\textbf{Fabrication tolerance of the electro-optic interaction.} Calculated normalized $(V_\pi L)_\mathrm{MZM}$ as a function of the lateral electrode offset relative to the optical waveguide (left) and the LiTaO$_3$ etch depth (right). The red markers indicate the nominal electrode alignment and the nominal $400~\mathrm{nm}$ etch depth.}
    \label{fig:SI_fabrication_tolerance_g0}
\end{figure}

We also evaluate the sensitivity of the transverse electro-optic interaction to the lateral electrode offset relative to the optical waveguide and to the LiTaO$_3$ etch depth. \Fref{fig:SI_fabrication_tolerance_g0} shows the calculated normalized $(V_\pi L)_\mathrm{MZM}$ for these two parameters. Across an electrode-offset range of $\pm 200~\mathrm{nm}$, $(V_\pi L)_\mathrm{MZM}$ changes by less than $0.14\%$. Across an etch-depth range of $350$--$450~\mathrm{nm}$, centred at the nominal $400~\mathrm{nm}$ etch depth, it varies from $92.2\%$ to $109.6\%$ of its nominal value. Since $g_0\propto 1/(V_\pi L)_\mathrm{MZM}$, the electrode-offset variation produces a negligible change in $g_0$, whereas the etch-depth variation corresponds to a change in $g_0$ from approximately $+8.5\%$ to $-8.8\%$.

\section{Device Fabrication}
Devices were fabricated on x-cut thin-film lithium tantalate-on-insulator (LTOI) wafers supplied by SIMIT, with a nominal lithium tantalate device-layer thickness of 600~nm.
Photonic integrated circuits were defined using a Diamond-like Carbon (DLC) hard mask~\cite{wangLithiumTantalatePhotonic2024b, li2023high}. Photonics patterns were written by deep-UV stepper lithography and transferred into the DLC via oxygen plasma etching.
Ion-beam etching was then used to transfer the patterns $400~\mathrm{nm}$ into the LiTaO$_3$, leaving a $200~\mathrm{nm}$ slab for dispersion control.
Wet cleaning with 40\% \ch{KOH} and 30\% \ch{H2O2} with volume ratio 3:1 at 80$^\circ C$ is employed to clean the redeposition.
The electrode layer is patterned with the same DUV tool using a $1.2~\mu m$ PECVD \ch{SiO2} as a sacrificial layer. 
The \ch{SiO2} layer was undercut for roughly $200\mathrm{nm}$ in buffered oxide etchant, after which a metal stack of 20~nm Al and 800~nm Au was deposited to lower Ohmic losses and improve RF performance.
Aluminum metal bridges (800~nm thick) with 20nm Titanium adhesion layer were then fabricated on top of the remaining oxide cladding.
Finally, chip separation was performed by dry etching the LiTaO$_3$ and deep reactive-ion etching of the silicon carrier, ensuring smooth side facets for laser coupling. Remaining photoresist was stripped with remover NI555 at $65 ^\circ \mathrm{C}$ overnight and cleaned with oxygen plasma.

\section{Raman suppression and the optical nonlinear power limit}
Stimulated Raman scattering can limit the intracavity pump power of a high-$Q$ electro-optic resonator. This section discusses the Raman-limited behavior of devices without a Raman-suppression structure, then describes the wavelength-selective dissipation engineering used to suppress Raman lasing, and finally discusses optical parametric oscillation (OPO) as the high-power limit for devices with Raman-suppression coupling sections.

\subsection{Raman-limited operation without Raman suppression}
In devices without Raman suppression, the resonantly enhanced optical pump can drive stimulated Raman scattering once the round-trip Raman gain exceeds the loss of a cavity mode within the Raman-gain band.
For X-cut lithium tantalate with a pump wavelength near $1550~\mathrm{nm}$, the main Raman-gain peak occurs near $1605~\mathrm{nm}$.
Above threshold, optical power is transferred from the pump to the Raman mode. This pump depletion clamps the intracavity pump population and therefore limits the electro-optic cooperativity and conversion efficiency in continuous-wave operation.

\begin{figure}[htbp]
	\centering
	\includegraphics[width=1.0\linewidth]{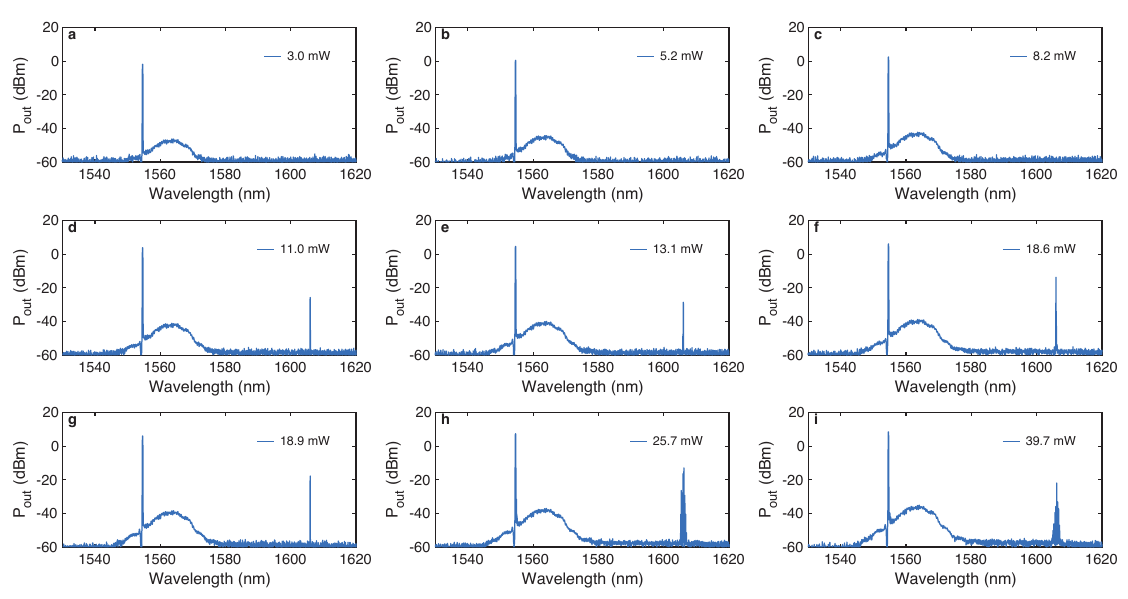}
	\caption{\textbf{Raman-limited operation without Raman suppression.} Output optical spectra measured without electro-optic modulation for incident off-chip pump powers from $3.0$ to $39.7~\mathrm{mW}$. A narrow Raman-lasing line emerges near $1605~\mathrm{nm}$ at approximately $11~\mathrm{mW}$ and increases with pump power.}
	\label{fig:SI_raman_without_suppression}
\end{figure}

\Fref{fig:SI_raman_without_suppression} shows output spectra measured without electro-optic modulation as the incident off-chip pump power is increased. A Raman-lasing line appears near $1605~\mathrm{nm}$ at approximately $11~\mathrm{mW}$ and grows rapidly at higher power.
For a resonant continuous-wave stimulated-Raman-scattering process, the input threshold can be written as
\begin{equation}
P_{\mathrm{th}}^{\mathrm{SRS}}
=\frac{\pi^2 n_0^2 f^{\mathrm{P}}f^{\mathrm{R}}}
{c^2g_{\mathrm{eff}}^{\mathrm{SRS}}(\delta)}
\frac{V_{\mathrm{eff}}}
{Q_{\mathrm{L}}^{\mathrm{P}}(f^{\mathrm{P}})
 Q_{\mathrm{L}}^{\mathrm{R}}(f^{\mathrm{R}})},
\qquad
V_{\mathrm{eff}}=L_{\mathrm{R}}A_{\mathrm{eff}}.
\label{eq:SI_raman_threshold}
\end{equation}
Here, $n_0$ is the refractive index, $f^{\mathrm{P}}$ and $f^{\mathrm{R}}=f^{\mathrm{P}}+\delta$ are the pump- and Raman-mode resonance frequencies, $g_{\mathrm{eff}}^{\mathrm{SRS}}(\delta)$ is the effective Raman gain at detuning $\delta$, and $V_{\mathrm{eff}}$ is determined by the optical round-trip length $L_{\mathrm{R}}$ and effective pump--Raman overlap area $A_{\mathrm{eff}}$. The loaded quality factors of the pump and Raman modes are denoted by $Q_{\mathrm{L}}^{\mathrm{P}}$ and $Q_{\mathrm{L}}^{\mathrm{R}}$, respectively. Equation~\eqref{eq:SI_raman_threshold} shows that the Raman threshold can be raised by selectively lowering the loaded quality factor (by increasing $\kappa_\mathrm{ex}$) of modes in the Raman-gain band while preserving the pump-mode quality factor. Optical-grade X-cut LiTaO$_3$ also exhibits a weaker principal Raman response than X-cut LiNbO$_3$ under comparable measurements~\cite{songStableGigahertzMmWaverepetitionrate2025}, making it a favorable material for high-power cavity electro-optics.

\subsection{Wavelength-selective Raman suppression}

\begin{figure}[htbp]
  \centering
  \includegraphics[width=0.95\linewidth]{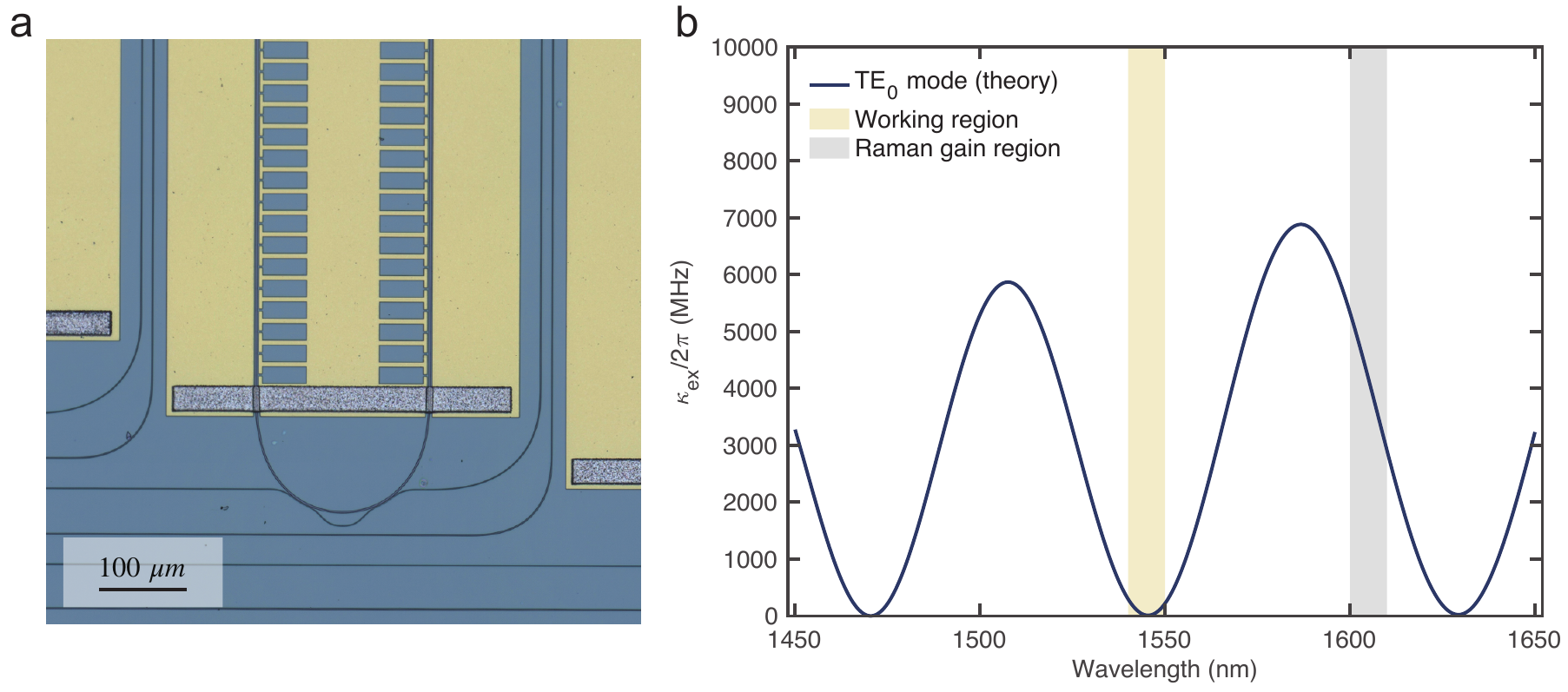}
  \caption{\textbf{Wavelength-selective Raman-suppression design.} (a) Optical micrograph of the Mach--Zehnder-interferometer-type coupling section between the primary racetrack and auxiliary ring. (b) Calculated external coupling rate $\kappa_{\mathrm{ex}}/2\pi$ of the fundamental TE mode. The coupling is weak in the working region containing the pump, Stokes, and anti-Stokes cavity modes and strong in the Raman-gain region, selectively reducing the loaded quality factor of Raman-band modes.}
  \label{fig:SI_raman_suppression_design}
\end{figure}

We implement wavelength-selective dissipation engineering~\cite{song2024octave} with a Mach--Zehnder-interferometer-type coupling section between the primary racetrack and an auxiliary ring. The interferometric phase is selected so that the external coupling remains weak in the working band around the pump, Stokes, and anti-Stokes resonances, while modes in the Raman-gain band are strongly coupled out. This design preserves resonant pump enhancement and electro-optic transduction in the working band but lowers $Q_{\mathrm{L}}^{\mathrm{R}}$ in Eq.~\eqref{eq:SI_raman_threshold}, thereby increasing the Raman-lasing threshold.
\Fref{fig:SI_raman_suppression_design} shows the device layout and calculated wavelength-dependent external coupling rate.

\begin{figure}[htbp]
  \centering
  \includegraphics[width=1.0\linewidth]{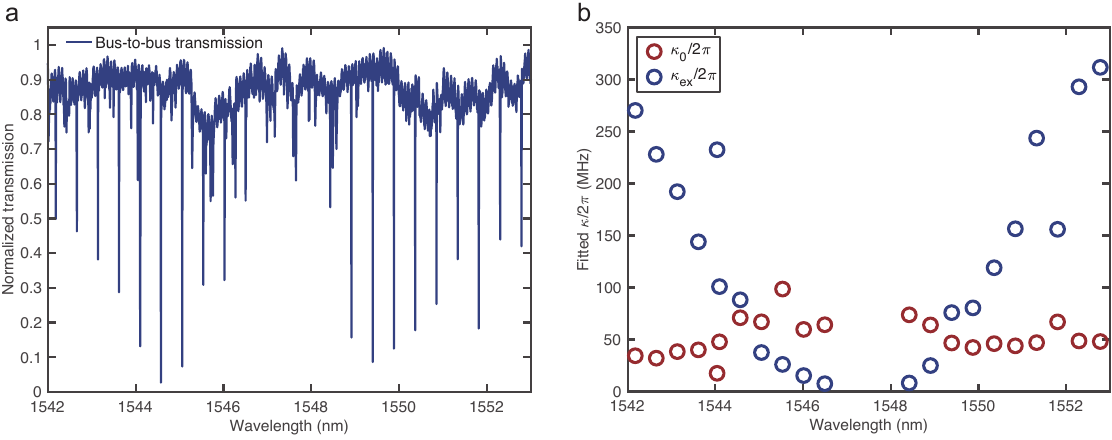}
  \caption{\textbf{Optical characterization of the Raman-suppression structure.} (a) Measured bus-to-bus transmission spectrum in the working band. (b) Intrinsic and external optical decay rates obtained by fitting the measured resonances. The external coupling minimum near the operating wavelength preserves pump-mode enhancement.}
  \label{fig:SI_raman_suppression_rates}
\end{figure}

\Fref{fig:SI_raman_suppression_rates} shows the optical resonances in the working band of the Raman-suppressed device. The fitted intrinsic loss rate remains relatively uniform, whereas the external coupling rate varies strongly with wavelength and reaches a minimum near the operating wavelength. This is consistent with the intended interferometric loading: the pump resonance retains a narrow loaded linewidth while the wavelength-dependent coupling provides substantially stronger loading in the Raman-gain band.

\subsection{OPO-limited operation with Raman suppression}

\begin{figure}[htbp]
  \centering
  \includegraphics[width=1.0\linewidth]{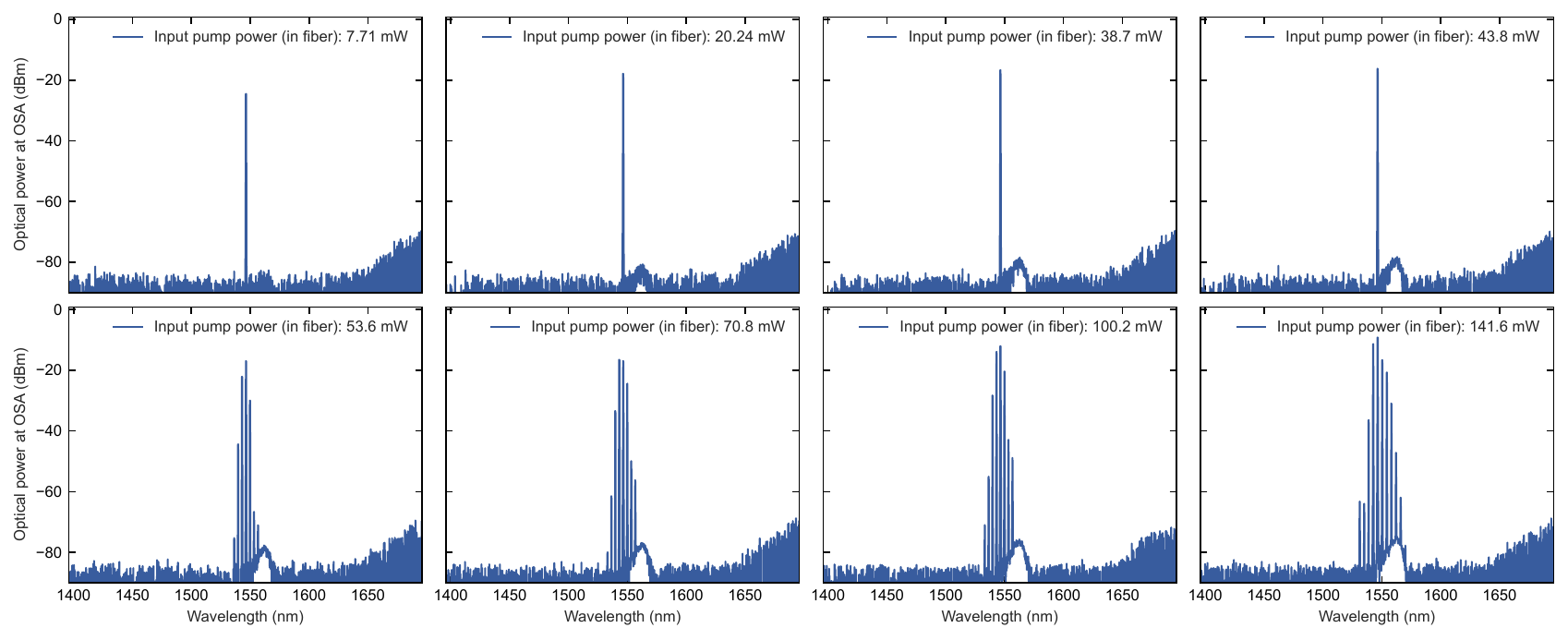}
  \caption{\textbf{Optical parametric oscillation in the Raman-suppressed device.} Discrete coherent sidebands are first observed at an in-fiber pump power of $53.6~\mathrm{mW}$ and become more pronounced at higher powers, showing the onset of Kerr-mediated OPO.}
  \label{fig:SI_opo_spectra}
\end{figure}

We measured the output optical spectra of the Raman-suppressed device at various in-fiber pump powers, as shown in \fref{fig:SI_opo_spectra}. Optical parametric oscillation (OPO) sidebands are first observed at $53.6~\mathrm{mW}$ and become more pronounced at higher optical pump powers.

\section{Experiment methods of transduction measurement}

Due to the symmetry nature of the Hamiltonian for cavity electro-optics, the transduction process is bidirectional, i.e., the microwave signal can be converted to an optical signal and vice versa.
The conversion efficiency for either up-conversion (microwave to optics) or down-conversion (optics to microwave) should be equal, and the conversion bandwidth is determined by the smaller linewidth in the system, i.e., the optical linewidth in our case.

\begin{figure}[htbp]
	\centering
	\includegraphics[width=1.0\linewidth]{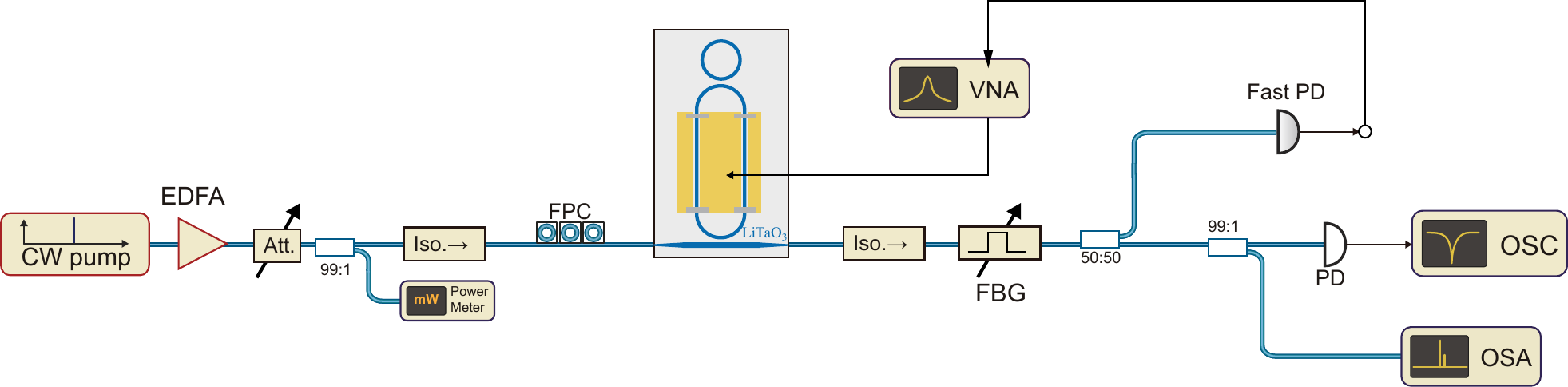}
	\caption{\textbf{Coherent up-conversion (mmWave-to-optical) measurement setup.} A continuous-wave optical pump is amplified and coupled into the electro-optic transducer, while a vector network analyzer (VNA) drives the microwave resonator. The up-converted anti-Stokes sideband beats with the residual pump on a fast photodiode, and the resulting electrical response is returned to the VNA. A fiber Bragg grating (FBG) suppresses the Stokes sideband and partially attenuates the pump to avoid photodiode saturation. An optical spectrum analyzer (OSA) and an auxiliary photodiode--oscilloscope channel monitor the optical output and calibrate the absolute photon-number conversion efficiency. EDFA, erbium-doped fiber amplifier; Att., attenuator; Iso., optical isolator; FPC, fiber polarization controller; PD, photodiode; OSC, oscilloscope.}
	\label{SI_setup_up_conversion}
\end{figure}

We perform coherent response measurement for both up-conversion and down-conversion using vector-network-analyzer (VNA).
\Fref{SI_setup_up_conversion} presents the detailed experiment setup for up-conversion. 
The up-conversion measurement is straightforward; the VNA output is directly connected to the microwave resonator, and the up-converted optical sideband is detected by a fast-photodiode. 
We use a fiber-Bragg grating (FBG) to furthur suppress the Stokes sideband, to only measure the beat-note between the up-converted anti-Stokes sideband and the pump laser.
The FBG also suppresses part of the pump laser to avoid the saturation of the photodiode.
The measured response in the VNA, i.e., the $S_{21}$, is calibrated by the optical spectrum analyzer (OSA) to know the absolute photon conversion efficiency.

\begin{figure}[htbp]
	\centering
	\includegraphics[width=1.0\linewidth]{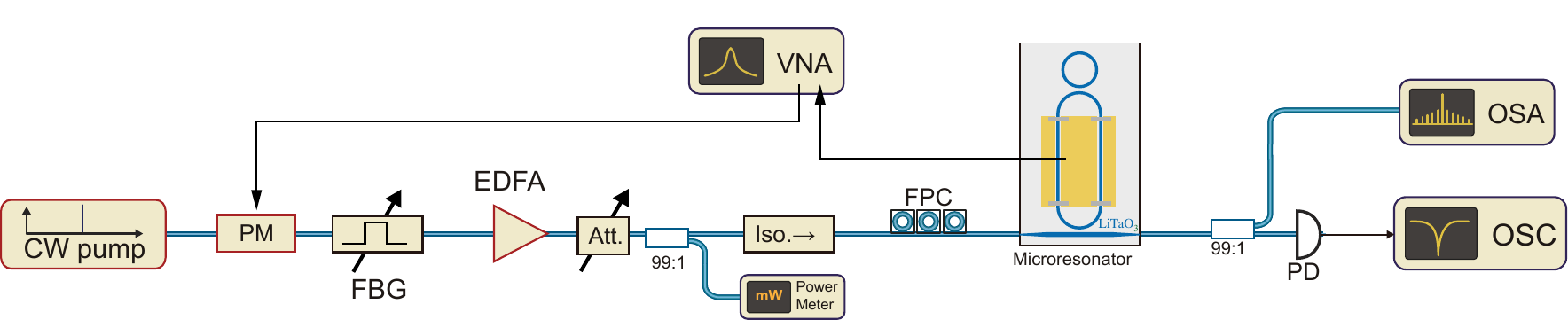}
	\caption{\textbf{Coherent down-conversion (optical-to-mmWave) measurement setup.} A vector network analyzer (VNA) drives a phase modulator (PM) to generate phase-locked optical sidebands from a continuous-wave pump. A fiber Bragg grating (FBG) suppresses the Stokes sideband and partially attenuates the pump. The remaining pump and anti-Stokes tone are amplified, power-adjusted, and coupled into the electro-optic transducer. The generated microwave signal is returned directly to the VNA for measurement. EDFA, erbium-doped fiber amplifier; Att., attenuator; Iso., optical isolator; FPC, fiber polarization controller; PD, photodiode; OSC, oscilloscope.}
	\label{SI_setup_down_conversion}
\end{figure}

\Fref{SI_setup_down_conversion} presents the detailed experiment setup for down-conversion (optical to mmWave), and the coherent response measurement is also performed using a vector-network-analyzer (VNA).
We use a phase modulator to generate phase-locked sidebands. Because of the higher $V_{\pi}$ at high frequencies and the limited VNA output power, the generated sideband is much weaker than the pump.
We use a fiber-Bragg grating (FBG) to filter out the Stokes part of the phase modulator sideband while suppressing part of the pump laser.
We use an optical spectrum analyzer and power meter to calibrate the pump and sideband power after the EDFA (not included in \fref{SI_setup_down_conversion}).
The pump and anti-Stokes sideband is then later input to the transducer chip, and the down-converted microwave signal is directly measured by the VNA. 
In both up and down-conversion response measurement, the probe landing position is chosen to critically couple the microwave resonator, so that the transduction efficiency is maximized.

\section{Experiment methods of intrinsic noise characterization and small signal detection}
\begin{figure}[htbp]
	\centering
	\includegraphics[width=1.0\linewidth]{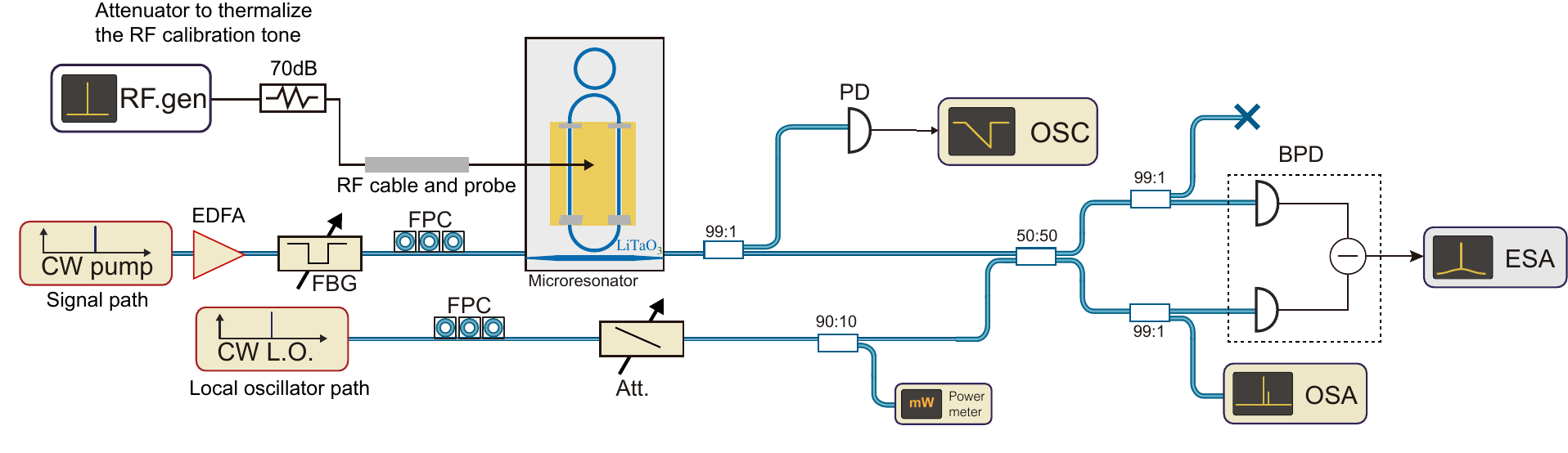}
	\caption{\textbf{{Heterodyne setup for intrinsic noise measurement and small mmWave signal detection.}} 
	RF gen: RF signal generator. PD: Photodiode. BPD: Balanced photodiode. LO: Local oscillator. FBG: Fiber Bragg grating. OSA: Optical spectrum analyzer. 
    ESA: Electrical spectrum analyzer. OSC: Oscilloscope. FPC: Fiber polarization controller. Att: Attenuator.
    A shot-noise limited heterodyne setup is used to probe the weak sideband generated from the up-conversion process. The up-conversion can either be generated from the weak mmWave signal or the intrinsic thermal noise of the microwave resonator. The heterodyne measurement is set to be dominated by the optical shot-noise by applying a large local oscillator (LO) power. The heterodyne response is calibrated by applying a known mmWave calibration tone to the photonics chip. 
    A broadband microwave attenuator with attenuation of 70dB is used to suppress the levitated noise background from the microwave signal generator.
    }
	\label{SI_setup_heterodyne}
\end{figure}

In this section, we detailed the experiment setup we used for intrinsic cavity thermal noise (Johnson-Nyquist noise, and then transduced into the optical domain by the transducer, and the small mmWave signal detection. 
The cavity electro-optic transducer transduces the mmWave signal into optical sidebands, the optical sidebands can then be detected with quantum-limited noise.
To obtain phase-sensitive and quantum-limited measurement, we used a heterodyne setup to detect the optical sideband. 
The complete setup is shown in \fref{SI_setup_heterodyne}. 
We use a total local oscillator (LO) laser power of 18mW incident on the two balanced-photodiode ports to make sure the balanced photodiode (BPD) is shot-noise limited. A fiber-Bragg grating is used before the chip to suppress input amplified spontaneous emission (ASE), while the near-critically coupled pump resonance provides on-chip pump rejection, eliminating the need for a post-chip FBG.

The heterodyne measurement response is calibrated by applying a known mmWave calibration tone to the photonics chip. 
The mmWave calibration tone is generated by the RF signal generator (a vector network analyzer operating in continuous-wave mode) and then passed through a 70 dB attenuator chain to thermalize the calibration tone.
The attenuator chain is necessary as the RF sources we used has noise background roughly 20 dB larger than the thermal background of -174 dBm/Hz at room temperature, so the attenuator chain is used to suppress the noise background from the RF sources, similar to the attenuator chain used in cryogenic measurements.
The absolute RF power is measured by a calibrated mmWave power meter, and the attenuator chain, together with all the cable losses, is measured separately by VNA.
For intrinsic noise measurement, the RF probe is lifted far away from the microwave resonator, so the system only measures the noise from the microwave resonator itself.
In experiments, an auxiliary RF source connected to an open-ended coaxial cable is used to couple an RF tone that matches the optical resonances difference wirelessly, to help allocate the local oscillator laser frequency to the correct optical resonance. The auxiliary RF tone is turned off and the open-ended coaxial cable is removed during the measurement.
\begin{figure}[htbp]
	\centering
	\includegraphics[width=1.0\linewidth]{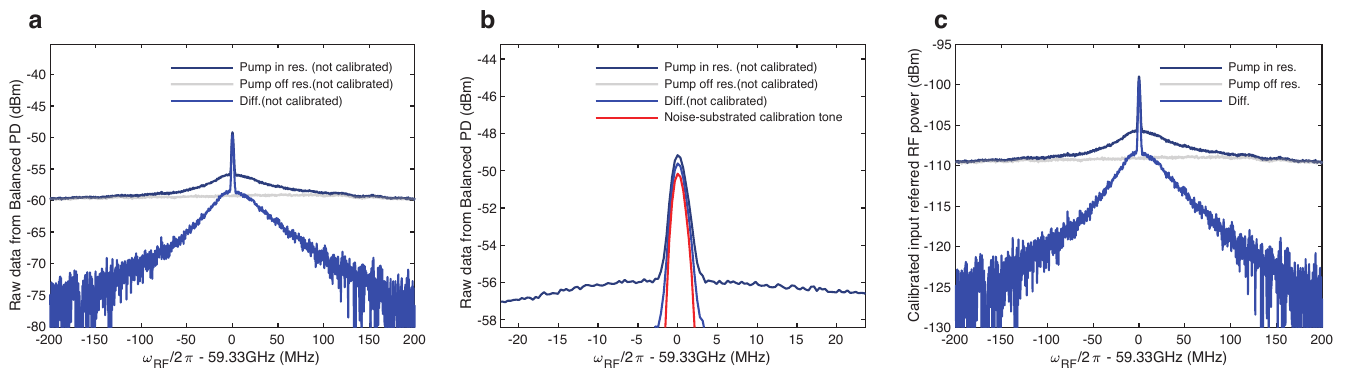}
	\caption{\textbf{{Example data for calibrated thermal noise measurement / small signal detection.}} 
	(a) The raw data from the balanced photodiode output, measured by an electronics spectrum analyzer with a resolution bandwidth of 2MHz (same for (b) and (c)).
    (b) The raw data with the noise background subtracted to obtain the absolute power of the calibration tone, which is then used to calibrate the response of the whole heterodyne measurement chain.
    (c) The final calibrated data.
    }
	\label{SI_example_cal_tone}
\end{figure}
\Fref{SI_example_cal_tone} illustrates the calibration process we used for intrinsic thermal noise measurement and small signal detection. 
Two spectra are measured: one with the pump laser tuned to the resonance, and the other with the pump laser far detuned from the resonance. 
As the shot-noise level is dominated by the local oscillator power, the spectrum acquired with the pump detuned far from resonance is therefore used as the shot-noise baseline. The difference between the pump-in-resonance data and the pump-off-resonance data are thus reflecting the thermal noise, whose spectrum is directly proportional to the intra-cavity microwave photon spectrum.
To obtain an absolute calibrated photon spectrum or the input-referred microwave spectrum, we use calibration tone to calibrated the response for the whole heterodyne measurement chain.

\section{Experiment methods for the photonic wireless communication demonstration}
\begin{figure}[htbp]
	\centering
	\includegraphics[width=0.8\linewidth]{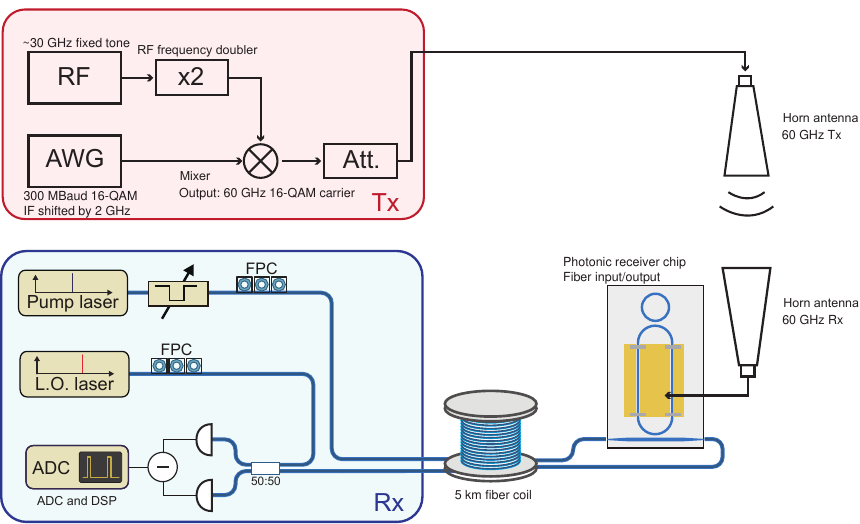}
	\caption{\textbf{Experimental setup for the photonic wireless communication demonstration.} At the transmitter (Tx), a 300~MBaud 16-QAM signal at a 2~GHz intermediate frequency (IF) is mixed with a frequency-doubled ${\sim}30~\mathrm{GHz}$ tone, attenuated, and radiated near 60~GHz by a horn antenna. At the receiver (Rx), an identical horn antenna feeds the photonic chip, which transduces the signal onto a 35~mW optical pump. After transmission through a 5~km fiber spool, the optical signal is combined with a local oscillator (LO) for balanced heterodyne detection and digital signal processing (DSP). AWG, arbitrary waveform generator; FPC, fiber polarization controller; ADC, analog-to-digital converter.}
	\label{fig:SI_wireless_setup}
\end{figure}

\Fref{fig:SI_wireless_setup} presents the experimental setup for the electronic transmitter (Tx) and the photonic receiver (Rx) frontend used in the wireless communication demonstration.

At the receiver, the optical backend (blue box) is identical to that used for the intrinsic-noise measurement. The only difference is that the optical power must be controlled to avoid nonlinearities in the 5~km fiber spool; therefore, we use a moderate direct laser output of 35~mW without an EDFA. The on-chip microwave input port is connected to a horn antenna (Gwave, GHA392596-WR19-25, 25~dBi gain) through a coaxial cable and an RF probe. Optical coupling to the photonic chip is provided by lensed fibers.

At the transmitter, an electronic chain synthesizes a 300~MBaud 16-QAM signal with a carrier frequency near 60~GHz, which is transmitted through an identical horn antenna. Owing to the limited space on the experimental bench, the two antennas are placed close together. An RF attenuator is therefore used to limit the Tx power until no nonlinearity is observable in the photonic receiver. In this experiment, the nonlinearity is limited by the input 1-dB compression point ($IP_{1\mathrm{dB}}$) and input third-order intercept point (IIP3) of the balanced photodiode.
The 300~MBaud 16-QAM signal, centered at an intermediate frequency (IF) of 2~GHz, is generated by an arbitrary waveform generator (Keysight M8195A) and sent to an RF mixer. The RF local oscillator is obtained by frequency-doubling the output of a 30~GHz RF generator (Rohde \& Schwarz SMB100A); its frequency is tuned to align the transmitted signal with the photonic receiver bandwidth. The mixing process also generates an idler, which lies outside the photonic receiver bandwidth.

\begin{figure}[htbp]
	\centering
	\includegraphics[width=1.0\linewidth]{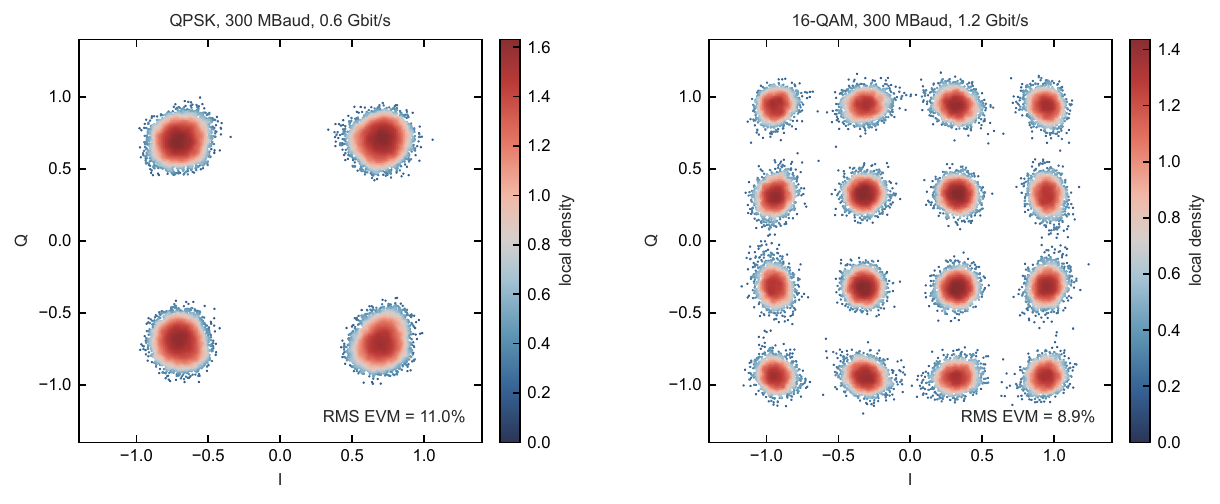}
	\caption{\textbf{Recovered constellations for the photonic wireless communication demonstration.} Recovered QPSK (left) and 16-QAM (right) constellations at 300~MBaud, corresponding to data rates of 0.6 and 1.2~Gbit/s, respectively. The measured RMS EVM values are 11.0\% for QPSK and 8.9\% for 16-QAM. Color indicates the local symbol density.}
	\label{fig:SI_wireless_recovered_constellations}
\end{figure}

Because the pump and optical local-oscillator lasers are not mutually phase-locked, the receiver-side heterodyne IF is not fixed. We therefore recover the IF from the fourth-power spectrum of the analytic signal:
\begin{equation}
\hat{f}_{\mathrm{IF}}=\frac{1}{4}f_{\mathrm{peak}}\!\left(\left|\mathrm{FFT}\!\left\{x_a^4[n]\right\}\right|^2\right).
\end{equation}
Here, $x_a[n]$ is the analytic signal constructed from the digitized balanced-photodiode waveform using a Hilbert transform, and $f_{\mathrm{peak}}(\cdot)$ denotes the frequency of the largest spectral peak.
For square 16-QAM, the nonzero fourth moment produces a spectral component at four times the IF. We use the recovered IF to digitally downconvert and low-pass filter the signal.
\Fref{fig:SI_wireless_recovered_constellations} shows the recovered QPSK and 16-QAM constellations.

We calculate the root-mean-square error vector magnitude (RMS EVM) from the recovered symbols and their nearest ideal constellation points as follows:
\begin{equation}
\mathrm{EVM}_{\mathrm{rms}}(\%)=100\sqrt{\frac{\sum_{k=1}^{N}\left|x_k-s_k\right|^2}{\sum_{k=1}^{N}\left|s_k\right|^2}}.
\end{equation}
Here, $x_k$ is a recovered symbol and $s_k$ is its nearest ideal constellation point. For Gray-coded QPSK and 16-QAM under an additive white Gaussian noise approximation, a pre-FEC BER threshold of $3.8\times10^{-3}$ corresponds to approximate RMS-EVM limits of $37.5\%$ and $17.4\%$, respectively. The measured values of $11.0\%$ and $8.9\%$ are well below these limits.

\bibliography{refs}
\bibliographystyle{apsrev4-2}